\documentclass[twocolumn,english,nofootinbib]{revtex4-1}
\usepackage{fourier}

\usepackage{luximono}

\usepackage[T1]{fontenc}
\usepackage[latin9]{inputenc}
\usepackage{amsmath}
\usepackage{amssymb}
\usepackage{graphicx}
\usepackage{esint}

\makeatletter

\providecommand{\tabularnewline}{\\}

\usepackage{amsmath}

\makeatother

\usepackage{babel}
\begin{document}

\title{{\LARGE{}Plexcitons: Dirac points and topological modes}}

\author{Joel Yuen-Zhou$^{1}$, Semion K. Saikin$^{2,3}$, Tony H. Zhu$^{4,5}$,
Mehmet C. Onbasli$^{6}$, Caroline A. Ross$^{6}$, Vladimir Bulovic$^{5,7}$,
and Marc A. Baldo$^{5,7}$}

\affiliation{$^{1}$Department of Chemistry and Biochemistry, University of California
San Diego, La Jolla, CA, USA.}

\email{joelyuen@ucsd.edu}

\selectlanguage{english}%

\affiliation{$^{2}$Department of Chemistry and Chemical Biology, Harvard University,
Cambridge, MA, USA.}

\affiliation{$^{3}$Department of Physics, Kazan Federal University, Kazan 420008,
Russian Federation}

\affiliation{$^{4}$Department of Physics, Massachusetts Institute of Technology,
Cambridge, MA, USA.}

\affiliation{$^{5}$Center for Excitonics, Research Laboratory of Electronics,
Massachusetts Institute of Technology, Cambridge, MA, USA.}

\affiliation{$^{6}$Department of Materials Science and Engineering, Massachusetts
Institute of Technology, Cambridge, MA, USA.}

\affiliation{$^{7}$Department of Electrical Engineering and Computer Science,
Massachusetts Institute of Technology, Cambridge, MA, USA.}
\begin{abstract}
Plexcitons are polaritonic modes that result from the strong coupling
between excitons and plasmons. We consider plexcitons emerging from
the interaction of excitons in an organic molecular layer with surface
plasmons in a metallic film. We predict the emergence of Dirac cones
in the two-dimensional bandstructure of plexcitons due to the inherent
alignment of the excitonic transitions in the organic layer. These
Dirac cones may open up in energy by simultaneously interfacing the
metal with a magneto-optical layer and subjecting the whole system
to a perpendicular magnetic field. The resulting energy gap becomes
populated with topologically protected one-way modes which travel
at the interface of this plexcitonic system. Our theoretical proposal
suggests that plexcitons are a convenient and simple platform for
the exploration of exotic phases of matter as well as of novel ways
to direct energy flow at the nanoscale.
\end{abstract}

\maketitle
When UV-visible light is absorbed by an organic molecular aggregate,
it promotes molecules from their ground to their excited electronic
states. The resulting excitations, known as excitons, can migrate
between molecules via a mixture of coherent and incoherent processes
\cite{semion_excitonics}. Understanding and controlling how this
migration of energy occurs is a fundamental problem of chemistry and
physics of condensed phases. Furthermore, it is also a technological
problem which is relevant to the development of efficient organic
solar cells and light-emitting devices as well as all-optical circuitry
\cite{baldo2009optical}. Many strategies to enhance the motion of
excitons exist, a particularly interesting one being where they couple
to surface plasmons (SPs) \cite{barnes}. In such strategy, the spatial
coherence of plasmons assists the transport of an exciton across lengthscales
that are dozens of times larger than regular exciton diffusion lengths.
When the coupling is strong, meaning that the energy exchange between
the exciton and plasmon is faster than the respective decay times
\cite{PhysRevLett.93.036404,govorov_nanoletters,sukharev,gonzalez_tudela,torma2015strong},
plexcitons (a class of polaritons) emerge \cite{halas,nordlander},
and energy can migrate ballistically over the coherence length of
the plasmon. Besides their usefulness in energy transport, organic
plexcitons are promised to be an exciting room-temperature ''laboratory''
for the study of light-matter and many-body interactions at the nanoscale
\cite{torma2015strong}. In this letter, we propose novel plexcitonic
phenomenona which should be readily realizable with current experimental
capabilities: Dirac cones and topologically nontrivial plexcitons
which travel along preferred directions at the edge of an organic
layer.

Topologically nontrivial states of matter have been a topic of great
interest in condensed matter physics owing to the discoveries of the
Quantum Hall Effect \cite{qhe}, and more recently, of topological
insulators \cite{kane_mele,Bernevig15122006}. The systems supporting
these states are characterized by topological invariants \cite{bernevigbook},
integer numbers that remain unchanged by weak perturbations. Physically,
a nontrivial topological invariant signals the presence of one-way
edge modes that are immune against moderate amount of disorder. Even
though these phenomena were first conceptualized for fermions in solids,
they have been successfully generalized to bosonic systems including
photons in waveguides \cite{haldane_raghu,khanikaev,lu_soljacic},
ring resonator arrays \cite{hafezi_2}, ultracold atoms in optical
lattices \cite{aidelsburger2015measuring}, and classical electric
circuits \cite{simon_circuit,PhysRevLett.114.173902}. Furthermore,
we have recently proposed an excitonic system consisting of a two-dimensional
porphyrin film which becomes topologically nontrivial in the presence
of a magnetic field \cite{yuen_topological}. A challenging feature
of that proposal is the requirement of large magnetic fields (>10
T) and cryogenic temperatures to preserve exciton coherence. Even
though we do not discourage the experimental implementation of the
latter, we consider a conceptually different platform which, by using
plexcitons, avoids the use of large magnetic fields and, under appropriate
circumstances, may work at room temperature. In the last year, Dirac
and topological polaritons have been proposed in other contexts, such
as optomechanical arrays and inorganic materials in optical cavities.
All of these works share a common goal to ours, which is the design
of exotic modes in strongly coupled light-matter systems. However,
there are substantial qualitative and quantitative differences arising
from the choices of material (organic exciton \emph{vs} inorganic
exciton \cite{dirac_polariton,karzig,karzig_garden} or mechanical
mode \cite{manquardt2,manquardt}) and electromagnetic (SP \emph{vs}
microcavity \cite{dirac_polariton,karzig,karzig_garden} or photonic
crystal \cite{manquardt2,manquardt}) excitations. Hence, the physics
involved in our plexciton system contrasts with the other proposals
in terms of the energy and lengthscales involved in the excitations,
the magnitude of the couplings, the generation of nontrivial topology,
and the experimental conditions for its realization. Organic excitons
differ from their inorganic counterparts in that they have large binding
energies and are associated with large transition dipole moments.
SP electromagnetic fields are strongly confined compared to their
microcavity counterparts. The combination of all these properties
in the organic plexciton context gives rise to strong light-matter
interactions even at room temperature and in an ``open cavity''
setup \cite{torma2015strong}.

\begin{figure}
\centering{}\includegraphics[scale=0.2]{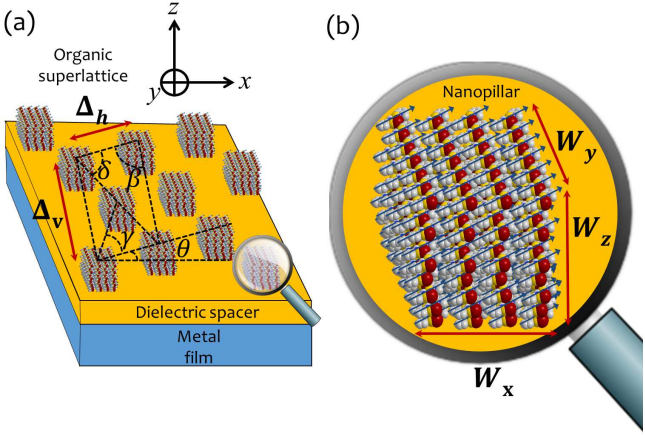}\protect\caption{\emph{Plexciton setup.} It consists of a plasmonic metal film, a dielectric
spacer, and an organic layer. The latter is taken to be a monoclinic
superlattice which makes an angle $\theta$ with respect to the $x$
axis, and is further characterized by angles $\beta,\gamma,\delta$
as well as distances between nanopillars $\Delta_{h}$ and $\Delta_{v}$.
Each element in the superlattice is a nanopillar of organic emitters
of dimensions $W_{i}$ along each axis. When the density of emitters
is big enough, the coupling between the excitons in the organic layer
and the surface plasmons (SPs) in the metal becomes larger than their
linewidths, giving rise to polaritonic eigenmodes that are superpositions
of excitons and plasmons, or more succintly, plexcitons. In this article,
we shall also consider the case where the dielectric spacer is a magneto-optical
(MO) material. \label{fig:Plexciton-setup.}}
\end{figure}

The setup of interest is depicted in Fig. \ref{fig:Plexciton-setup.}.
It consists of three layers, from bottom to top: a plasmonic metal
modelled with a Drude permittivity ($\epsilon_{m}(\omega)=\epsilon_{\infty}-\frac{\omega_{P}^{2}}{\omega^{2}}$,
$\epsilon_{\infty}\sim4$, $\omega_{P}\sim9$~eV, representative
parameters for Ag), an $a=10\,\mbox{nm}$ thick dielectric spacer
($\epsilon_{d}\sim1$), and an organic layer ($\epsilon_{org}\sim1$).
The spacer is placed to avoid quenching of the excitons by the surface
plasmons (SPs) upon close contact, but as we shall see, we will also
consider the case where it is a magneto-optical (MO) material.

A quantum mechanical description of this setup is given by a Hamiltonian,

\begin{equation}
H=H_{exc}+H_{SP}+H_{exc-SP},\label{eq:total_hamiltonian}
\end{equation}
where each of the terms denotes the energetic contributions from the
excitons in the organic layer, the SPs, and the coupling between them.
More specifically ($\hbar=1$),

\begin{subequations}\label{eq:eqntriad}
\begin{eqnarray}
H_{exc} & = & \sum_{\boldsymbol{n},s}\omega_{\boldsymbol{n}}\sigma_{\boldsymbol{n}}^{\dagger}\sigma_{\boldsymbol{n}}+\sum_{\boldsymbol{n}\neq\boldsymbol{n'}}(J_{\boldsymbol{n}\boldsymbol{n'}}\sigma_{\boldsymbol{n}}^{\dagger}\sigma_{\boldsymbol{n'}}+\mbox{h.c.}),\label{eq:H_mol}\\
H_{SP} & = & \sum_{\boldsymbol{k}}\omega(\boldsymbol{k})a_{\boldsymbol{k}}^{\dagger}a_{\boldsymbol{k}},\label{eq:H_SP}\\
H_{exc-SP} & = & \sum_{\boldsymbol{k},\boldsymbol{n}}\mathcal{J}_{\boldsymbol{k}\boldsymbol{n}}a_{\boldsymbol{k}}\sigma_{\boldsymbol{n}}^{\dagger}e^{i\boldsymbol{k}\cdot\boldsymbol{r_{n}}}+\mbox{h.c.}\label{eq:H_mol_SP}
\end{eqnarray}

\label{eq:eqntriad}\end{subequations}\noindent Here, in order to
obtain a specific shape of the exciton energy dispersion (see below),
we have taken the organic layer to be an oblique superlattice of organic
nanopillars. The superlattice is defined by the distances $\Delta_{h}$
(horizontal) and $\Delta_{v}$ (vertical), as well as angles $\beta$,
$\gamma$, $\delta$, and $\theta$ (Fig. \ref{fig:Plexciton-setup.}a);
the nanopillars are in turn rectangular parallelepipeds of densely
packed organic chromophores (assuming a van der Waals distance between
chromophores of 0.3 nm, $\rho_{np}=$37 chromophores/nm$^{3}$) with
volume $V_{np}=W_{x}W_{y}W_{z}$ (Fig. \ref{fig:Plexciton-setup.}b),
obtained from growing a J-aggregate film \cite{kuhn_chapter,bradley_bulovic}.
Given this preamble, $\sigma_{\boldsymbol{n}}^{\dagger}(\sigma_{\boldsymbol{n}})$
and $a_{\boldsymbol{k}}^{\dagger}(a_{\boldsymbol{k}})$ label the
creation (annihilation) operators for the collective exciton at the
$\boldsymbol{n}$-th nanopillar and the $\boldsymbol{k}$-th SP mode,
respectively, where $\boldsymbol{n}$ and $\boldsymbol{k}$ are (two-dimensional)
in-plane vectors denoting a position and a wavevector, respectively.
J-aggregation of chromophores results in a collective transition dipole
$\boldsymbol{\mu}_{\boldsymbol{n}}$ at an excitation energy $\omega_{\boldsymbol{n}}$,
while the dispersion energy of the $\boldsymbol{k}$-th SP mode is
denoted $\omega(\boldsymbol{k})$. Dipolar interactions $J_{\boldsymbol{n}\boldsymbol{n'}}$
couple the various nanopillars. The coupling between the exciton and
the SP depends on the average in-plane location $\boldsymbol{r}_{\boldsymbol{n}}$
of the $\boldsymbol{n}$-th nanopillar, and is also dipolar in nature,

\begin{equation}
\mathcal{J}_{\boldsymbol{k}\boldsymbol{n}}=\sqrt{\frac{\omega(\boldsymbol{k})}{2\epsilon_{0}SL_{\boldsymbol{k}}}}e^{-\alpha_{org0}(\boldsymbol{k})\bar{z}(\boldsymbol{k})}\boldsymbol{\mu}_{\boldsymbol{n}}\cdot\boldsymbol{E}(\boldsymbol{k}).\label{eq:J_kn}
\end{equation}
Here, $L_{\boldsymbol{k}}$ denotes a vertical ($z$-direction) mode-length
of the SP which guarantees that the total energy of a SP prepared
at the $\boldsymbol{k}$th mode is quantized at the energy $\omega(\boldsymbol{k})$,
$\boldsymbol{E}(\boldsymbol{k})$ is an appropriately scaled electric
field of the corresponding mode, and $e^{-\alpha_{org0}(\boldsymbol{k})\bar{z}(\boldsymbol{k})}$
yields a mean-field average of the interaction of the evanescent SP
field (with decay constant $\alpha_{org0}$ in the organic layer)
over the chromophores at different vertical positions of the nanopillar;
it optimizes the interaction such that one may assume the nanopillar
is a point-dipole located at the mean height $\bar{z}(\boldsymbol{k})$.
The latter average renders the originally 3D system into an effectively
2D one. Detailed derivations of Eqs. (\ref{eq:eqntriad}) and (\ref{eq:J_kn})
are available in the Supplementary Information (SI)-III.

Assuming perfect periodicity of the superlattice ($\omega_{\boldsymbol{n}}=\bar{\omega}$,
$\boldsymbol{\mu}_{\boldsymbol{n}}=\boldsymbol{\mu}$) and only nearest
and next-nearest neighbor (NN and NNN) dipolar interactions, we can
re-express $H_{exc}$ (Eq. (\ref{eq:H_mol})) in terms of $\boldsymbol{k}$
modes. As explained in SI-IIIA, it is possible to approximate $H_{exc}$
up to $O(|\boldsymbol{k}|^{2})$ as arising from an \emph{effective}
simplified rectangular (rather than monoclinic) lattice aligned along
the $x$, $y$ axes, and with NN interactions only. This is a reasonable
thing to do, as the topological effects we are interested arise at
relatively long wavelengths. Within this approximation, we may construct
Fourier modes $\sigma_{\boldsymbol{n}}^{\dagger}=\frac{1}{\sqrt{N_{x}N_{y}}}\sum_{\boldsymbol{k'}}\sigma_{\boldsymbol{k'}}^{\dagger}e^{-i\boldsymbol{k'}\cdot\boldsymbol{r_{n}}}$
(here, $N_{i}$ is the effective number of nanopillars along the $i$-th
direction) and rewriting Eq. (\ref{eq:total_hamiltonian}) in reciprocal
space, we obtain $H=\sum_{\boldsymbol{k}}H_{\boldsymbol{k}}$,

\begin{eqnarray}
H_{\boldsymbol{k}} & = & \underbrace{\omega_{exc,\boldsymbol{k}}\sigma_{\boldsymbol{k}}^{\dagger}\sigma_{\boldsymbol{k}}}_{\equiv H_{exc,\boldsymbol{k}}}+\underbrace{\omega(\boldsymbol{k})a_{\boldsymbol{k}}^{\dagger}a_{\boldsymbol{k}}}_{\equiv H_{SP,\boldsymbol{k}}}+\underbrace{\Big[\mathcal{J}(\boldsymbol{k})a_{\boldsymbol{k}}\sigma_{\boldsymbol{k}}^{\dagger}+\mbox{h.c.}\Big]}_{\equiv H_{exc-SP,\boldsymbol{k}}},\label{eq:H_k}
\end{eqnarray}
where,

\begin{subequations}\label{eq:H_k_components}
\begin{eqnarray}
\omega_{exc,\boldsymbol{k}} & = & \bar{\omega}_{eff}+2J_{x}\mbox{cos}(k_{x}\Delta_{x})+2J_{y}\mbox{cos}(k_{y}\Delta_{y}),\label{eq:H_exc_k}\\
\mathcal{J}(\boldsymbol{k}) & = & \sqrt{\Bigg(\frac{N_{x}N_{y}}{S}\Bigg)\Bigg(\frac{\omega(\boldsymbol{k})}{2\epsilon_{0}L_{\boldsymbol{k}0}}\Bigg)}e^{-\alpha_{org0}(\boldsymbol{k})\bar{z}(\boldsymbol{k})}\boldsymbol{\mu}\cdot\boldsymbol{E}(\boldsymbol{k}),\label{eq:J(k)_bis}
\end{eqnarray}

\label{eq:H_and_J}\end{subequations}\noindent where $J_{i}$ and
$\Delta_{i}$ are effective NN hoppings and spacings along the $i$-th
axis, respectively, $\bar{\omega}_{eff}$ is the effective nanopillar
site energy, and we have taken $\boldsymbol{\mu}_{\boldsymbol{n}}=\boldsymbol{\mu}$
for all $\boldsymbol{n}$. Eq. (\ref{eq:J(k)_bis}) denotes the $\boldsymbol{k}$-dependent
coupling between a plasmonic mode and a collective exciton state throughout
the organic layer. It features the interaction between nanopillars
and the plasmonic mode, which is proportional to the square root of
the total number of nanopillars $\sqrt{N_{x}N_{y}}$ coherently coupled
to the plasmon (compare with Eq. (\ref{eq:J_kn})) \cite{herrera2014quantum,coles2014strong,spano2015optical}.

\begin{figure*}
\centering{}\includegraphics[scale=0.3]{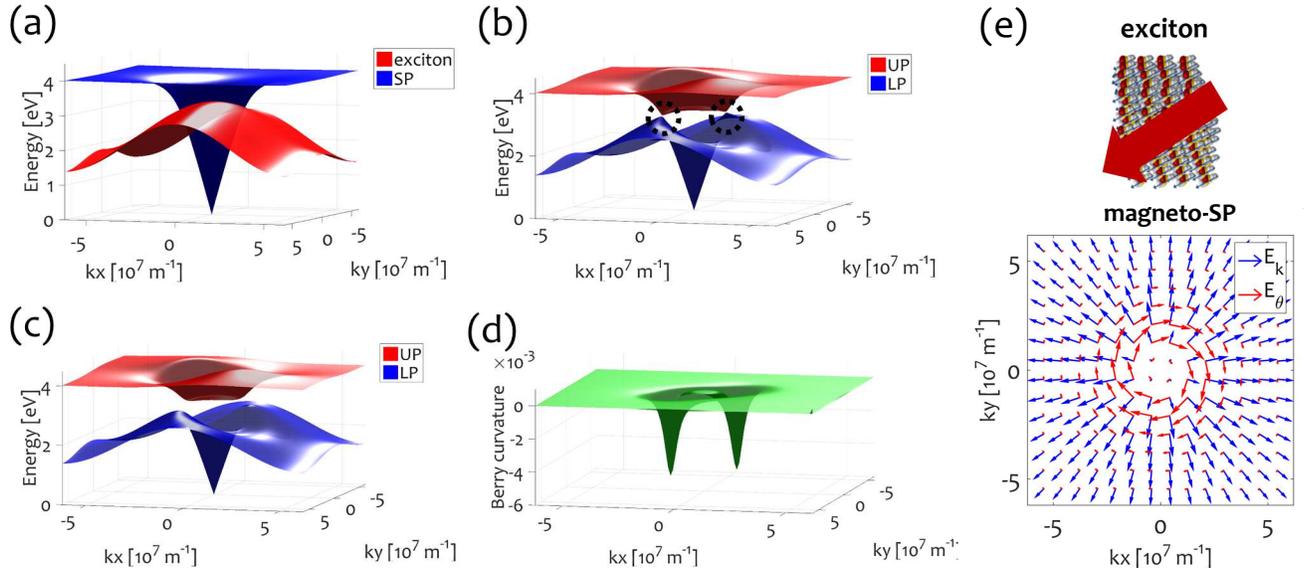}\protect\caption{\emph{Bulk plexciton properties. }Dispersion relations:\emph{ }(a)
for SP and exciton (organic layer) modes independently, (b) when they
couple in the absence of the MO effect, yielding lower (LP) and upper
(UP) plexciton branches which feature two Dirac cones (dashed black
circles), and (c) when they couple in the presence of the MO effect
($g=0.3$), lifting the Dirac cones. (d) Berry curvature associated
with the LP in (c). (e) Physical mechanism for the appearance of plexciton
Dirac points: (top) collective (in-plane) exciton transition dipole
for a nanopillar; (bottom) magnitude of the electric field of magneto-SP
modes as a function of wavevector. In the absence of the MO effect,
only the wavevector-parallel components $\boldsymbol{E}_{\boldsymbol{k}}$(blue)
are present. Thus, the nanopillars experience no coupling with modes
whose wavevectors are perpendicular to the transition dipole. Along
these directions, degeneracies between the SP and the exciton modes
are not lifted, yielding two plexciton Dirac points. Nonzero tangential
components $\boldsymbol{E}_{\boldsymbol{\theta}}$ (red) emerge upon
inclusion of the MO effect, lifting these degeneracies. \label{fig:dispersions}}
\end{figure*}

For our simulation, we choose the length parameters $\Delta_{h}=100$,
$\Delta_{v}=88$, $W_{x}=7.5$, $W_{y}=50$, $W_{z}=70$ nm, and take
$\beta=13.1^{o}$ (see SI-IIIA for explanation of choice of parameters).
Denoting the transition dipole $\boldsymbol{\mu}=\mu\hat{\boldsymbol{\boldsymbol{\mu}}}$,
we estimate $\mu=\sqrt{N_{np}}\times10\,\mbox{D}=9855\,\mbox{D}$,
where $N_{np}=\rho_{np}V_{np}$ is the number of chromophores in the
nanopillar; $\hat{\boldsymbol{\boldsymbol{\mu}}}$ is the \emph{in-plane}
unit vector making an angle of $\alpha=223^{o}$) with respect to
$\hat{\boldsymbol{x}}$, that is, $\hat{\boldsymbol{\boldsymbol{\mu}}}=\mbox{cos}\alpha\hat{\boldsymbol{x}}+\mbox{sin}\alpha\hat{\boldsymbol{y}}$.
Choosing the simulation values for $H_{exc}$ in Eq. (\ref{eq:H_exc_k})
to be $\Delta_{x}=\Delta_{y}=50\,\mbox{nm}$, we get the effective
parameters $\bar{\omega}_{eff}=2.15\,\mbox{eV}$, $J_{x}=362\,\mbox{meV}$,
and $J_{y}=107\,\mbox{meV}$. Fig. \ref{fig:dispersions}a shows superimposed
dispersion curves for $H_{exc,\boldsymbol{k}}$ and $H_{SP,\boldsymbol{k}}$
independently (taking $H_{exc-SP}=0$). The superlattice has been
intentionally constructed to obtain $J_{x},\,J_{y}>0$ yielding a
``dome''-like dispersion for $H_{exc,\boldsymbol{k}}$, \emph{i.e.}
it features a maximum at $\boldsymbol{k}=0$, behaving as a 2D ``H-aggregate''
\cite{semion_excitonics}. As for $H_{SP,\boldsymbol{k}}$, its dispersion
has the shape of a rotationally symmetric ``fountain,'' and is nothing
more than the 2D rendering of the standard 1D textbook result \cite{novotny},
featuring a linear dependence of the energy at short wavevectors and
a plateau at large ones, indicating excitations that are qualitatively
closer to light or to charge oscillations in the metal, respectively.

Fig. \ref{fig:dispersions}b shows the two-plexciton-branch bandstructure
arising from the diagonalization of Eq. (\ref{eq:H_k}). We notice
that anticrossing gaps are noticeably opened in the vicinity of where
the dispersion curves for $H_{exc,\boldsymbol{k}}$ and $H_{SP,\boldsymbol{k}}$
used to cross in Fig. \ref{fig:dispersions}a. This is a signature
of exciton-SP coupling $H_{exc,SP,\boldsymbol{k}}$. Given a fixed
wavevector direction, whenever these anticrossings occur, the lower-plexciton
(LP) branch starts off as being mostly SP at short $\boldsymbol{k}$
values, but parametrically morphs into mostly exciton at large ones;
the opposite happens with the upper-plexciton (UP) branch. However,
probably the most striking feature of Fig. \ref{fig:dispersions}b
is the appearance of two Dirac cones (see dashed circles) at critical
wavevectors $\boldsymbol{k}^{*}$ in which anticrossings do not happen.
Their onset coincides with the directions at which $\boldsymbol{k}$
is orthogonal to $\boldsymbol{\mu}$. Their physical origin is explained
in Fig. \ref{fig:dispersions}e which, in its top panel, shows the
in-plane electric field for the $\boldsymbol{k}$-th SP mode, $\boldsymbol{E}_{\perp}(\boldsymbol{k})\equiv\boldsymbol{E}(\boldsymbol{k})-\boldsymbol{E}(\boldsymbol{k})\cdot\hat{\boldsymbol{z}}\hat{\boldsymbol{z}}$
and is purely parallel to $\boldsymbol{k}$, $\boldsymbol{E}_{\perp}(\boldsymbol{k})=\boldsymbol{E}_{\boldsymbol{k}}(\boldsymbol{k})$
(blue vector field). If all the dipoles in the organic layer are aligned
(in-plane) along $\boldsymbol{\mu}$, their projection onto the SP
electric field, which gives rise to the exciton-SP coupling (see Eq.
(\ref{eq:J(k)_bis})), will wax and wane as a function of the azimuthal
angle $\varphi$ between the fixed dipole and the varying SP wavevector
according to $\boldsymbol{\mu}\cdot\boldsymbol{E}(\boldsymbol{k})\propto\mbox{cos}\varphi$.
Clearly, this projection will vanish if $\boldsymbol{k}$ happens
to be orthogonal to $\boldsymbol{\mu}$, that is, at the special angles
$\varphi=\frac{\pi}{2},\frac{3\pi}{2}$, so that any degeneracy between
the exciton and the SP modes will remain unlifted along these directions.
From this physical picture, we can extract the two essential ingredients
for the emergence of the plexciton Dirac cones. First, the dipoles
need to be aligned to create an anisotropic exciton-SP coupling as
a function of $\varphi$. Second, this alignment needs to be horizontal,
as a vertical component of the dipole will couple to the vertical
component of the electric field $\boldsymbol{E}(\boldsymbol{k})\cdot\hat{\boldsymbol{z}}\hat{\boldsymbol{z}}$,
and this coupling, unlike its horizontal counterpart, does not vanish
for any $\varphi$. It is important to note that neither of these
requirements requires the use of the superlattice, which will be exploited
for a different purpose (the engineering of topological edge modes,
as explained in the next paragraph). Therefore, a standard organic
molecular crystal with aligned transition dipoles lying on the horizontal
$xy$ plane will suffice. These plexciton Dirac cones which, to our
knowledge, have not been reported in the past, should be easily detectable
by collecting the reflected light spectra upon excitation of the plexcitonic
system in a grating, Otto, or Kretschmann configurations \cite{novotny},
by systematically scanning across $|\boldsymbol{k}|$ and $\varphi$
values. For a general $(|\boldsymbol{k}|,\varphi)$, the spectrum
should consist of two dips as a function of dispersed energy, each
associated with the corresponding eigenenergies of the LP and UP.
Importantly, however, the two dips merge at the Dirac cones. In a
standard plexciton dispersion measurement, one only scans across $|\boldsymbol{k}|$.
Since we are interested in a two-dimensional dispersion, the scan
must also be performed across $\varphi$.

Having elucidated the mechanism for the formation of plexciton Dirac
cones, we proceed to entertain a more ambitious goal. We aim to engineer
topologically protected plexcitons by opening the Dirac cones using
a time-reversal symmetry breaking (TRSB) perturbation \cite{bernevigbook}.
To accomplish this, we now assume that the dielectric spacer has magneto-optical
(MO) properties; that is, upon application of a perpendicular magnetic
field, its permittivity becomes anisotropic, $\epsilon_{d}\to\overleftrightarrow{\epsilon}_{MO}$,

\begin{equation}
\overleftrightarrow{\epsilon}_{MO}=\left[\begin{array}{ccc}
\epsilon_{d} & ig & 0\\
-ig & \epsilon_{d} & 0\\
0 & 0 & \epsilon_{d}
\end{array}\right].\label{eq:e_tensor}
\end{equation}
Materials associated with this dielectric tensor exhibit Faraday effect
and are of great interest in the fabrication of optical isolators
\cite{dionne2005circular}. For the near-IR and visible, yttrium iron
garnets ($\mbox{Y}_{3}\mbox{Fe}_{5}\mbox{O}_{12}$, YIG) substituted
with Bi (BiYIG), Ce (CeYIG) or other rare earths provide high Faraday
rotation with low optical absorption \cite{kuzmiak,Onbasli:14}; alternatively,
one may use an MO active Co-alloy film \cite{temnov2010active} or
multilayer with an insulating layer to avoid quenching of the excitons.
In this article, we are interested in the new SP modes, denoted as
magneto-SP modes, arising at the interface of the plasmonic metal
and the MO layer. The solution for the latter is highly non-trivial,
and we refer the reader to our perturbative solution in the SI-I,II,
which builds a perturbation theory based upon an initial calculation
due to Chiu and Quinn \cite{chiu_quinn}. Fig. \ref{fig:dispersions}c
shows essentially the same calculation as Fig. \ref{fig:dispersions}b,
except for the inclusion of the MO effect. Interestingly, we notice
that the Dirac cones have been lifted. A physical understanding of
the latter phenomenon can be obtained by appealing to Fig. \ref{fig:dispersions}e
again. Within our perturbation theory, the magneto-SP modes differ
from their original SP counterparts in that there are additional tangential
components (red) to the electric field. The clockwise vortex vector
field is a signature of TRSB; it becomes counterclockwise upon change
of direction of the magnetic field. This tangential electric field
is the sole responsible for opening the Dirac cones at the critical
angles $\varphi=\frac{\pi}{2},\frac{3\pi}{2}$, where the original
field (blue) ceased to couple to the excitons. Hence, we have concocted
a situation where anticrossings occur for all azimuthal angles $\varphi$.
To characterize the topology of the resulting bandstructure, we numerically
compute the Berry curvature for each plexciton branch \cite{fukui2005chern};
we show that of the LP in Fig. \ref{fig:dispersions}d. Its integral
with respect to the Brillouin zone is the so-called Chern number $C$,
an integer which, if nonzero, signals a topologically nontrivial phase.
Fig. \ref{fig:dispersions}d clearly shows that this integral is non-vanishing,
and in fact, adds up to $C=-1$ (by the sum rule of Chern numbers,
the upper branch necessarily has $C=1$). Intuitively, it is also
clear that most of the nontrivial topology, and hence, Berry curvature,
is concentrated in the vicinity of what used to be the Dirac cones.
In passing, we note that considerable attention has recently been
given to magneto-SPs where the magnetic field is applied parallel
(instead of perpendicular) to the metal film itself, yielding dispersion
relations which are nonreciprocal \cite{PhysRevLett.100.023902}.
Curiously, this arrangement does not give us the topological states
we are looking for, although it might be intriguing to explore the
connection between these magneto-SPs and the ones exploited in our
present work, arising from a perpendicular magnetic field.

\begin{figure}
\centering{}\includegraphics[scale=0.25]{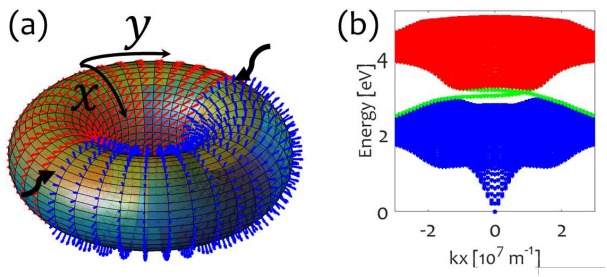}\protect\caption{\emph{Topologically protected edge modes.} (a) Simulation of edge
modes where magneto-SPs are computed in the torus geometry. Two domains
of organic layers are placed on top of it (just like ``two icings''
on a donut). In-plane (red) and out-of-plane (blue) transition dipoles
yield topologically nontrivial and trivial plexcitons, respectively.
Topologically protected one-way plexcitons appear at the interfaces
(thick black arrows). Each interface features a different plexciton
direction of motion. (b) 1D dispersion relation $\omega(k_{x})$ for
the setup in (a). Bulk LPs (blue) and UPs (red) separated by edge
modes (green) featuring positive and negative dispersions, respectively,
and localized along each interface. \label{fig:Topologically-protected-edge}}
\end{figure}

\begin{figure}
\centering{}\includegraphics[scale=0.25]{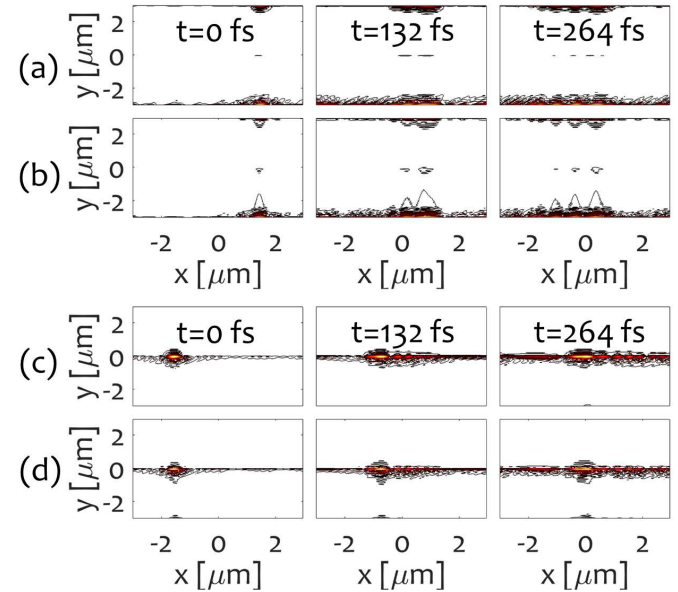}\protect\caption{\emph{Snapshots of dynamics of topologically-protected edge modes.}
(a,b) and (c,d) depict the dynamics of an initially localized plexciton
along the $y=0$ and $y=\pm\frac{L_{y}}{2}=\pm3\,\mu m$ interfaces
of our smiulation. (a,c) and (b,d) show the exciton and magneto-SP
components, respectively. These modes, which are robust to disorder,
have substantial excitonic and magneto-SP components, travel in opposite
directions along different interfaces (see Fig. 3b), and are robust
against disorder. \label{fig:Snapshots-of-dynamics}}
\end{figure}

So far, all the described calculations have been carried out in the
bulk. By virtue of the bulk-boundary correspondence \cite{bernevigbook},
we expect topologically protected one-way edge modes associated with
this setup. In order to compute them, it is convenient to keep periodic
boundary conditions for the magneto-SP modes, yet consider two domains
of excitons on top (Fig. \ref{fig:Topologically-protected-edge}a),
one with (in-plane) dipoles pointing along $\hat{\boldsymbol{\mu}}$
(red dipoles) and the other one with vertical dipoles along $\hat{\boldsymbol{z}}$
(blue dipoles). Pictorially, this setup resembles a \textquotedbl{}donut
with two icings,'' where the donut is the metal with toroidal geometry,
and the two icings are the domains of excitons separated by two interfaces
located at $y=\pm\frac{L_{y}}{2}$ and $y=0$, where $L_{i}$ is the
total width of the simulated sample along $i$ (in our calculations,
we take $L_{x}=L_{y}=6\,\mu\mbox{m}$). The Chern numbers associated
with the bulk LP branch of each domain are $C=-1$ and $C=0$, respectively.
Hence, the plexcitons for the blue domain are topologically trivial.
This can be understood by recalling that no plexciton Dirac points
occur when dipoles are vertically aligned, regardless of the MO effect.
In the limit of no disorder along the $x$-direction, $k_{x}$ is
still a good quantum number, and Fig. \ref{fig:Topologically-protected-edge}b
shows the corresponding plexciton dispersion relation. This bandstructure
is essentially a projection of the gapped 2D bulk bandstructures of
both domains of plexcitons onto one axis $k_{x}$ with additional
states spanning the topological gap between the LP and UP branches.
Inspection of the nature of these mid-gap states reveals that they
have substantial exciton and magneto-SP character, and that they are
precisely the edge states we are searching for: one band has positive
(negative) dispersion and is localized along $y=0$ ($y=\pm\frac{L_{y}}{2}$).
Thus, by preparing a plexciton wavepacket localized along one of the
interfaces, and making sure it is composed of energy states within
this topological gap, one ensures that transport occurs robustly without
much probability of backscattering. The reason being that backscattering
requires coupling between counterpropagating modes which are separated
by a distance $\frac{L_{y}}{2}$, which is large compared to the width
of the corresponding wavefunctions along $\hat{\boldsymbol{y}}$.
Fig. \ref{fig:Snapshots-of-dynamics}(a,b) shows snapshots of the
dynamics associated with these edge states. Panels (a,b) and (c,d)
show a plexciton that starts localized at $x=\frac{L_{x}}{4}$ and
$x=-\frac{L_{x}}{4}$, respectively, and tracks the one-way (to the
left or to the right) nature of their motion within the femtosecond
timescale.

A useful ingredient guaranteeing the robustness of one-way transport
of these edge modes is that they appear within a global gap, the latter
of which is a consequence of our 2D H-aggregate superlattice design.
It is easy to check that if $J_{x}\leq0$ or $J_{y}\leq0$, this global
gap is not guaranteed anymore, and edge modes may become degenerate
with bulk modes. Hence, these two types of modes could readily hybridize,
yielding channels connecting one edge to the other, allowing for backscattering.
Importantly however, the bandstructure remains topologically nontrivial
even in the absence of such global gap, so even if perfect one-way
transport is not observed in these cases, signatures of the latter
may remain. Furthermore, we have shown that a superlattice with in-plane
dipoles and NN and NNN couplings between nanopillars gives rise to
an effective 2D H-aggregate. It is unclear whether this superlattice
is necessary for our goal, or whether a simple organic molecular crystal
can yield a similar behavior once we take into account all the dipolar
contributions, from short to long range. These issues will be explored
in future work. In the mean time, it suffices to note that, as a proof
of concept, a global gap which hosts topologically protected edge
states can be obtained by using an organic superlattice.

Notice that the dispersion of the edge plexcitons is such that a subset
falls within the light cone ($\omega=ck_{x}/\sqrt{\epsilon_{d}}$)
so far field excitation and detection of this fraction is possible
via interaction with the organic layer itself. The rest of the plexcitons
can be probed using the already mentioned SP measurement techniques,
by launching plexcitons exciting the metallic layer itself. Furthermore,
the ballistic and one-way nature of these modes can in principle be
elucidated using fluorescence microscopy \cite{akselrod}. It is important
to note that, owing to the topological nature of these states, perfect
lattices are not required, so this phenomenon holds as long as orientational
and site energy disorder induce perturbations which are smaller than
the topological anticrossings. We tested these ideas by simulating
lattices with disorder in the site energies ($\bar{\omega}_{eff}\to\bar{\omega}_{eff}+\Delta_{\bar{\omega}_{eff}}$)
as well as in the orientations of the dipoles ($\hat{\boldsymbol{\boldsymbol{\mu}}}\to\mbox{cos}\Delta_{\phi}[\mbox{cos}(\alpha+\Delta_{\alpha})\hat{\boldsymbol{x}}+\mbox{sin}(\alpha+\Delta_{\alpha})\hat{\boldsymbol{y}}]+\mbox{sin}\Delta_{\phi}\hat{\boldsymbol{z}}$),
where $\Delta_{j}$ are chosen to be Gaussian random variables centered
at 0 and having disorder widths $\sigma_{j}$ for each $j=\bar{\omega}_{eff},$
$\alpha$, $\phi$. By systematically varying these widths independently
and keeping track of the presence of the one-way edge states, we noticed
that the latter survive under large amounts of disorder, whose thresholds
are approximately located at $\Delta_{\bar{\omega}_{eff}}\sim0.25\,\mbox{eV}$,
$\Delta_{\alpha}\sim30^{o}$, and $\Delta_{\phi}\sim\mbox{15}^{o}$.

To clearly illustrate the nature of the topologically protected edge
states, we have taken $\epsilon_{d}=1$, $g=0.3$, yielding a minimum
gap between plexciton branches (at the wavevectors $\boldsymbol{k}^{*}$
of the original Dirac points) of 0.48 eV. The crossing of the SP and
exciton dispersion curves happens at 3.1 eV. Given typical linewidths
associated with the various dissipative mechanisms at room temperature
($\gamma_{exc,rel}\sim5\,\mbox{meV},\gamma_{exc,deph}\sim40\,\mbox{meV}$,
$\gamma_{SP,rel}\sim10\,\mbox{meV}$, where $rel$ and $deph$ stand
for relaxation and dephasing), we anticipate that exciton-magneto-SP
couplings need to be at least $\Big|H_{exc-SP}(\boldsymbol{k}^{*})\Big|>\sim10\,\mbox{meV}$
in order for the topological plexciton edge states to be meaningful
(\emph{i.e.}, be in the regime of strong coupling \cite{gonzalez_tudela}).
Since our perturbative theory is linear in $g$, this means that keeping
all other parameters fixed, we require $g>0.03$. Even though this
value of MO effect is very reasonable for BiYIG films, their polarizability
is much larger than what we have used in our calculations; in fact,
$\epsilon_{d}\sim6.25$ and $g\sim0.1$ when the perpendicular magnetic
field is about 0.01 T \cite{kuzmiak}, yielding $\Big|H_{exc-SP}(\boldsymbol{k}^{*})\Big|\sim10\,\mbox{meV}$,
which would render the topological edge states difficult to detect,
i.e., we need to reach the plexciton regime . A possible solution
to this problem is to consider novel MO garnet compositions which
maximize the $g/\epsilon_{d}$ ratio. In SI-III-4, we summarize our
understanding of the optimization of $\Big|H_{exc-SP}(\boldsymbol{k}^{*})\Big|$
for the parameter space comprised by $\epsilon_{d}$, $g$, and $a$.
In our simulations, we have assumed that the organic-layer has been
embedded into the MO layer, since the calculation with the proper
three-layer setup yielded couplings $\Big|H_{exc-SP}(\boldsymbol{k}^{*})\Big|$
that did not surpass the dissipative linewidths. However, as far as
we are aware, the weakness of the MO effect obtained with MO garnets
is not fundamentally limited. Once we identify the physics that controls
the strength of $\Big|H_{exc-SP}(\boldsymbol{k}^{*})\Big|$, we can
optimize the MO layer by appealing to different engineering strategies
to increase $g$ and decrease $\epsilon_{d}$ (e.g. by using of MO
garnet sphere arrays \cite{PhysRevB.89.214410}, plasmonic/magnetic
metal nanostructures \cite{temnov2010active}, Ce substituted YIGs
\cite{ross_acsphotonics}, Eu nanocrystals \cite{hasegawa2013magnetic},
etc.).

In terms of the fabrication of the plexciton setup, we warn that the
creation of BiYIG layers typically need high temperature and oxygen,
which is incompatible with deposition on Ag or organic materials.
Hence, the MO layer should first be deposited on garnet substrates
such as GGG ($\mbox{Gd}_{3}\mbox{Ga}_{5}\mbox{O}_{12}$) (111), and
subsequently floated off by dissolving or polishing the substrate.
One should then transfer the film on a Ag-coated substrate and the
organic layer may be deposited and patterned on garnet.

To summarize, we have described the design of exotic plexcitons via
a judicious choice of material and electromagnetic excitation modes.
We showed that Dirac cones and topologically protected edge states
emerge from relatively simple hybrid organic/inorganic nanostructures.
Even though we have not precisely identified an explicit MO material
which fully satisfies our requirements, we believe its design is within
reach, and is the subject of our present investigations. It is also
worth noting that the physical origin of the described edges states
is different from that of edge plasmons in disk geometries \cite{fetter1986edge,song2015chiral},
although the connections are intriguing. The possibility of directed
migration of excitation energy at the nano- and mesoscale offers exciting
prospects in light-harvesting and all-optical circuit architectures.
Furthermore, given the recent experimental discovery of nonlinear
many-body effects such as Bose-Einstein condensation of organic cavity-polaritons
\cite{daskalakis2014nonlinear,bittner2012thermodynamics,bittner_silva}
and plexcitons \cite{rodriguez2013thermalization} at room temperature,
the introduction of the novel features described in this letter enriches
the scope of these materials as a testbed for novel many-body quantum
phenomena.

\emph{ACKNOWLEDGEMENTS.} J.Y.-Z thanks J.C.W. Song for a discussion on edge magnetoplasmons. J.Y.-Z
acknowledges partial support from startup funds at UC San Diego. J.Y-Z.,
T.Z., V.B., and M.B. were supported by an Energy Frontier Research
Center funded by the US Department of Energy, Office of Science, Office
of Basic Energy Sciences under Award Number DESC0001088. S.K.S. was
supported by the Defense Threat Reduction Agency grant HDTRA1-10-1-0046.
S.K.S. is also grateful to the Russian Government Program of Competitive
Growth of Kazan Federal University. Finally, CAR and MCO acknowledge
support of the Solid-State Solar-Thermal Energy Conversion Center
(S3TEC), award DE-SC0001299, and FAME, a STARnet Center of SRC supported
by DARPA and MARCO.



\newpage

 \onecolumngrid

\begin{center}
{\LARGE{}Supplementary Information for ``Plexcitons: Dirac points
and topological modes''}
\par\end{center}{\LARGE \par}

\global\long\def\theequation{S\arabic{equation}}
 \setcounter{equation}{0}

\global\long\def\thefigure{S\arabic{figure}}
 \setcounter{figure}{0}

\author{Joel Yuen-Zhou$^{1}$, Semion K. Saikin$^{2,3}$, Tony H. Zhu$^{4,5}$,
Mehmet C. Onbasli$^{6}$, Caroline A. Ross$^{6}$, Vladimir Bulovic$^{5,7}$,
and Marc A. Baldo$^{5,7}$}

\affiliation{$^{1}$Department of Chemistry and Biochemistry, University of California
San Diego, La Jolla, CA, USA.}

\email{joelyuen@ucsd.edu}

\selectlanguage{english}%

\affiliation{$^{2}$Department of Chemistry and Chemical Biology, Harvard University,
Cambridge, MA, USA.}

\affiliation{$^{3}$Department of Physics, Kazan Federal University, Kazan 420008,
Russian Federation}

\affiliation{$^{4}$Department of Physics, Massachusetts Institute of Technology,
Cambridge, MA, USA.}

\affiliation{$^{5}$Center for Excitonics, Research Laboratory of Electronics,
Massachusetts Institute of Technology, Cambridge, MA, USA.}

\affiliation{$^{6}$Department of Materials Science and Engineering, Massachusetts
Institute of Technology, Cambridge, MA, USA.}

\affiliation{$^{7}$Department of Electrical Engineering and Computer Science,
Massachusetts Institute of Technology, Cambridge, MA, USA.}

~

In this Supplementary Information, we derive some results that are
used in the main text of the article. We compute the electromagnetic
fields associated with the magneto-surface-plasmon (magneto-SP) modes
arising in two- and three-layer setups (Sec. \ref{sec:Interface-between-a}
and Sec. \ref{sec:3layers}). Finally, we derive the effective 2D
Hamiltonian for the plexciton system by considering the dipolar couplings
between chromophores in the organic layer and between the latter and
the magneto-SPs (Sec. \ref{sec:Coupling-of-excitons}).

\tableofcontents{}


\section{Magneto-SPs at a metal-dielectric interface\label{sec:Interface-between-a}}

\subsection{Maxwell's equations\label{sub:Maxwell's-equations}}

Maxwell's equations in arbitrary media read as follows,

\begin{eqnarray}
\nabla\cdot\vec{D} & = & 0,\label{eq:M1}\\
\nabla\cdot\vec{B} & = & 0,\label{eq:M2}\\
\nabla\times\vec{E} & = & -\partial_{t}\vec{B},\label{eq:M3}\\
\nabla\times\vec{H} & = & \partial_{t}\vec{D}.\label{eq:M4}
\end{eqnarray}
We are interested in the case where the electric displacement $\vec{D}$
and the magnetic field $\vec{H}$ are related to the corresponding
electric fields $\vec{E}$ and the magnetic inductions $\vec{B}$
via the linear constitutive relations $\vec{D}=\epsilon_{0}\overleftrightarrow{\epsilon}\vec{E}$
and $\vec{H}=\mu_{0}^{-1}\overleftrightarrow{\mu}{}^{-1}\vec{B}$,
where $\epsilon_{0}=8.854\times10^{-12}\frac{\mbox{C}}{\mbox{V}\,\mbox{m}}$
and $\mu_{0}=1.257\times10^{-6}\frac{\mbox{kg}\,\mbox{m}}{\mbox{C}^{2}}$
are the permittivity and permeability of vacuum, and $\overleftrightarrow{\epsilon}$
and $\overleftrightarrow{\mu}$ are the corresponding scaling (unitless)
tensors for the medium in question. By taking the curl of Eq. (\ref{eq:M3}),
we obtain the wave equation,

\begin{eqnarray}
\nabla(\nabla\cdot\vec{E})-\nabla^{2}\vec{E} & = & -\frac{\overleftrightarrow{\epsilon}}{c^{2}}\partial_{t}^{2}E,\label{eq:curl2}
\end{eqnarray}
where we have introduced the free space speed of light $c=(\mu_{0}\epsilon_{0})^{-\frac{1}{2}}$.
A completely analogous equation holds for $H$ by taking the curl
of Eq. (\ref{eq:M4}), but the latter suffices for our purposes.

\subsection{Permittivities}

\begin{figure}
\centering{}\includegraphics[scale=0.35]{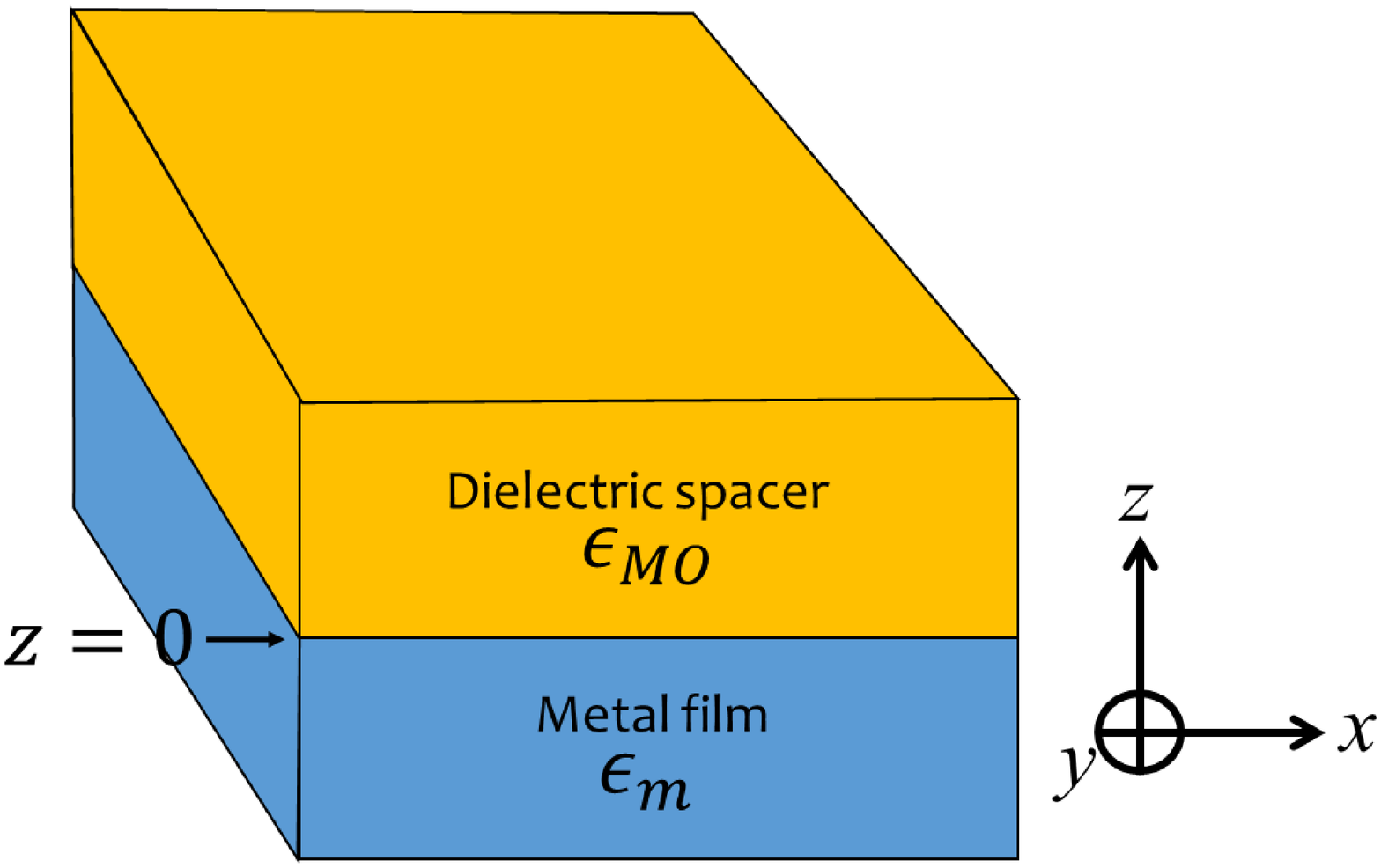}\protect\caption{\emph{Metal-MO dielectric interface at $z=0$. }We solve for the surface
plasmon (SP) modes arising at the metal-dielectric interface, in particular,
when the permittivity of the dielectric $\protect\overleftrightarrow{\epsilon}_{MO}$
is anisotropic due to the application of a perpendicular external
magnetic field.\emph{ }The SP modes in this case are referred to as
magneto-SP modes. As a first approximation to the exciton-SP coupling,
we assume that the organic layer is embedded within the dielectric
spacer medium. \label{fig:Metal-dielectric interface}}
\end{figure}

We are interested in (magneto-)SPs arising at the interface between
a plasmonic metal (silver or gold) and a dielectric medium which is
endowed with magneto-optical (MO) properties (see Fig. \ref{fig:Metal-dielectric interface}).
The latter can be a mix of a magnetic oxide \cite{kuzmiak,Onbasli:14,dionne2009magnetic}
dissolved in a polymer. We assume that the magnetization of both layers
is null, $\overleftrightarrow{\mu}=1$.

The permittivity of the metal is taken to be isotropic $\overleftrightarrow{\epsilon}=\epsilon_{m}\overleftrightarrow{I}$,
where
\begin{equation}
\epsilon_{m}(\omega)=\epsilon_{\infty}-\frac{\omega_{P}^{2}}{\omega^{2}+i\omega\gamma}\label{eq:drude}
\end{equation}
is of the Drude form (the parameters for Ag (Au) are $\epsilon_{\infty}=3.7(6.9)$,
$\omega_{P}=9.2(8.9)\,\mbox{eV}$, and $\gamma=0.01(0.07)\,\mbox{eV}$
\cite{drude_parameters}). Throughout this work, we will set $\gamma=0$
in order to keep the formalism simple. Physically, as long as the
relevant energy scales of interest are larger than $\gamma$ (see
main text), this is a good approximation. Since we are interest in
the plexciton (strong exciton-SP coupling) regime \cite{PhysRevLett.93.036404,govorov_nanoletters,bartlett,PhysRevB.76.035420,sukharev,gonzalez_tudela,torma2015strong},
this should be a good approximation to take. Otherwise, the quantization
of the problem becomes much more complicated.

For the MO layer, the permittivity is anisotropic: upon interaction
with an external magnetic field in the perpendicular $z$-direction,
it acquires the form,

\begin{equation}
\overleftrightarrow{\epsilon}_{MO}=\left[\begin{array}{ccc}
\epsilon_{d} & ig & 0\\
-ig & \epsilon_{d} & 0\\
0 & 0 & \epsilon_{d}
\end{array}\right],\label{eq:e_tensor_supp}
\end{equation}
where the tensor has been written in Cartesian coordinates ($\hat{\boldsymbol{x}},\hat{\boldsymbol{y}},\hat{\boldsymbol{z}}$),
and we take $\epsilon_{d}=1$ and $g=0.1$. Here, the off-diagonal
term is proportional to the Faraday rotation that a linearly polarized
plane-wave electric field experiences as it passes through the material;
$g$ changes sign upon change of magnetic field direction. Typically,
MO magnetic oxides like Bismuth- and Yttrium-Iron Garnets (BIG, YIG)
\cite{kuzmiak,Onbasli:14} have permittivities of $\epsilon_{d}\sim6$
\cite{kuzmiak,Onbasli:14,dionne2009magnetic}, which imply a severe
index mismatch with the metal and organic layers of interest. Hence,
we are implicitly assuming that we have a Maxwell garnet blend with
a low index polymer or aerogel \cite{aerogel} at our disposition,
which yields an effective $\epsilon_{d}=1$. On the other hand, $g\sim0.1$
is a reasonable parameter for MO garnets under a magnetic field of
0.1 Tesla \cite{kuzmiak}. Just as with $\epsilon_{m}$, we ignore
imaginary (absorptive) contributions to $\epsilon_{d}$. Subsec. \ref{sub:Representative coupling values}
discusses other MO materials that could be used for the purposes of
our study.

Chiu and Quinn \cite{chiu_quinn} have solved a slightly different
problem, namely, the magneto-SPs arising from a metal under a strong
magnetic field coupled to a isotropic non-MO dielectric. In their
study, the anisotropy arises in the metal permittivity rather than
in the one corresponding to the dielectric. The resulting equations
are, as expected, very similar, and one could translate their equations
to our setup by some careful changes of variables. However, for clarity
of presentation and in order to develop the three-layer calculation
of Sec. (\ref{sec:3layers}), which is a generalization of the two-layer
case, we shall outline the entire procedure here. Importantly, in
doing so, we manage to go further than Chiu and Quinn and construct
a perturbation theory in the small parameter $g$. This allows us
to develop explicit expressions for the electromagnetic modes which,
as far as we are aware, have not appeared in the literature before.

\subsection{Electromagnetic modes for each layer}

The problem is rotationally symmetric about the vertical $z$-direction,
so it is convenient to adopt a cylindrical coordinate system. Let
us search for SP modes labeled by $\boldsymbol{k}$ which propagate
in-plane and decay along $\hat{\boldsymbol{z}}$ (this is precisely
the condition for SP modes),

\begin{subequations}
\begin{eqnarray}
\vec{E}(\boldsymbol{k}) & = & \boldsymbol{E}(\boldsymbol{k})e^{i(kr_{k}+k_{z}z-\omega t)},\label{eq:mode_E}\\
\vec{B}(\boldsymbol{k}) & = & \boldsymbol{B}(\boldsymbol{k})e^{i(kr_{k}+k_{z}z-\omega t)}.\label{eq:mode_B}
\end{eqnarray}

\end{subequations} For a given direction of $\boldsymbol{k}$, we
shall write vectors in the right-handed cylindrical coordinate system
spanned by the unit vectors $\hat{\boldsymbol{k}},\hat{\boldsymbol{\theta}}_{\boldsymbol{k}},\hat{\boldsymbol{z}}$
such that $\hat{\boldsymbol{k}}\times\hat{\boldsymbol{\theta}}_{\boldsymbol{k}}=\hat{\boldsymbol{z}}$.
for instance, $\boldsymbol{E}=(E_{k},E_{\theta_{k}},E_{z})$, where
$E_{i}=\boldsymbol{E}\cdot\hat{\boldsymbol{i}}$ (beware that we have
defined the tangential direction of $\hat{\boldsymbol{\theta}}_{\boldsymbol{k}}$
with respect to $\hat{\boldsymbol{k}}$ and not to $\hat{\boldsymbol{r}}$).
Physically, $\boldsymbol{k}$ and $\omega=\omega(\boldsymbol{k})$
denote the in-plane (propagating) wavevector and frequency of the
monochromatic wave, respectively, $r_{k}=\boldsymbol{r}\cdot\hat{\boldsymbol{k}}$
is the projection of the position vector $\boldsymbol{r}=(r_{k},r_{\theta},z)$
along the $\hat{\boldsymbol{k}}$ direction, and $k_{z}$ is the imaginary
wavevector associated with the evanescent wave along the perpendicular
direction. Inserting these modes into Eq. (\ref{eq:curl2}) yields
an anisotropic wave equation for $\boldsymbol{E}$,

\begin{equation}
\sum_{ljm}[(\delta_{il}\delta_{jm}-\delta_{im}\delta_{jl})k_{j}k_{l}+\Omega^{2}\epsilon_{im}]E_{m}=0,\label{eq:wave_equation}
\end{equation}
where we have used the notation $\Omega(\boldsymbol{k})=\frac{\omega(\boldsymbol{k})}{c}$
corresponding to the free space wavevector. Here, the $i,j,l,m$ indices
run through $k,\theta,z$ and, formally, we may write $\boldsymbol{k}=(k_{k},k_{\theta},k_{z})=(k,0,k_{z})$.
Due to rotational symmetry, $\overleftrightarrow{\epsilon}_{MO}$
has the same form in the $\hat{\boldsymbol{k}},\hat{\boldsymbol{\theta}}_{\boldsymbol{k}},\hat{\boldsymbol{z}}$
coordinates as Eq. (\ref{eq:e_tensor_supp}).

\subsubsection{MO layer ($z>0$)\label{sub:BIG-layer-()}}

Inserting the dielectric tensor associated with the MO layer (Eq.
(\ref{eq:e_tensor_supp})) into Eq. (\ref{eq:wave_equation}) yields
a matrix equation $\mathbb{M}_{MO}\boldsymbol{E}_{MO}=0$ which explicitly
reads as,

\begin{equation}
\left[\begin{array}{ccc}
\epsilon_{d}-\Omega^{-2}k_{z,MO}^{2} & ig & \Omega^{-2}kk_{z,MO}\\
-ig & \epsilon_{d}-\Omega^{-2}(k^{2}+k_{z,MO}^{2}) & 0\\
kk_{z,MO}\Omega^{-2} & 0 & \epsilon_{d}-\Omega^{-2}k^{2}
\end{array}\right]\begin{bmatrix}1\\
E_{\theta,MO}\\
E_{z,MO}
\end{bmatrix}=\begin{bmatrix}0\\
0\\
0
\end{bmatrix}.\label{eq:ME=00003D0}
\end{equation}
At this point, we have chosen the arbitrary normalization condition
$E_{k,MO}=1$. The secular equation corresponding to Eq. (\ref{eq:ME=00003D0})
is,

\begin{equation}
\left(\frac{k_{z}}{\Omega}\right)^{4}+\mathbb{B}\left(\frac{k_{z}}{\Omega}\right)^{2}+\mathbb{C}=0,\label{eq:quadratic}
\end{equation}
where,

\begin{subequations}

\begin{eqnarray}
\mathbb{B} & = & 2\left[\left(\frac{k}{\Omega}\right)^{2}-\epsilon_{d}\right],\label{eq:B}\\
\mathbb{C} & = & \left[\left(\frac{k}{\Omega}\right)^{2}-\epsilon_{d}\right]\left[\left(\frac{k}{\Omega}\right)^{2}-\frac{\epsilon_{d}^{2}-g^{2}}{\epsilon_{d}}\right].\label{eq:C}
\end{eqnarray}

\end{subequations}The bi-quadratic Eq. (\ref{eq:quadratic}) yields
solutions $k_{z,MO}=i\alpha_{MO}^{\pm},-i\alpha_{MO}^{\pm}$, where
$\alpha_{MO}^{\pm}$ are two different evanescent decay or exponentially
rising constants given by,

\begin{eqnarray}
\alpha_{MO}^{\pm} & = & \Omega\sqrt{\frac{\mathcal{\mathbb{B}}}{2}\pm\sqrt{\frac{\mathbb{B}{}^{2}}{4}-\mathbb{C}}}.\label{eq:soln_quad_eqn}
\end{eqnarray}
Note that these (in general, complex-valued) constants $\alpha_{MO}^{\pm}$
must have positive real part for $e^{-\alpha_{MO}^{\pm}z}$ to decay
or for $e^{\alpha_{MO}^{\pm}z}$ to rise, respectively. In the first
two-layer setup we are considering, we will assume that the MO layer
extends indefinitely for $z>0$ so the field in this region must be
a superposition of the two evanescent fields; exponentially rising
fields will become important when we add an additional interface at
$z=a$ (see Sec. \ref{sec:3layers}). The tangential and perpendicular
components of the electric field (given $E_{k,MO}=1$) can be obtained
from Eq. (\ref{eq:ME=00003D0}),

\begin{subequations}

\begin{eqnarray}
E_{\theta,MO}^{\pm} & = & \frac{-i\Omega^{2}g}{k^{2}-(\alpha_{MO}^{\pm})^{2}-\Omega^{2}\epsilon_{d}},\label{eq:E_theta_BIG}\\
E_{z,MO}^{\pm} & = & \frac{ik\alpha_{MO}^{\pm}}{k^{2}-\Omega^{2}\epsilon_{d}}.\label{eq:E_z_BIG}
\end{eqnarray}

\end{subequations}

\subsubsection{Metal layer ($z<0$)\label{sub:metal_layer}}

In the metal, $\mathbb{M}_{m}\boldsymbol{E}_{m}=0$ corresponds to,

\begin{equation}
\left[\begin{array}{ccc}
\epsilon_{m}-\Omega^{-2}k_{z,m}^{2} & 0 & \Omega^{-2}kk_{z,m}\\
0 & \epsilon_{m}-\Omega^{-2}(k^{2}+k_{z,m}^{2}) & 0\\
kk_{z,m}\Omega^{-2} & 0 & \epsilon_{m}-\Omega^{-2}k^{2}
\end{array}\right]\begin{bmatrix}1\\
E_{\theta,m}\\
E_{z,m}
\end{bmatrix}=\begin{bmatrix}0\\
0\\
0
\end{bmatrix},\label{eq:secular_metal}
\end{equation}
where again we have chosen $E_{k,m}=1$. This leads to the secular
equation which yields the exponentially decaying field for $z<0$;
by letting $k_{z,m}=-i\alpha_{m}$,

\begin{equation}
\alpha_{m}=\sqrt{k^{2}-\Omega^{2}\epsilon_{m}}.\label{eq:soln_2nd_quad_eqn}
\end{equation}
Eq. (\ref{eq:secular_metal}) reveals that

\begin{eqnarray}
E_{z,m} & = & -\frac{ik}{\alpha_{m}}\label{eq:E_z_m}
\end{eqnarray}
but does not inform us about the tangential component $E_{\theta,m}$
as the entry $(\mathbb{M}_{m})_{\theta\theta}=\epsilon_{m}-\Omega^{-2}(k^{2}+k_{z,m}^{2})=0$.
This missing component will be deduced by matching the electromagnetic
fields at the boundary $z=0$.

\subsection{Matching the modes at the boundary ($z=0$)\label{sub:Matching-the-boundary}}

We are looking for magneto-SP modes labeled by a propagating wavevector
$k$ and frequency $\omega(\boldsymbol{k})$, but, in general, different
$k_{z}$ evanescent wavevectors for each layer, which we have denoted
$k_{z,MO}^{\pm}=i\alpha_{MO}^{\pm}$ and $k_{z,m}=-i\alpha_{m}$.
The energy $\omega(\boldsymbol{k})$ (and therefore $\Omega$) is
unknown.

To summarize, for each pair $(k,\omega)$, there are two possible
modes in the MO layer associated with different decay constants $\alpha_{MO}^{\pm}$
(see Eq. (\ref{eq:soln_quad_eqn}), $z>0$),

\begin{subequations}

\begin{eqnarray}
\vec{E}_{MO}^{\pm} & = & \boldsymbol{E}_{MO}^{\pm}\eta e^{-\alpha_{MO}^{\pm}z}\nonumber \\
 & = & (1,E_{\theta,MO}^{\pm},E_{z,MO}^{\pm})\eta e^{-\alpha_{MO}^{\pm}z},\label{eq:E_BIG}\\
\vec{B}_{MO}^{\pm} & = & \boldsymbol{B}_{MO}^{\pm}\eta e^{-\alpha_{MO}^{\pm}z}\nonumber \\
 & = & \frac{-i}{\omega}(\alpha^{\pm}E_{\theta,MO}^{\pm},-ikE_{z,MO}^{\pm}-\alpha_{MO}^{\pm},ikE_{\theta,MO}^{\pm})\eta e^{-\alpha_{MO}^{\pm}z},\label{eq:B_BIG}
\end{eqnarray}
\end{subequations}where the magnetic induction in Eq. (\ref{eq:B_BIG})
has been deduced from Eq. (\ref{eq:E_BIG}) and Maxwell's Eq. (\ref{eq:M3}).
The vectors have been written in cylindrical coordinates and we have
defined $\eta\equiv e^{ikr-i\omega t}$. Hence, the total fields in
the MO layer read,

\begin{subequations}

\begin{eqnarray}
\vec{E}_{MO} & = & t_{MO}^{+}\vec{E}_{MO}^{+}+t_{MO}^{-}\vec{E}_{MO}^{-},\label{eq:E_BIG_superposition}\\
\vec{B}_{MO} & = & t_{MO}^{+}\vec{B}_{MO}^{+}+t_{MO}^{-}\vec{B}_{MO}^{-}.\label{eq:B_BIG_superposition}
\end{eqnarray}

\end{subequations}where $t^{\pm}$ are coefficients to be determined.
Similarly, for the metal layer ($z<0$),

\begin{subequations}

\begin{eqnarray}
\vec{E}_{m} & = & \boldsymbol{E}_{m}\eta e^{\alpha_{m}z}\nonumber \\
 & = & (1,E_{\theta,m},E_{z,m})\eta e^{\alpha_{m}z},\label{eq:E_m}\\
\vec{B}_{m} & = & \boldsymbol{B}_{m}\eta e^{\alpha_{m}z}\nonumber \\
 & = & \frac{-i}{\omega}(-\alpha_{m}E_{\theta,m},-ikE_{z,m}+\alpha_{m},ikE_{\theta,m})\eta e^{\alpha_{m}z}.\label{eq:B_m}
\end{eqnarray}
\end{subequations}Here, we keep the arbitrary normalization where
$E_{k,m}=1$. Later on, we shall fix this normalization via quantization
of the energy of the modes (see \ref{sub:Quantization}). Furthermore,
it is also safe to arbitrarily assume $E_{k,MO}^{\pm}=1$ because
Eqs. (\ref{eq:E_BIG_superposition}) and (\ref{eq:B_BIG_superposition})
contain scaling coefficients $t^{\pm}$ which will be fixed by the
boundary conditions at the metal-MO interface.

We are ready to match the fields at the interface at $z=0$. The in-plane
electric field and the perpendicular electric displacement each need
to be continuous across the boundary: $E_{i,MO}=E_{i,m}$ for $i=r,\theta$,
while $\epsilon_{d}E_{z,MO}=\epsilon_{m}E_{z,m}$. Furthermore, the
magnetic field, and because $\overleftrightarrow{\mu}=1$, its induction,
are all continuous throughout, $B_{i,MO}=B_{i,m}$. These constraints
altogether read,

\begin{subequations}

\begin{eqnarray}
1 & = & t_{MO}^{+}+t_{MO}^{-},\label{eq:constraint1}\\
E_{\theta,m} & = & t_{MO}^{+}E_{\theta,MO}^{+}+t_{MO}^{-}E_{\theta,MO}^{-},\label{eq:constraint2}\\
\epsilon_{m}E_{z,m} & = & \epsilon_{d}(t_{MO}^{+}E_{z,MO}^{+}+t_{MO}^{-}E_{z,MO}^{-}),\label{eq:constraint3}\\
\alpha_{m}E_{\theta,m} & = & -t_{MO}^{+}\alpha_{MO}^{+}E_{\theta,MO}^{+}-t_{MO}^{-}\alpha_{MO}^{-}E_{\theta,MO}^{-},\label{eq:constraint4}\\
kE_{z,m}+i\alpha_{m} & = & t_{MO}^{+}(kE_{z,MO}^{+}-i\alpha_{MO}^{+})+t_{MO}^{-}(kE_{z,MO}^{-}-i\alpha_{MO}^{-}),\label{eq:constraint5}\\
E_{\theta,m} & = & t_{MO}^{+}E_{\theta,MO}^{+}+t_{MO}^{-}E_{\theta,MO}^{-}.\label{eq:constraint6}
\end{eqnarray}

\end{subequations} These constraints read similarly to the ones derived
by Chiu and Quinn in \cite{chiu_quinn} (see their Eqs. (39)--(44)),
except for the different coordinate conventions. Clearly, Eqs. (\ref{eq:constraint2})
and (\ref{eq:constraint6}) are identical. Furthermore, Eqs. (\ref{eq:constraint3})
and (\ref{eq:constraint5}) contain the same information, as can be
shown by using Eqs. (\ref{eq:soln_2nd_quad_eqn}) (\ref{eq:E_z_BIG}),
(\ref{eq:E_z_m}), and (\ref{eq:constraint1}). The remaining constraints
yield the equation,

\begin{eqnarray}
k^{2}-\Omega^{2}\epsilon_{d}+\alpha_{MO}^{+}\alpha_{MO}^{-}+(\alpha_{MO}^{+}+\alpha_{MO}^{-})\alpha_{m}](k^{2}-\Omega^{2}\epsilon_{d})\epsilon_{m}\nonumber \\
+\alpha_{m}\epsilon_{d}\{\alpha_{MO}^{+}\alpha_{MO}^{-}(\alpha_{MO}^{+}+\alpha_{MO}^{-})+\alpha_{m}[(\alpha_{MO}^{+})^{2}+\alpha_{MO}^{+}\alpha_{MO}^{-}+(\alpha_{MO}^{-})^{2}]-\alpha_{m}(k^{2}-\Omega^{2}\epsilon_{d})\} & = & 0.\label{eq:final_equation}
\end{eqnarray}
By inserting Eqs. (\ref{eq:soln_quad_eqn}), (\ref{eq:soln_2nd_quad_eqn})
into Eq. (\ref{eq:final_equation}), we obtain a nonlinear equation
in $\Omega$ for every value of $\boldsymbol{k}$. This equation can
be numerically solved, at least in principle. $\Omega$ can be then
used as input to Eqs. (\ref{eq:E_theta_BIG}), (\ref{eq:E_z_BIG}),
(\ref{eq:E_z_m}), (\ref{eq:E_BIG_superposition}), (\ref{eq:B_BIG_superposition}),
and (\ref{eq:constraint1})--(\ref{eq:constraint6}) to solve for
the electromagnetic modes. As we shall see, the most important qualitative
feature of the solution of this problem is that the magneto-SP fields
acquire tangential components (see Eqs. (\ref{eq:E_theta_BIG}) and
(\ref{eq:constraint2})) which are absent when $g=0$ \cite{maier,novotny},
that is, in the absence of an external magnetic field. Eq. (\ref{eq:final_equation})
is identical to Eq. (45) in \cite{chiu_quinn} upon carrying out the
substitutions $\epsilon_{xx}=\epsilon_{zz}=\epsilon_{d}$.

\subsection{Perturbation expansion on $g$\label{sec:Perturbation-expansion-on}}

Chiu and Quinn reported a dispersion relation $\Omega$ \emph{vs}
$\boldsymbol{k}$ by numerically solving Eq. (\ref{eq:final_equation})
for a very similar setup to the one of our interest. However, a detailed
description of the resulting electromagnetic modes was not presented
in that work. As explained, one may in principle solve for the profile
of the electromagnetic modes once this dispersion is known. However,
a numerical attempt at the problem using a standard nonlinear solver
yielded spurious results for the modes.

Since the solution to the SP problem with no magnetic field ($g=0$)
is a well-known textbook result, and $g$ is anyway much smaller than
$\epsilon_{d}$ in a realistic setup, we may use a perturbation expansion
of the equation in powers of $g$. Our goal is to obtain the electric
fields up to $O(g)$, so that we can compute the magnitude of the
exciton-SP coupling to that same order. To accomplish such objective,
we first need to solve for $\Omega$ as well as the coefficients $t_{MO}^{\pm}$
up to $O(g)$. As we shall see, however, knowledge of $t_{MO}^{\pm}$
requires information about $\Omega$ up to $O(g^{2})$. Once this
is done, we simply Taylor expand the fields (\ref{eq:E_BIG}), 9\ref{eq:B_BIG}),
(\ref{eq:E_m}), (\ref{eq:B_m}) and collect the results according
to Eqs. (\ref{eq:E_BIG_superposition})--(\ref{eq:B_m})). In retrospect,
the original problem we faced by trying to directly solve Eq. (\ref{eq:final_equation})
originated from the fact that we were using the very small $O(g^{2})$
corrections to $\Omega$ as an input to solve for the $O(g)$ electromagnetic
modes. This requires an accurate solution of $\Omega$, which is complicated
by the highly nonlinear dependence of Eq. (\ref{eq:final_equation})
on $\Omega$. As a future consideration, it might be worth exploring
numerical methodologies to attack this problem beyond the perturbative
regime, although for our purposes, the latter suffices.

Even though the algebra below seems involved, it is straightforward
to derive using a symbolic algebra package such as Wolfram Mathematica(c).

\subsubsection{Solving for $\Omega$\label{sub:Solving-for-Omega}}

Given that the right hand side of Eq. (\ref{eq:final_equation}) is
zero, the polynomials at each power of $g$ must each vanishing identically.

\begin{figure}
\begin{centering}
\includegraphics[scale=0.5]{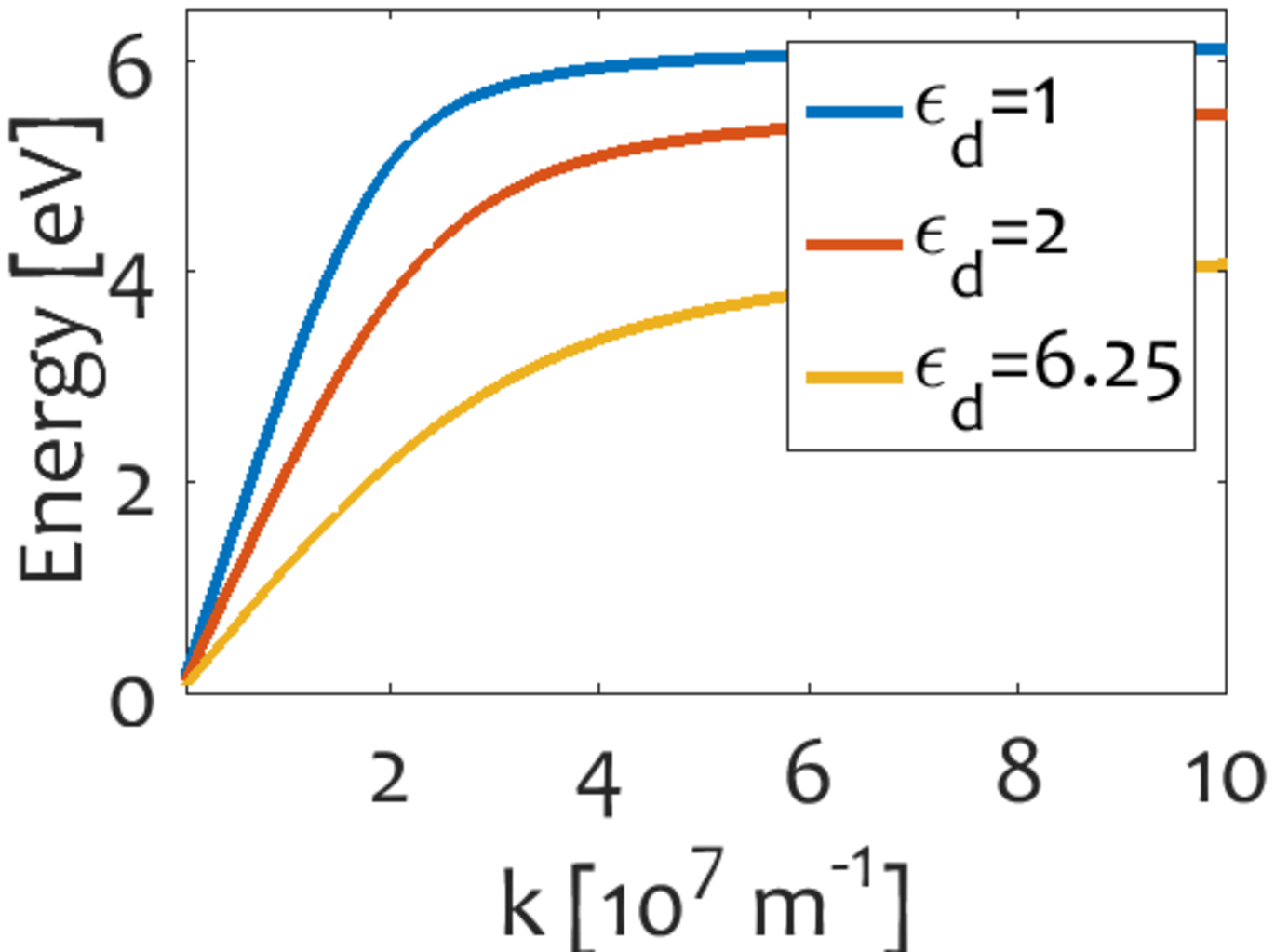}
\par\end{centering}

\protect\caption{\emph{SP dispersion energy as a function of $k=|\boldsymbol{k}|$
for the two-layer (metal-MO dielectric) setup, assuming $g=0$}. Plots
generated using Eq. (\ref{eq:Omega_0}) with Drude parameters for
Ag, $\epsilon_{\infty}\sim4$, $\omega_{P}\sim9$~eV, and varying
the dielectric permittivity $\epsilon_{d}$. Since $\Omega=\Omega_{0}+O(g^{2})$,
the plots are correct up to $O(g)$, which is our perturbation order
of interest. Notice that the dispersion curves start-off linearly
(light-like excitations), but plateau to constant values at large
wavevectors (charge oscillations in the metal), which become smaller
as the dielectric permittivity increases.\label{fig:SP-dispersion-energy}}
\end{figure}

To start with and as a consistency check, at zeroth order in $g$,
Eq. (\ref{eq:final_equation}) becomes

\begin{eqnarray}
[2\alpha_{d0}^{2}+2\alpha_{d0}\alpha_{m0}]\alpha_{d0}^{2}\epsilon_{m0}+\alpha_{m0}\epsilon_{d0}\{\alpha_{d0}^{2}(2\alpha_{d0})+\alpha_{m0}[3\alpha_{d0}^{2}]-\alpha_{m}\alpha_{d0}^{2}\} & =\nonumber \\
{}[2\alpha_{d0}^{3}+2\alpha_{d0}^{2}\alpha_{m0}](\alpha_{d0}\epsilon_{m0}+\alpha_{m0}\epsilon_{d0}) & = & 0,\label{eq:O(g^0)}
\end{eqnarray}
where the 0-subscripted variables denote the corresponding functions
in Eq. (\ref{eq:drude}), (\ref{eq:soln_quad_eqn}), (\ref{eq:soln_2nd_quad_eqn})
taking $\Omega=\Omega_{0}$,

\begin{subequations}

\begin{eqnarray}
\epsilon_{m0} & = & \epsilon_{\infty}-\frac{\Omega_{P}^{2}}{\Omega_{0}^{2}},\label{eq:em0}\\
\alpha_{d0} & = & \sqrt{k^{2}-\epsilon_{d}\Omega_{0}^{2}},\label{eq:alphad0}\\
\alpha_{m0} & = & \sqrt{k^{2}-\epsilon_{m0}\Omega_{0}^{2}},\label{eq:alpham0}
\end{eqnarray}

\end{subequations}where $\Omega_{P}\equiv\frac{\omega_{P}}{c}$.
Eq. (\ref{eq:O(g^0)}) implies that either

\begin{equation}
2\alpha_{d0}^{3}+2\alpha_{d0}^{2}\alpha_{m0}=0,\label{eq:cond1}
\end{equation}
or

\begin{equation}
\alpha_{d0}\epsilon_{m0}+\alpha_{m0}\epsilon_{d}=0.\label{eq:cond2}
\end{equation}
The condition in Eq. (\ref{eq:cond1}) requires that $\alpha_{d0}=-\alpha_{m0}$,
contradicting the very nature of the SP solution we are looking for,
where $\alpha_{d0},\,\alpha_{m0}>0$ represent evanescent fields.
However, the condition in Eq. (\ref{eq:cond2}) is simply the standard
equation for the dispersion relation of a SP at the interface of an
unmagnetized MO sample and a metal film \cite{maier,novotny}. It
can be readily solved yielding,

\begin{equation}
k=\Omega_{0}\sqrt{\frac{\epsilon_{m0}\epsilon_{d}}{\epsilon_{m0}+\epsilon_{d}}},\label{eq:k_Omega_0}
\end{equation}
or more explicitly,

\begin{eqnarray}
\Omega_{0} & = & \sqrt{\frac{\epsilon_{d}\Omega_{P}^{2}+k^{2}(\epsilon_{d}+\epsilon_{\infty})-\sqrt{[\epsilon_{d}\Omega_{P}^{2}+k^{2}(\epsilon_{d}+\epsilon_{\infty})]^{2}-4\epsilon_{\infty}\epsilon_{d}k^{2}\Omega_{P}^{2}}}{2\epsilon_{\infty}\epsilon_{d}}}.\label{eq:Omega_0}
\end{eqnarray}
At short $k$, the (linear) dispersion is very light-like, $\Omega_{0}=\frac{k}{\sqrt{\epsilon_{d}}}$,
and at large $k$, it plateaus to $\Omega_{0}\to\frac{\Omega_{P}}{\sqrt{\epsilon_{d}+\epsilon_{\infty}}}$,
corresponding to collective charge oscillations in the metal (see
Fig. \ref{fig:SP-dispersion-energy}).

Moving on to the $g^{1}$ terms of Eq. (\ref{eq:final_equation})
yields an equation of the form $\Omega^{(1)}f=0$ where $f(\alpha_{d0},\alpha_{m0})$
is a nonzero polynomial in $\alpha_{d0}$ and $\alpha_{m0}$, implying
that $\Omega^{(1)}=0$. This result can be quickly derived as follows:
the $g^{1}$ terms stem only from the power expansion of $\alpha_{MO}^{+}$,
$\alpha_{MO}^{-}$, $\alpha_{m}$, and $\epsilon_{m}$. Some $g^{1}$
contributions from $\alpha_{MO}^{+}$, $\alpha_{MO}^{-}$ are proportional
to $\Omega_{1}$ but some are not. Regardless, the ones from $\alpha_{MO}^{+}$
come with the opposite sign to the ones from $\alpha_{MO}^{-}$; hence,
they vanish identically as Eq. (\ref{eq:final_equation}) is symmetric
in $\alpha_{MO}^{+}$ and $\alpha_{MO}^{-}$. On the other hand, every
$g^{1}$ term for $\alpha_{m}$ and $\epsilon_{m}$ is strictly proportional
to $\Omega_{1}$ requiring $\Omega_{1}=0$ for the all the $g^{1}$
terms to cancel. Hence, the lowest order correction to $\Omega_{0}$
of $\Omega$ arises at $O(g^{2})$, that is,
\begin{equation}
\Omega\approx\Omega_{0}+g^{2}\Omega_{2}.\label{eq:Omega_approx}
\end{equation}
As mentioned, we are solely interested in the calculation of the electric
fields in each layer up to $O(g)$. However, as we shall see in the
next subsection, these corrections depend on $\Omega_{2}$. To obtain
this coefficient, we expand $\alpha_{MO}^{+}$, $\alpha_{MO}^{-}$,
$\alpha_{m}$, and $\epsilon_{m}$ up to $O(g^{2})$, but not beyond
that,

\begin{subequations}

\begin{eqnarray}
\alpha_{MO}^{\pm} & \approx & \alpha_{d0}\pm g\alpha_{d1}+g^{2}(\alpha_{d20}+\alpha_{d22}\Omega_{2}),\label{eq:alpha_BIG_expansion}\\
\alpha_{m} & \approx & \alpha_{m0}+g^{2}\alpha_{m22}\Omega_{2},\label{eq:alpha_m_expansion}\\
\epsilon_{m} & \approx & \epsilon_{m0}+g^{2}\epsilon_{m22}\Omega_{2},\label{eq:e_m_expansion}
\end{eqnarray}

\end{subequations}where the coefficients in the expansions take the
form,

\begin{subequations}

\begin{eqnarray}
\alpha_{d1} & = & \frac{i\Omega_{0}}{2\sqrt{\epsilon_{d}}},\label{eq:alphad1}\\
\alpha_{d20} & = & \frac{\Omega_{0}^{2}}{8\epsilon_{d}\alpha_{d0}},\label{eq:alphad20}\\
\alpha_{d22} & = & \frac{\Omega_{0}\epsilon_{d}}{\alpha_{d0}},\label{eq:alphad22}\\
\alpha_{m22} & = & -\frac{\Omega_{0}\epsilon_{i}}{\alpha_{m0}},\label{eq:alpham22}\\
\epsilon_{m22} & = & \frac{2\Omega_{P}^{2}}{\Omega_{0}^{3}}.\label{eq:alpham22-1}
\end{eqnarray}

\end{subequations}Notice that in order to ultimately solve for $\Omega_{2}$,
we have separated the $g^{2}$ terms into two categories: those which
are proportional to $\Omega_{2}$ ($g^{2}\alpha_{d22}\Omega_{2}$,
$g^{2}\alpha_{m22}\Omega_{2}$, $g^{2}\epsilon_{m22}\Omega_{2}$)
and those that are not ($g^{2}\alpha_{d20}$). We substitute these
expressions into Eq. (\ref{eq:final_equation}) and collect the $g^{2}$
terms, which ought to cancel. The manipulations yield in a linear
equation for $\Omega_{2}$ which gives,

\begin{equation}
\Omega_{2}=\frac{\Omega_{0}^{5}\epsilon_{m0}^{2}(3\epsilon_{d}-\epsilon_{m0})}{8\epsilon_{d}(\epsilon_{d}-\epsilon_{m0})\Big[\Omega_{P}^{2}(2\alpha_{d0}^{2}\epsilon_{m0}+\Omega_{0}^{2}\epsilon_{d}^{2})+\Omega_{0}^{4}\epsilon_{d}\epsilon_{m0}(\epsilon_{d}-\epsilon_{m0})\Big]},\label{eq:omega_2}
\end{equation}
where we have also used Eq. (\ref{eq:cond2}) in the form of $\alpha_{m0}=-\frac{\alpha_{d0}\epsilon_{m0}}{\epsilon_{d}}$
to simplify the final expression.

\subsubsection{Solving for $t_{MO}^{\pm}$\label{sub:Solving-for-t}}

Next, we expand the coefficient $t_{MO}^{+}$, which denotes the contribution
of the $+$ fields in Eqs. (\ref{eq:E_BIG_superposition}) and (\ref{eq:B_BIG_superposition})
($t_{MO}^{-}$ can be subsequently found using the constraint in Eq.
(\ref{eq:constraint1})),

\begin{eqnarray}
t_{MO}^{+} & \approx & t_{d0}+gt_{d1}.\label{eq:tBIG+}
\end{eqnarray}
Here, $t_{d0}$ and $t_{d1}$ are unknown, but we shall solve for
them using the boundary condition on the perpendicular electric fields
(see Eq. (\ref{eq:constraint3})). We expand $E_{z,MO}^{\pm}$ and
$E_{zm}$ in Eqs. (\ref{eq:E_z_BIG}) and (\ref{eq:E_z_m}) up to
$O(g^{2})$ \footnote{Here, as opposed to Eqs. (\ref{eq:alpha_BIG_expansion})--(\ref{eq:e_m_expansion}),
we have not separated the $g^{2}$ terms in $E_{z,MO}^{\pm}$ and
$E_{zm}$ ($g^{2}E_{zd2}$ and $g^{2}E_{zm2}$) into those contributions
which are proportional to $\Omega_{2}$ and those which are not, as
$\Omega_{2}$ has already been determined in Eq. (\ref{eq:omega_2}).},

\begin{subequations}

\begin{eqnarray}
E_{z,MO}^{\pm} & \approx & E_{z,d0}\pm gE_{z,d1}+g^{2}E_{z,d2},\label{eq:Ez_BIG_expansion}\\
E_{z,m} & \approx & E_{z,m0}+g^{2}E_{z,m2},\label{eq:Ezm_expansion}
\end{eqnarray}

\end{subequations}where,

\begin{subequations}

\begin{eqnarray}
E_{z,d0} & = & \frac{ik}{\alpha_{d0}},\label{eq:Ezd0}\\
E_{z,m0} & = & -\frac{ik}{\alpha_{m0}},\label{eq:Ezm0}\\
E_{z,d1} & = & -\frac{k\Omega_{0}}{2\alpha_{d0}^{2}\sqrt{\epsilon_{d}}},\label{eq:Ezd1}\\
E_{z,d2} & = & \frac{i(\alpha_{d0}\alpha_{d20}k+\alpha_{d0}\alpha_{d22}k\Omega_{2}+2k\Omega_{0}\Omega_{2}\epsilon_{d})}{\alpha_{d0}^{3}},\label{eq:Ezd2}\\
E_{z,m2} & = & -\frac{ik\Omega_{0}\Omega_{2}\epsilon_{i}}{\alpha_{m0}^{3}}.\label{eq:Ezm2}
\end{eqnarray}

\end{subequations}We plug these expressions into Eq. (\ref{eq:constraint3}).
At zeroth-order in $g$, we get $\epsilon_{m0}E_{zm0}=\epsilon_{d}E_{zd0}$,
which is equivalent to Eq. (\ref{eq:cond2}) and does not give us
information on $t_{d0}$ or $t_{d1}$ (this indeterminacy ultimately
reveals why we need expansions up to $O(g^{2})$ to obtain fields
up to $O(g)$). At $O(g)$ we get the intuitive result,
\begin{equation}
t_{d0}=\frac{1}{2},\label{eq:t10}
\end{equation}
and at $O(g^{2})$ we obtain,
\begin{eqnarray}
t_{d1} & = & \frac{-E_{z,d2}\epsilon_{d}+E_{z,m0}\Omega_{2}\epsilon_{m22}+E_{z,m2}\epsilon_{m0}}{2E_{zd1}\epsilon_{d}}\nonumber \\
 & = & i\underbrace{\Bigg(\frac{16\alpha_{d0}^{3}\alpha_{m0}^{2}\Omega_{2}\Omega_{P}^{2}+8\alpha_{d0}^{3}\Omega_{0}^{4}\Omega_{2}\epsilon_{\infty}\epsilon_{m0}+\alpha_{m0}^{3}\Omega_{0}^{5}+8\alpha_{m0}^{3}\Omega_{0}^{4}\Omega_{2}\epsilon_{d}^{2}}{8\alpha_{d0}\alpha_{m0}^{3}\Omega_{0}^{4}\sqrt{\epsilon_{d}}}\Bigg)}_{\equiv\tau_{d1}},\label{eq:t11}
\end{eqnarray}
which, together with $\Omega_{2}$ in Eq. (\ref{eq:omega_2}), can
be readily evaluated with zeroth-order parameters. We notice that
$t_{d1}$ is purely imaginary-valued, so we have written it in terms
of a purely real-valued $\tau_{d1}$. Given Eq. (\ref{eq:tBIG+}),
it is clear from Eq. (\ref{eq:constraint1}) that
\begin{eqnarray}
t_{MO}^{-} & \approx & t_{d0}-gt_{d1}.\label{eq:tBIG-}
\end{eqnarray}

\subsubsection{Collecting the expressions for the fields \label{sub:Collecting-the-expressions}}

We now have all the ingredients to evaluate $\vec{E}_{MO}$ and $\vec{B}_{MO}$
(see Eqs. (\ref{eq:E_BIG_superposition})--(\ref{eq:B_BIG_superposition}))
up to $O(g)$,

\begin{subequations}

\begin{eqnarray}
\vec{E}_{MO} & \approx & \vec{E}_{d0}+g\vec{E}_{d1},\label{eq:E_BIG_expansion}\\
\vec{B}_{MO} & \approx & \vec{B}_{d0}+g\vec{B}_{d1},\label{eq:B_BIG-expansion}
\end{eqnarray}

\end{subequations}where, at zeroth-order, in the absence of external
magnetic field, we have the standard SP mode which is a transverse
magnetic (TM) mode \cite{maier,novotny},

\begin{subequations}

\begin{eqnarray}
\vec{E}_{d0} & = & \underbrace{\Big(1,0,\frac{ik}{\alpha_{d0}}\Big)}_{\equiv\boldsymbol{E}_{d0}}\eta_{0}e^{-\alpha_{d0}z},\label{eq:Ed0}\\
\vec{B}_{d0} & = & \frac{-i}{\Omega_{0}c}\Big(0,\frac{k^{2}}{\alpha_{d0}}-\alpha_{d0},0\Big)\eta_{0}e^{-\alpha_{d0}z}\nonumber \\
 & = & \underbrace{-\frac{i\epsilon_{d}\Omega_{0}}{\alpha_{d0}c}(0,1,0)}_{\equiv\boldsymbol{B}_{d0}}\eta_{0}e^{-\alpha_{d0}z}.\label{eq:Bd0}
\end{eqnarray}

\end{subequations}In going from the first to the second line of Eq.
(\ref{eq:Bd0}), we have used Eq. (\ref{eq:alphad0}). Eqs. (\ref{eq:Ed0})--(\ref{eq:Bd0})
feature an elliptically-polarized electric field with no tangential
component, and a purely tangential and imaginary-valued magnetic induction.
The opposite is true for the first-order correction: it consists of
a purely tangential and imaginary-valued electric field (recall that
$t_{d1}$ is purely imaginary, see Eq. (\ref{eq:t11})) and an elliptically
polarized electric field with no tangential part,

\begin{subequations}

\begin{eqnarray}
\vec{E}_{d1} & = & \underbrace{\frac{i\Omega_{0}(4\tau_{d1}\sqrt{\epsilon_{d}}-\Omega_{0}z)}{2\alpha_{d0}}(0,1,0)}_{\equiv\boldsymbol{E}_{d1}}\eta_{0}e^{-\alpha_{d0}z},\label{eq:Ed1}\\
\vec{B}_{d1} & = & \underbrace{\frac{1}{2\alpha_{d0}c}\Bigg(-4\alpha_{d0}\tau_{d1}\sqrt{\epsilon_{d}}+\Omega_{0}(1-\alpha_{d0}z),0,ik(4\tau_{d1}\sqrt{\epsilon_{d}}-\Omega_{0}z)\Bigg)}_{\equiv\boldsymbol{B}_{d1}}\eta_{0}e^{-\alpha_{d0}z}.\label{eq:Bd1}
\end{eqnarray}

\end{subequations}In deriving Eqs. (\ref{eq:E_BIG_expansion})--(\ref{eq:Ed1}),
we have used the $O(g)$ Taylor expansion for the fields (see Eqs.
(\ref{eq:E_theta_BIG})--(\ref{eq:E_z_BIG}), (\ref{eq:B_BIG})).
Notice that besides the exponentially decreasing dependence of the
fields, we also obtain a polynomial contribution in $z$. We already
computed some of the relevant expansion coefficients in Eqs. (\ref{eq:Ez_BIG_expansion})--(\ref{eq:Ezd1});
the remaining ones that we used are,

\begin{subequations}

\begin{eqnarray}
E_{\theta,MO}^{\pm} & \approx & \pm\frac{\Omega_{0}\sqrt{\epsilon_{d}}}{\alpha_{d0}},\label{eq:E_th_BIG_expansion}\\
B_{r,MO}^{\pm} & \approx & \mp\frac{i\sqrt{\epsilon_{d}}}{c}+g\frac{\Omega_{0}}{2\alpha_{d0}c},\label{eq:BrBIG}\\
B_{\theta,MO}^{\pm} & \approx & -\frac{i\Omega_{0}\epsilon_{d}}{\alpha_{d0}c}+g\frac{\Omega_{0}^{2}\sqrt{\epsilon_{d}}}{2\alpha_{d0}^{2}c},\label{eq:BthBIG}\\
B_{z,MO}^{\pm} & \approx & \pm\frac{k\sqrt{\epsilon_{d}}}{\alpha_{d0}c},\label{eq:BzBIG}
\end{eqnarray}

\end{subequations}as well as those for the plane wave components,

\begin{eqnarray}
\eta e^{-\alpha_{MO}^{\pm}z} & = & e^{i(kr-\omega t)-\alpha_{MO}^{\pm}z}\nonumber \\
 & \approx & \underbrace{e^{i(kr-c\Omega_{0}t)}}_{\equiv\eta_{0}}e^{-\alpha_{d0}z}\Bigg(1\mp g\frac{i\Omega_{0}}{2\sqrt{\epsilon_{d}}}z\Bigg).\label{eq:eta_exp}
\end{eqnarray}
Finally, given $\vec{E}_{MO}$, $\vec{E}_{m}$ can be readily obtained
from the boundary conditions for the fields at $z=0$ (see Eqs. (\ref{eq:constraint1})--(\ref{eq:constraint6}))
as well as the original ansatz for their functional forms (see Eq.
(\ref{eq:E_m})--(\ref{eq:B_m})),

\begin{eqnarray}
\vec{E}_{m} & \approx & \vec{E}_{m0}+g\vec{E}_{m1},\label{eq:E_BIG_expansion-1}
\end{eqnarray}
where, at zeroth-order we have the standard SP electric field as if
there were no magnetic field present,

\begin{subequations}

\begin{eqnarray}
\vec{E}_{m0} & = & \Big(1,0,\frac{\epsilon_{d}}{\epsilon_{m0}}\frac{ik}{\alpha_{d0}}\Big)\eta_{0}e^{\alpha_{m0}z}=\underbrace{\Big(1,0,-\frac{ik}{\alpha_{m0}}\Big)}_{=\boldsymbol{E}_{m0}}\eta_{0}e^{\alpha_{m0}z},\label{eq:Em0}\\
\vec{B}_{m0} & = & \frac{i\epsilon_{m0}\Omega_{0}}{\alpha_{m0}c}(0,1,0)\eta_{0}e^{\alpha_{m0}z}=\underbrace{-\frac{i\epsilon_{d}\Omega_{0}}{\alpha_{d0}c}(0,1,0)}_{\equiv\boldsymbol{B}_{m0}}\eta_{0}e^{\alpha_{m0}z}.\label{eq:Bm0}
\end{eqnarray}

\end{subequations}Here, we have used Eq. (\ref{eq:cond2}) in both
lines. Similarly, the first order correction to $\vec{E}_{m0}$ is,

\begin{eqnarray}
\vec{E}_{m1} & = & \underbrace{\frac{2i\tau_{d1}\Omega_{0}\sqrt{\epsilon_{d}}}{\alpha_{d0}}(0,1,0)}_{\equiv\boldsymbol{E}_{m1}}\eta_{0}e^{\alpha_{m0}z},\label{eq:Ed1-1}\\
\vec{B}_{m1} & = & \underbrace{\frac{2\tau_{d1}}{\alpha_{d0}c}\Bigg(\Omega_{0}-\alpha_{d0}\sqrt{\epsilon_{d}},0,ik\sqrt{\epsilon_{d}}\Bigg)}_{\equiv\boldsymbol{B}_{m1}}\eta_{0}e^{\alpha_{m0}z}.\nonumber
\end{eqnarray}

\subsection{Quantization and normalization of modes\label{sub:Quantization}}

The previous section shows how to (perturbatively) compute the frequency
$\Omega$, the electric field, and the magnetic induction of an SP
mode with wavevector $\boldsymbol{k}$. Note that, so far, we have
invoked an arbitrary normalization (setting $E_{r}=1$). To fix this,
we first ought to compute the energy associated with the unnormalized
modes. Consider placing electric field amplitude $\mathcal{A}_{\boldsymbol{k}}$
into the radial component of the $\boldsymbol{k}$ mode. The energy
in this mode is quadratic in the fields\footnote{Note that even if the electric field is much larger than the magnetic
induction, the prefactors of $\epsilon_{0}$ and $\frac{1}{\mu_{0}}$
weigh them in a way that their contributions to the energy density
are on the same order of magnitude} \cite{novotny,archambault},

\begin{equation}
H_{SP,\boldsymbol{k}}=\frac{1}{2}\sum_{i}\int dV\Bigg[\epsilon_{0}\sum_{j}\frac{d(\omega\epsilon_{ij}^{*}(\omega))}{d\omega}(\vec{E})_{j}^{*}(\vec{E})_{i}+\frac{1}{\mu_{0}\mu}|(\vec{B})_{i}|^{2}\Bigg]|\mathcal{A}_{\boldsymbol{k}}|^{2},\label{eq:energy_sp}
\end{equation}
where $i,j\in\{r,\theta,z\}$, and electric field and magnetic inductions
in each $\boldsymbol{k}$ mode (throughout $z$) are conveniently
written as,

\begin{subequations}

\begin{eqnarray}
\vec{E} & = & \Theta(-z)\vec{E}_{m}+\Theta(z)\vec{E}_{MO},\label{eq:totalE}\\
\vec{B} & = & \Theta(-z)\vec{B}_{m}+\Theta(z)\vec{B}_{MO},\label{eq:totalB}
\end{eqnarray}

\end{subequations} and $(\vec{E})_{i}$, $(\vec{B})_{i}$ denote
the $i$-th components of the respective fields (which include the
plane wave exponential factors, see (\ref{eq:E_BIG_superposition})--(\ref{eq:B_m})).
In Eq. (\ref{eq:energy_sp}), we have absorbed the electric field
units into $\mathcal{A}_{\boldsymbol{k}}$, so $E_{i}$ is taken to
be a unitless quantity. The integration $\int dV$ is carried out
over all 3D space.

Assuming a finite size box of in-plane area $S$ and infinite perpendicular
dimension and plugging in Eqs. (\ref{eq:E_BIG_superposition}), (\ref{eq:B_BIG_superposition}),
(\ref{eq:E_m}), and (\ref{eq:B_m}) into Eq. (\ref{eq:energy_sp}),
we obtain,

\begin{eqnarray}
H_{SP,\boldsymbol{k}} & = & \frac{S|\mathcal{A}_{\boldsymbol{k}}|^{2}}{2}\sum_{i}\int_{-\infty}^{\infty}dz\Big(\epsilon_{0}\sum_{j}\frac{d(\omega\epsilon_{ij}^{*}(\omega))}{d\omega}(\vec{E})_{j}^{*}(\vec{E})_{i}+\frac{1}{\mu_{0}\mu}|(\vec{B})_{i}|^{2}\Big)\nonumber \\
 & = & \frac{S|\mathcal{A}_{\boldsymbol{k}}|^{2}}{2}\sum_{i}\sum_{\gamma,\delta\in\{+,-\}}(t_{MO}^{\gamma})^{*}t_{MO}^{\delta}\Bigg[\epsilon_{0}\sum_{j}\epsilon_{MO,ij}^{*}(E_{j,MO}^{\gamma})^{*}E_{i,MO}^{\delta}+\frac{1}{\mu_{0}\mu}(B_{i,MO}^{\gamma})^{*}B_{i,MO}^{\delta}\Bigg]\nonumber \\
 &  & \,\,\,\,\,\,\,\,\,\,\,\,\,\,\,\,\,\,\,\,\,\,\,\,\,\,\,\,\times\Bigg[\int_{0}^{\infty}dze^{-[(\alpha_{MO}^{\gamma})^{*}+\alpha_{MO}^{\delta}]z}\Bigg]\nonumber \\
 &  & +\frac{S|\mathcal{A}_{\boldsymbol{k}}|^{2}}{2}\sum_{i}\Bigg[\epsilon_{0}\frac{d(\omega\epsilon_{m}(\omega))}{d\omega}|E_{i,m}|^{2}+\frac{1}{\mu_{0}\mu}|B_{i,m}|^{2}\Bigg]\Bigg[\int_{-\infty}^{0}dze^{(\alpha_{0}^{*}+\alpha_{0})z}\Bigg]\nonumber \\
 & = & S|\mathcal{A}_{\boldsymbol{k}}|^{2}\sum_{i}\sum_{\gamma,\delta\in\{+,-\}}(t_{MO}^{\gamma})^{*}t_{MO}^{\delta}\Bigg[\epsilon_{0}\sum_{j}\epsilon_{MO,ij}^{*}(E_{j,MO}^{\gamma})^{*}E_{i,MO}^{\delta}+\frac{1}{\mu_{0}\mu}(B_{i,MO}^{\gamma})^{*}B_{i,MO}^{\delta}\Bigg]\frac{1}{2(\alpha_{\gamma}^{*}+\alpha_{\delta})}\nonumber \\
 &  & +S|\mathcal{A}_{\boldsymbol{k}}|^{2}\sum_{i}\Bigg[\epsilon_{0}\frac{d(\omega\epsilon_{m}(\omega))}{d\omega}|E_{i,m}|^{2}+\frac{1}{\mu_{0}\mu}|B_{i,m}|^{2}\Bigg]\frac{1}{2(\alpha_{m}^{*}+\alpha_{m})}\nonumber \\
 & \equiv & S\Bigg(\frac{\epsilon_{0}L_{\boldsymbol{k}}}{4}\Bigg)\Big(2|\mathcal{A}_{\boldsymbol{k}}|^{2}\Big)\nonumber \\
 & = & S\frac{\epsilon_{0}L_{\boldsymbol{k}}}{4}(\mathcal{A}_{\boldsymbol{k}}\mathcal{A}_{\boldsymbol{k}}^{*}+\mathcal{A}_{\boldsymbol{k}}^{*}\mathcal{A}_{\boldsymbol{k}}),\label{eq:H_SP_k_supp}
\end{eqnarray}
where, following \cite{archambault}, we have defined the vertical
mode length as,

\begin{eqnarray}
L_{\boldsymbol{k}} & = & \sum_{i}\sum_{\gamma,\delta\in\{+,-\}}(t_{MO}^{\gamma})^{*}t_{MO}^{\delta}\Bigg[\epsilon_{0}\sum_{j}\epsilon_{MO,ij}^{*}(E_{j,MO}^{\gamma})^{*}E_{i,MO}^{\delta}+\frac{1}{\mu_{0}\mu}(B_{i,MO}^{\gamma})^{*}B_{i,MO}^{\delta}\Bigg]\frac{1}{\epsilon_{0}(\alpha_{\gamma}^{*}+\alpha_{\delta})}\nonumber \\
 &  & +\sum_{i}\Bigg[\epsilon_{0}\frac{d(\omega\epsilon_{m}(\omega))}{d\omega}|E_{i,m}|^{2}+\frac{1}{\mu_{0}\mu}|B_{i,m}|^{2}\Bigg]\frac{1}{\epsilon_{0}(\alpha_{m}^{*}+\alpha_{m})}.\label{eq:C_w}
\end{eqnarray}
Since $E_{i}$ is taken to be unitless, $L_{\boldsymbol{k}}$ effectively
has units of length. Physically, $L_{\boldsymbol{k}}$ defines a mode
volume $SL_{\boldsymbol{k}}$ with a quantized amount of energy corresponding
to the frequency $\omega_{\boldsymbol{k}}$. Importantly Eq. (\ref{eq:H_SP_k_supp})
is quadratic in the electric field amplitude $\mathcal{A}_{\boldsymbol{k}}$,
which implies that each $\boldsymbol{k}$ mode corresponds to a harmonic
oscillator. If we wish to quantize the energy in quanta of $\omega_{\boldsymbol{k}}$,

\begin{equation}
H_{SP,\boldsymbol{k}}=\frac{\omega(\boldsymbol{k})}{2}(\alpha_{\boldsymbol{k}}\alpha_{\boldsymbol{k}}^{*}+\alpha_{\boldsymbol{k}}^{*}\alpha_{\boldsymbol{k}}),\label{eq:quantization}
\end{equation}
we can define $\alpha_{\boldsymbol{k}}$ so that,

\begin{eqnarray}
\mathcal{A}_{\boldsymbol{k}} & = & \sqrt{\frac{\omega(\boldsymbol{k})}{SL_{\boldsymbol{k}}}}\alpha_{\boldsymbol{k}}.\label{eq:A_in_terms_of_alpha}
\end{eqnarray}
Other multiplicative phase factors in this amplitude definition (U(1)
gauge choice) do not affect the quantization. Promoting the complex
amplitudes to operators, $\alpha_{\boldsymbol{k}}\to a_{\boldsymbol{k}}$
and $\alpha_{\boldsymbol{k}}^{*}\to a_{\boldsymbol{k}}^{\dagger}$
with $[a_{\boldsymbol{k}},a_{\boldsymbol{k}}^{\dagger}]=1$,

\begin{eqnarray}
H_{SP,\boldsymbol{k}} & = & \frac{\omega(\boldsymbol{k})}{2}(a_{\boldsymbol{k}}a_{\boldsymbol{k}}^{\dagger}+a_{\boldsymbol{k}}^{\dagger}a_{\boldsymbol{k}})\nonumber \\
 & = & \omega(\boldsymbol{k})\Big(a_{\boldsymbol{k}}a_{\boldsymbol{k}}^{\dagger}+\frac{1}{2}\Big).\label{eq:quantization_final}
\end{eqnarray}
To summarize, we have normalized each $\boldsymbol{k}$ mode in Eqs.
(\ref{eq:totalE}) and (\ref{eq:totalB}) by associating energy quanta
$\omega_{\boldsymbol{k}}$ to a SP excitation in such mode. Finally,
the final electric field and magnetic induction are superpositions
of amplitudes in such modes,

\begin{subequations}

\begin{eqnarray}
\vec{\mathcal{E}} & = & \sum_{\boldsymbol{k}}\mathcal{A}_{\boldsymbol{k}}\vec{E}(\boldsymbol{k}).\label{eq:superposition_E}\\
\vec{\mathcal{B}} & = & \sum_{\boldsymbol{k}}\mathcal{A}_{\boldsymbol{k}}\vec{B}(\boldsymbol{k}).\label{eq:superposition_B}
\end{eqnarray}
\end{subequations} Promoting these amplitudes to operators (in the
Heisenberg picture) and using Eq. (\ref{eq:A_in_terms_of_alpha}),

\begin{subequations}

\begin{eqnarray}
\hat{\vec{\mathcal{E}}}(\boldsymbol{r},t) & = & \sum_{\boldsymbol{k}}2\sqrt{\frac{\omega(\boldsymbol{k})}{2\epsilon_{0}SL_{\boldsymbol{k}}}}a_{\boldsymbol{k}}\vec{E}(\boldsymbol{k})\nonumber \\
 & = & \sum_{\boldsymbol{k}}\sqrt{\frac{\omega(\boldsymbol{k})}{2\epsilon_{0}SL_{\boldsymbol{k}}}}a_{\boldsymbol{k}}\vec{E}(\boldsymbol{k})+\sqrt{\frac{\omega(\boldsymbol{k})}{2\epsilon_{0}SL_{\boldsymbol{k}}}}a_{\boldsymbol{k}}^{\dagger}\vec{E}^{*}(\boldsymbol{k}),\label{eq:operator_E}\\
\hat{\vec{\mathcal{B}}}(\boldsymbol{r},t) & = & \sum_{\boldsymbol{k}}\sqrt{\frac{\omega(\boldsymbol{k})}{2\epsilon_{0}SL_{\boldsymbol{k}}}}a_{\boldsymbol{k}}\vec{B}(\boldsymbol{k})+\sqrt{\frac{\omega(\boldsymbol{k})}{2\epsilon_{0}SL_{\boldsymbol{k}}}}a_{\boldsymbol{k}}^{\dagger}\vec{B}^{*}(\boldsymbol{k}),\label{eq:operator_B}
\end{eqnarray}
\end{subequations}where we have used the fact that $\vec{\mathcal{E}}$
and $\vec{\mathcal{B}}$ are real valued to write the operators in
a more conventionally symmetric form.

\subsection{Perturbation expansion of $L_{\boldsymbol{k}}$}

The fields deduced in Sec. \ref{sec:Perturbation-expansion-on} are
only correct up to $O(g)$, so it is important that we keep $L_{\boldsymbol{k}}$
up to that order too. Importantly, since $(\overleftrightarrow{\epsilon}_{MO})_{r\theta}=ig$
and $E_{\theta d}\propto g$, the lowest order contribution to $L_{\boldsymbol{k}}$
appears at $O(g^{2})$. Hence, we do not need to take the anisotropic
effects of $\epsilon_{MO}$ into account, and we can set $g=0$ in
Eq. (\ref{eq:e_tensor_supp}), i.e. $\overleftrightarrow{\epsilon}_{MO}\approx\epsilon_{d}\mathbb{I}$,
where $\mathbb{I}$ is the $3\times3$ identity matrix. Therefore,
the calculation for $L_{\boldsymbol{k}}\approx L_{\boldsymbol{k}0}$
reduces to that of the standard SP mode in the absence of external
magnetic field ($\alpha_{d0}$ and $\alpha_{m0}$ are purely real)
\cite{archambault},

\begin{eqnarray}
L_{\boldsymbol{k}0} & = & \sum_{i}\Bigg\{\Bigg[\epsilon_{0}\epsilon_{d}|E_{i,d0}|^{2}+\frac{1}{\mu_{0}\mu}|B_{i,d0}|^{2}\Bigg]\frac{1}{2\epsilon_{0}\alpha_{d0}}+\Bigg[\epsilon_{0}\frac{d(\omega\epsilon_{m0}(\omega))}{d\omega}\Bigg|_{\omega=\frac{\Omega_{0}}{c}}|E_{i,m0}|^{2}+\frac{1}{\mu_{0}\mu}|B_{i,m0}|^{2}\Bigg]\frac{1}{2\epsilon_{0}\alpha_{m0}}\Bigg\}\nonumber \\
 & = & \Bigg[\epsilon_{d}\Bigg(1+\frac{k^{2}}{\alpha_{d0}^{2}}\Bigg)+\Bigg(\frac{\epsilon_{d}\Omega_{0}}{\alpha_{d0}}\Bigg)^{2}\Bigg]\frac{1}{2\alpha_{d0}}+\Bigg[\frac{d(\omega\epsilon_{m0}(\omega))}{d\omega}\Bigg|_{\omega=\frac{\Omega_{0}}{c}}\Bigg(1+\frac{k^{2}}{\alpha_{m0}^{2}}\Bigg)+\Bigg(\frac{\epsilon_{d}\Omega_{0}}{\alpha_{d0}}\Bigg)^{2}\Bigg]\frac{1}{2\alpha_{m0}}\nonumber \\
 & = & \frac{-\epsilon_{m0}}{\alpha_{d0}}+\frac{1}{2\alpha_{m0}}\Bigg[\frac{d(\omega\epsilon_{m0}(\omega))}{d\omega}\Bigg|_{\omega=\frac{\Omega_{0}}{c}}\Bigg(\frac{\epsilon_{m0}-\epsilon_{d}}{\epsilon_{m0}}\Bigg)-\epsilon_{m0}-\epsilon_{d}\Bigg].\label{eq:Lk_calculation}
\end{eqnarray}
In this derivation, we have used $\mu=1$, $\epsilon_{0}\mu_{0}=c^{-2}$,
Eqs. (\ref{eq:k_Omega_0}), (\ref{eq:Ezd0}), and (\ref{eq:Ezm0})
as well as the expressions for (\ref{eq:alphad0}), and (\ref{eq:alpham0}).
\footnote{The final result in Eq. (\ref{eq:Lk_calculation}) is twice of what
is reported in the Supplementary Material of \cite{archambault}.
We believe our derivation has the correct prefactors; however, the
use of either result gives the same order of magnitude of the effects
we are interested in.}

\section{Magneto-SPs in a three-layer setup\label{sec:3layers}}

\begin{figure}
\centering{}\includegraphics[scale=0.35]{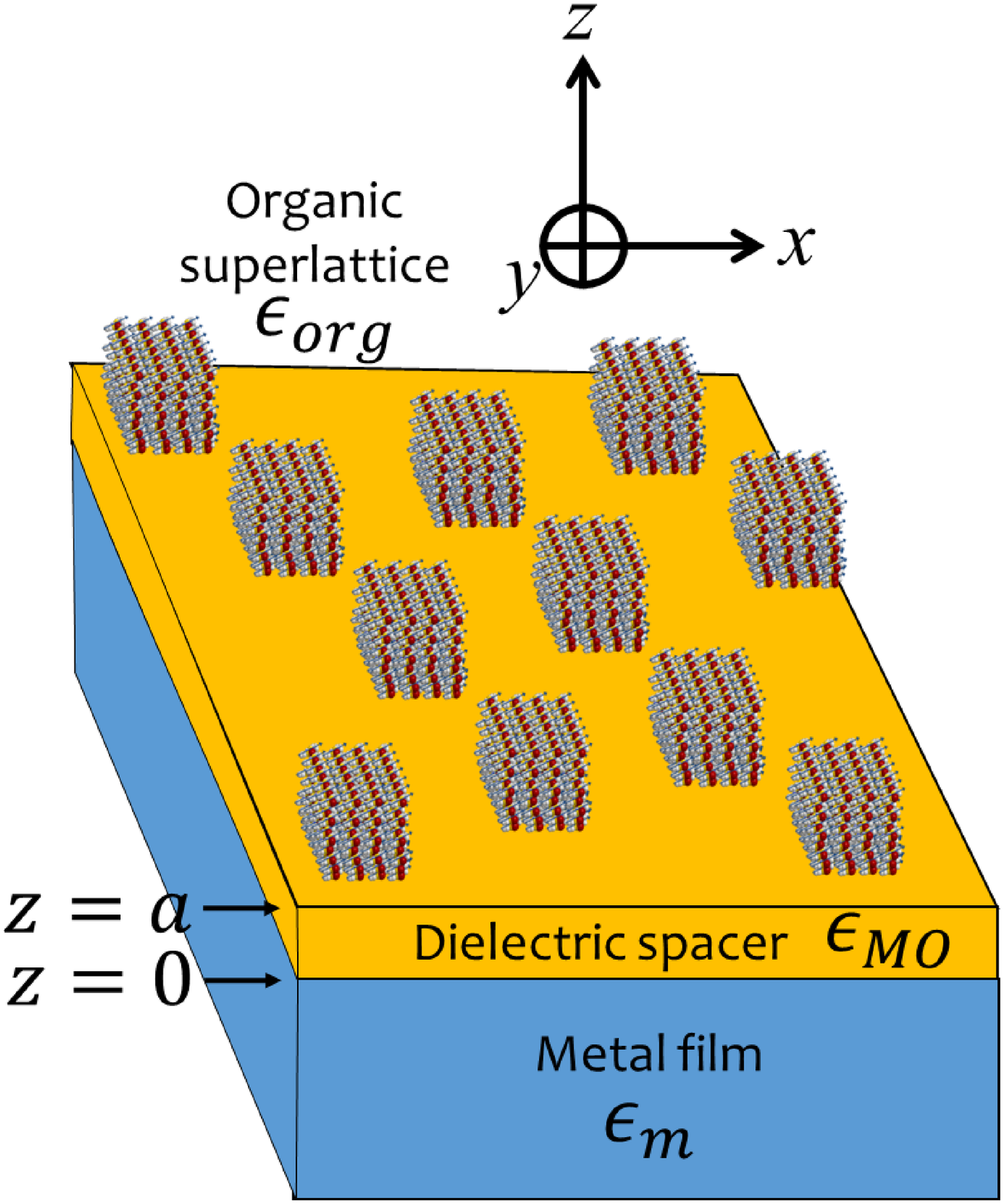}\protect\caption{\emph{Three-layer (metal-MO dielectric-organic) setup. }We are interested
in the (magneto)-SP modes arising at the dielectric-organic interface
($z=a$) upon application of a perpendicular external magnetic field.\emph{
}\label{fig:Three-layer-setup}}
\end{figure}

We shall now adapt the results from the previous sections to the situation
where the MO layer has a finite height $a$, and an organic layer
of isotropic dielectric $\epsilon_{org}$ is placed on top of it (see
Fig. \ref{fig:Three-layer-setup}). As far as we are aware, the resulting
expressions for the corresponding magneto-SPs have not appeared before
in the literature.

\subsection{Electromagnetic modes for each layer}

\subsubsection{Organic layer ($z>a$)}

Just as we did for the MO and the metal layers (see\emph{ }Eqs. (\ref{eq:ME=00003D0})
and (\ref{eq:secular_metal})), Eq. (\ref{eq:wave_equation}) for
the organic layer can be expressed in matrix form $\mathbb{M}_{org}\boldsymbol{E}_{org}=0$,

\begin{equation}
\left[\begin{array}{ccc}
\epsilon_{m}-\Omega^{-2}k_{z,org}^{2} & 0 & \Omega^{-2}kk_{z,org}\\
0 & \epsilon_{m}-\Omega^{-2}(k^{2}+k_{z,org}^{2}) & 0\\
kk_{z,org}\Omega^{-2} & 0 & \epsilon_{m}-\Omega^{-2}k^{2}
\end{array}\right]\begin{bmatrix}E_{k,org}\\
E_{\theta,org}\\
E_{z,org}
\end{bmatrix}=\begin{bmatrix}0\\
0\\
0
\end{bmatrix}.\label{eq:secular_metal-1}
\end{equation}
Importantly, we do not fix the radial component $E_{k,org}$ to $1$
because of the boundary conditions at the organic crystal and MO interface
at $z=a$\footnote{In our (arbitrary) normalization before quantization, we may set only
one of the field components in one of the layers to 1, and our convention
is to choose $E_{r,m}=1$ as in the two-layer case. The rest of the
fields are not arbitrary and satisfy the wave equation Eq. (\ref{eq:wave_equation})
as well as boundary conditions at each of the interfaces. }. The corresponding secular equation for the decaying field for $z>a$,
with $k_{z,org}=i\alpha_{org}$ yields

\begin{equation}
\alpha_{org}=\sqrt{k^{2}-\Omega^{2}\epsilon_{org}},\label{eq:alpha_org}
\end{equation}
as expected (see Eq. (\ref{eq:soln_2nd_quad_eqn})). Finally, just
like in the metal layer, Eq. (\ref{eq:secular_metal-1}) tells us
that

\begin{eqnarray}
E_{z,org} & = & \frac{ik}{\alpha_{org}}E_{r,org},\label{eq:E_z_m-1}
\end{eqnarray}
but does not inform us about the tangential component $E_{\theta,org}$
(nor about $E_{k,org}$), The missing components will be deduced by
matching the boundary at $z=a$. Altogether, the fields in the organic
layer read like those in Eqs. (\ref{eq:E_BIG}) and (\ref{eq:B_BIG}),

\begin{subequations}

\begin{eqnarray}
\vec{E}_{org} & = & \boldsymbol{E}_{org}\eta e^{-\alpha_{org}z}\nonumber \\
 & = & (E_{k,org},E_{\theta,m},E_{z,m})\eta e^{\alpha_{m}z},\label{eq:E_m-1-1}\\
\vec{B}_{org} & = & \boldsymbol{B}_{org}\eta e^{-\alpha_{org}z}\nonumber \\
 & = & \frac{-i}{\omega}(\alpha_{org}E_{\theta,org},-ikE_{z,org}-\alpha_{org}E_{k,org},ikE_{\theta,org})\eta e^{-\alpha_{org}z},\label{eq:B_m-1-1}
\end{eqnarray}

\end{subequations}denoting exponentially decreasing fields.

\subsubsection{MO layer ($a>z>0$)}

For the MO layer, we translate Eqs. (\ref{eq:E_BIG}) and (\ref{eq:B_BIG})
to this setup,

\begin{subequations}

\begin{eqnarray}
\vec{E}_{MO\downarrow}^{\pm} & = & \boldsymbol{E}_{MO\downarrow}^{\pm}\eta e^{-\alpha^{\pm}z}\nonumber \\
 & = & (1,E_{\theta,MO\downarrow}^{\pm},E_{z,MO\downarrow}^{\pm})\eta e^{-\alpha^{\pm}z},\label{eq:E_BIG_down}\\
\vec{B}_{MO\downarrow}^{\pm} & = & \boldsymbol{B}_{MO\downarrow}^{\pm}\eta e^{-\alpha^{\pm}z}\nonumber \\
 &  & \frac{-i}{\omega}(\alpha^{\pm}E_{\theta,MO\downarrow}^{\pm},-ikE_{z,MO\downarrow}^{\pm}-\alpha^{\pm},ikE_{\theta,MO\downarrow}^{\pm})\eta e^{-\alpha^{\pm}z},\label{eq:B_BIG_down}
\end{eqnarray}

\end{subequations}where $\vec{E}_{MO\downarrow}^{\pm}$ and $\vec{B}_{MO\downarrow}^{\pm}$
indicate exponentially decreasing fields ($k_{z}=i\alpha_{MO}^{\pm}$),
and by slightly adapting these expressions,

\begin{subequations}

\begin{eqnarray}
\vec{E}_{MO\uparrow}^{\pm} & = & \boldsymbol{E}_{MO\uparrow}^{\pm}\eta e^{\alpha^{\pm}z}\nonumber \\
 & = & (1,E_{\theta,MO\uparrow}^{\pm},E_{z,MO\uparrow}^{\pm})\eta e^{\alpha^{\pm}z},\label{eq:E_BIG_up}\\
\vec{B}_{MO\uparrow}^{\pm} & = & \boldsymbol{B}_{MO\uparrow}^{\pm}\eta e^{\alpha^{\pm}z}\nonumber \\
 &  & \frac{-i}{\omega}(-\alpha^{\pm}E_{\theta,MO\uparrow}^{\pm},-ikE_{z,MO\uparrow}^{\pm}+\alpha^{\pm},ikE_{\theta,MO\uparrow}^{\pm})\eta e^{\alpha^{\pm}z},\label{eq:B_BIG_up}
\end{eqnarray}

\end{subequations}where $\vec{E}_{MO\uparrow}^{\pm}$ and $\vec{B}_{MO\uparrow}^{\pm}$
denote exponentially increasing fields ($k_{z}=-i\alpha_{MO}^{\pm}$).
In previous sections where MO was considered to fill up all the space
$z>0$, the latter fields were not considered, the reason being that
$e^{\alpha_{MO}^{\pm}z}$ was unbounded as $z\to\infty$; this is
not the case when the largest value of $z$ is $a$ \footnote{A more intuitive way to describe this situation is that $\vec{E}_{BIG\downarrow}^{\pm}$
and $\vec{B}_{BIG\downarrow}^{\pm}$ denote fields that exponentially
decrease starting from $z=0$ going upwards; similarly, $\vec{E}_{BIG\uparrow}^{\pm}$
and $\vec{B}_{BIG\uparrow}^{\pm}$ describe exponentially decreasing
fields starting from $z=a$ going downwards.}. Hence, the analogous expressions to Eqs. (\ref{eq:E_BIG_superposition})
and (\ref{eq:B_BIG_superposition}) are,

\begin{subequations}

\begin{eqnarray}
\vec{E}_{MO} & = & t_{\downarrow}^{+}\vec{E}_{MO\downarrow}^{+}+t_{\downarrow}^{-}\vec{E}_{MO\downarrow}^{-}+t_{\uparrow}^{+}\vec{E}_{MO\uparrow}^{+}+t_{\uparrow}^{-}\vec{E}_{MO\uparrow}^{-},\label{eq:E_BIG_superposition-1}\\
\vec{B}_{MO} & = & t_{\downarrow}^{+}\vec{B}_{MO\downarrow}^{+}+t_{\downarrow}^{-}\vec{B}_{MO\downarrow}^{-}+t_{\uparrow}^{+}\vec{B}_{MO\uparrow}^{+}+t_{\uparrow}^{-}\vec{B}_{MO\uparrow}^{-}.\label{eq:B_BIG_superposition-1}
\end{eqnarray}

\end{subequations}where $t_{\downarrow}^{\pm}$ and $t_{\uparrow}^{\pm}$
are the unknown coefficients. Here, $E_{\theta,MO\downarrow}^{\pm}=E_{\theta,MO\uparrow}^{\pm}=E_{\theta,MO}^{\pm}$
but $E_{z,MO\downarrow}^{\pm}=-E_{z,MO\uparrow}^{\pm}=E_{z,MO}^{\pm}$.
These identities are easy to check as $\vec{E}_{MO\downarrow}^{\pm}$
is associated with $k_{z}=i\alpha_{MO}^{\pm}$ and $\vec{E}_{MO\uparrow}^{\pm}$
with $k_{z}=-i\alpha_{MO}^{\pm}$, but Eqs. (\ref{eq:E_theta_BIG})
and (\ref{eq:E_z_BIG}) were derived for $k_{z}=i\alpha_{MO}^{\pm}$.

\subsubsection{Metal layer ($z<0$)}

Finally, for the metal layer, all the expressions we derived in Sec.
\ref{sec:Interface-between-a} hold, in particular Eqs. (\ref{eq:E_m})
and (\ref{eq:B_m}); we still assume $E_{k,m}=1$.

\subsection{Matching the modes at the boundaries ($z=0$ and $z=a$)}

Given this prelude, the analogous boundary conditions to Eqs. (\ref{eq:constraint1})--(\ref{eq:constraint6})
for $z=0$ are,

\begin{subequations}

\begin{eqnarray}
1 & = & t_{MO\downarrow}^{+}+t_{MO\downarrow}^{-}+t_{MO\uparrow}^{+}+t_{MO\uparrow}^{-},\label{eq:constraint1-1}\\
E_{\theta,m} & = & t_{MO\downarrow}^{+}E_{\theta,MO\downarrow}^{+}+t_{MO\downarrow}^{-}E_{\theta,MO\downarrow}^{-}+t_{MO\uparrow}^{+}E_{\theta,MO\uparrow}^{+}+t_{MO\uparrow}^{-}E_{\theta,MO\uparrow}^{-},\label{eq:constraint2-1}\\
\epsilon_{m}E_{z,m} & = & \epsilon_{d}(t_{MO\downarrow}^{+}E_{z,MO\downarrow}^{+}+t_{MO\downarrow}^{-}E_{z,MO\downarrow}^{-}+t_{MO\uparrow}^{+}E_{z,MO\uparrow}^{+}+t_{MO\uparrow}^{-}E_{z,MO\uparrow}^{-}),\label{eq:constraint3-1}\\
\alpha_{m}E_{\theta,m} & = & -(\alpha_{MO}^{+}t_{MO\downarrow}^{+}E_{\theta,MO\downarrow}^{+}+\alpha_{MO}^{-}t_{MO\downarrow}^{-}E_{\theta,MO\downarrow}^{-})+(\alpha_{MO}^{+}t_{MO\uparrow}^{+}E_{\theta,MO\uparrow}^{+}+\alpha_{MO}^{-}t_{MO\uparrow}^{-}E_{\theta,MO\uparrow}^{-}),\label{eq:constraint4-1}\\
kE_{z,m}+i\alpha_{m} & = & k(t_{MO\downarrow}^{+}E_{z,MO\downarrow}^{+}+t_{MO\downarrow}^{-}E_{z,MO\downarrow}^{-}+t_{MO\uparrow}^{+}E_{zMO\uparrow}^{+}+t_{MO\uparrow}^{-}E_{z,MO\uparrow}^{-})\nonumber \\
 &  & -i\alpha_{MO}^{+}(t_{MO\downarrow}^{+}-t_{MO\uparrow}^{+})-i\alpha_{MO}^{-}(t_{MO\downarrow}^{-}-t_{MO\uparrow}^{-}),\label{eq:constraint5-1}\\
E_{\theta,m} & = & t_{MO\downarrow}^{+}E_{\theta,MO\downarrow}^{+}+t_{MO\downarrow}^{-}E_{\theta,MO\downarrow}^{-}+t_{MO\uparrow}^{+}E_{\theta,MO\uparrow}^{+}+t_{MO\uparrow}^{-}E_{\theta,MO\uparrow}^{-},\label{eq:constraint6-1}
\end{eqnarray}

\end{subequations}whereas for $z=a$ they are,

\begin{subequations}

\begin{eqnarray}
E_{k,org} & = & t_{MO\downarrow}^{+}\chi^{+}+t_{MO\downarrow}^{-}\chi^{-}+t_{MO\uparrow}^{+}\frac{1}{\chi^{+}}+t_{MO\uparrow}^{-}\frac{1}{\chi^{-}},\label{eq:constraint1-2}\\
E_{\theta,org} & = & t_{MO\downarrow}^{+}E_{\theta,MO\downarrow}^{+}\chi^{+}+t_{MO\downarrow}^{-}E_{\theta,MO\downarrow}^{-}\chi^{-}+t_{MO\uparrow}^{+}E_{\theta,MO\uparrow}^{+}\frac{1}{\chi^{+}}+t_{MO\uparrow}^{-}E_{\theta,MO\uparrow}^{-}\frac{1}{\chi^{-}},\label{eq:constraint2-2}\\
\epsilon_{org}E_{z,org} & = & \epsilon_{d}\Bigg(t_{MO\downarrow}^{+}E_{z,MO\downarrow}^{+}\chi^{+}+t_{MO\downarrow}^{-}E_{z,MO\downarrow}^{-}\chi^{-}+t_{MO\uparrow}^{+}E_{z,MO\uparrow}^{+}\frac{1}{\chi^{+}}+t_{MO\uparrow}^{-}E_{z,MO\uparrow}^{-}\frac{1}{\chi^{-}}\Bigg),\label{eq:constraint3-2}\\
-\alpha_{org}E_{\theta,org} & = & -\Bigg(\alpha_{MO}^{+}t_{MO\downarrow}^{+}E_{\theta,MO\downarrow}^{+}\chi^{+}+\alpha_{MO}^{-}t_{MO\downarrow}^{-}E_{\theta,MO\downarrow}^{-}\chi^{-}\Bigg)\nonumber \\
 &  & +\Bigg(\alpha_{MO}^{+}t_{MO\uparrow}^{+}E_{\theta,MO\uparrow}^{+}\frac{1}{\chi^{+}}+\alpha_{MO}^{-}t_{MO\uparrow}^{-}E_{\theta,MO\uparrow}^{-}\frac{1}{\chi^{-}}\Bigg),\label{eq:constraint4-2}\\
kE_{z,org}-i\alpha_{org}E_{r,org} & = & k\Bigg(t_{MO\downarrow}^{+}E_{z,MO\downarrow}^{+}\chi^{+}+t_{MO\downarrow}^{-}E_{z,MO\downarrow}^{-}\chi^{-}+t_{MO\uparrow}^{+}E_{z,MO\uparrow}^{+}\frac{1}{\chi^{+}}+t_{MO\uparrow}^{-}E_{z,MO\uparrow}^{-}\frac{1}{\chi^{-}}\Bigg)\nonumber \\
 &  & -i\alpha_{MO}^{+}\Bigg(t_{MO\downarrow}^{+}\chi^{+}-t_{MO\uparrow}^{+}\frac{1}{\chi^{+}}\Bigg)-i\alpha_{MO}^{-}\Bigg(t_{MO\downarrow}^{-}\chi^{-}-t_{MO\uparrow}^{-}\frac{1}{\chi^{-}}\Bigg),\label{eq:constraint5-2}\\
E_{\theta,org} & = & t_{MO\downarrow}^{+}E_{\theta,MO\downarrow}^{+}\chi^{+}+t_{MO\downarrow}^{-}E_{\theta,MO\downarrow}^{-}\chi^{-}+t_{MO\uparrow}^{+}E_{\theta,MO\uparrow}^{+}\frac{1}{\chi^{+}}+t_{MO\uparrow}^{-}E_{\theta,MO\uparrow}^{-}\frac{1}{\chi^{-}}.\label{eq:constraint6-2}
\end{eqnarray}

\end{subequations}Here, $\chi^{\pm}=e^{-\alpha_{MO}^{\pm}a}$ embodies
the vertical thickness dependence of the problem. Eqs. (\ref{eq:constraint1-1})--(\ref{eq:constraint6-2})
can be manipulated to yield an entirely analogous expression to Eq.
(\ref{eq:final_equation}). This resulting expression can then be
perturbatively expanded in $g$ and the analogous procedure of Sec.
\ref{sec:Perturbation-expansion-on} follows, with the caveat that
the algebra becomes much more laborious. We bypass the latter by automatizing
such work with Wolfram Mathematica(c). We summarize the results in
the following subsections.

\subsection{Perturbation expansion on $g$}

\subsubsection{Solving for $t_{\downarrow}^{\pm}$ and $t_{\uparrow}^{\pm}$}

\begin{figure}
\begin{centering}
\includegraphics[scale=0.5]{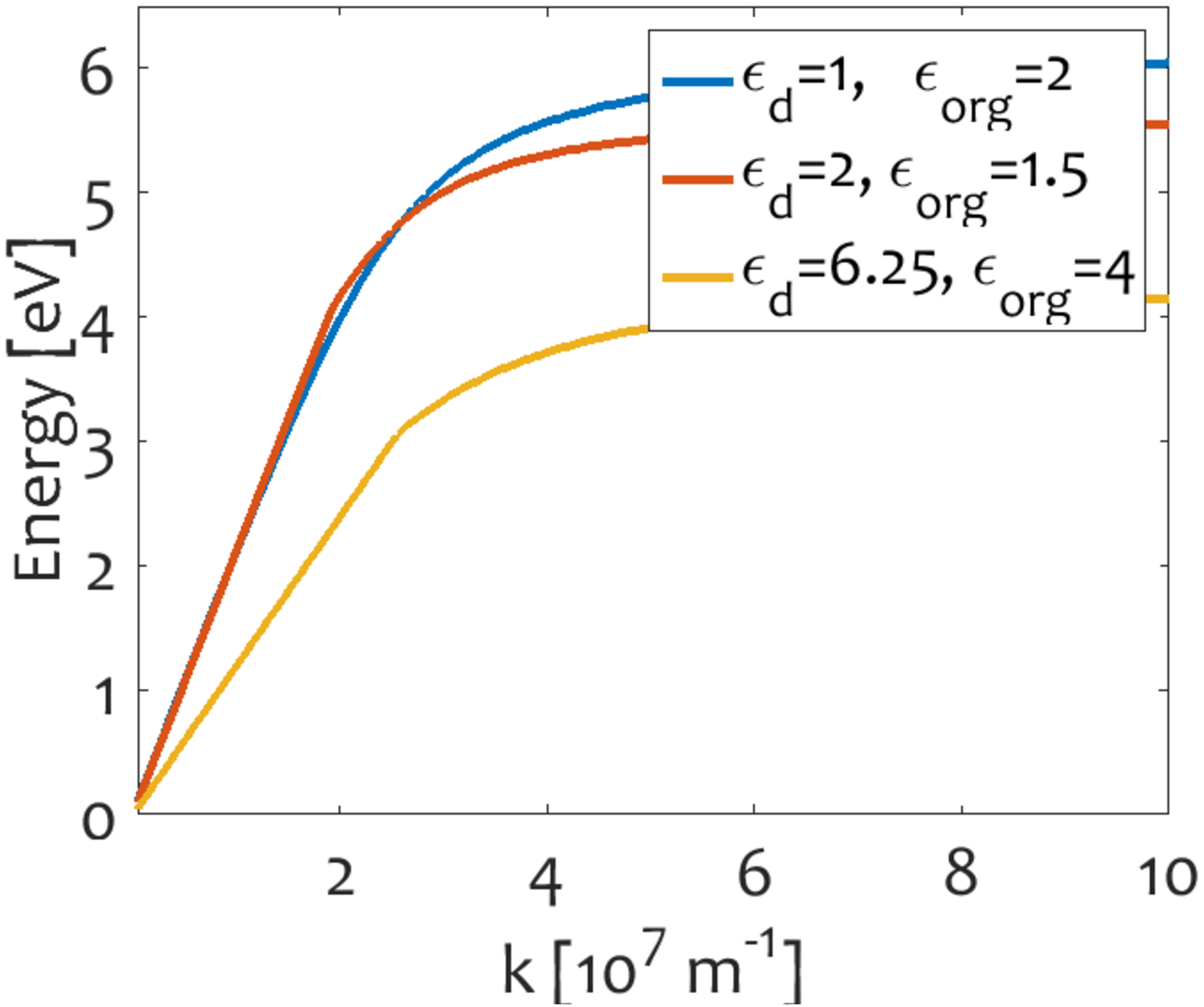}
\par\end{centering}

\protect\caption{\emph{SP dispersion energy as a function of $k=|\boldsymbol{k}|$
for the three-layer (metal-MO dielectric-organic) setup, assuming
$g=0$}. Plots generated using Eq. (\ref{eq:chi_0}) with Drude parameters
for Ag, $\epsilon_{\infty}\sim4$, $\omega_{P}\sim9$~eV, and varying
the dielectric permittivity $\epsilon_{d}$ as well as that of the
organic layer. Since $\Omega=\Omega_{0}+O(g^{2})$, the plots are
correct up to $O(g)$, which is our perturbation order of interest.
Just like with Fig. (\ref{fig:SP-dispersion-energy}), notice that
the dispersion curves start-off linearly (light-like excitations),
but plateau to constant values at large wavevectors (charge oscillations
in the metal), which become smaller as the dielectric permittivity
increases.\label{fig:SP-dispersion-energy-three-layers}}
\end{figure}

It is clear that provided that the frequencies $\Omega_{0}$ and $\Omega_{2}$
are modified accordingly (one finds $\Omega_{1}=0$ again), the expansions
for the dielectric constants and the wavevectors can all be recycled
from Sec. \ref{sec:Perturbation-expansion-on}. On top of these quantities,
following Sec. \ref{sec:Perturbation-expansion-on}, we ought to expand
$\alpha_{org}$ and $\chi^{\pm}$ up to $O(g^{2})$ in Eq. (\ref{eq:alpha_org}),

\begin{subequations}

\begin{eqnarray}
\alpha_{org} & \approx & \alpha_{org0}+g^{2}\alpha_{org22}\Omega_{2},\label{eq:alpha_org_approx}\\
\alpha_{org0} & = & \sqrt{k^{2}-\Omega_{0}^{2}\epsilon_{org}},\label{eq:alpha_org_0}\\
\alpha_{org22} & = & -\frac{\Omega_{0}\epsilon_{org}}{\alpha_{org0}},\label{eq:alpha_org_22}
\end{eqnarray}

\end{subequations}

\begin{equation}
\chi^{\pm}\approx\chi_{0}\Bigg\{1\mp ag\alpha_{d1}-g^{2}\Bigg[a^{2}\frac{\alpha_{d1}^{2}}{2}+a(\alpha_{d20}+\alpha_{d22}\Omega_{2})\Bigg]\Bigg\},\label{eq:chi_pm_approx}
\end{equation}
but we only need $t_{MO\downarrow}^{\pm}$ and $t_{MO\uparrow}^{\pm}$
up to $O(g)$,

\begin{subequations}

\begin{eqnarray}
t_{MO\downarrow}^{\pm} & \approx & t_{d\downarrow0}\pm gt_{d\downarrow1},\label{eq:tdown_pm}\\
t_{MO\uparrow}^{\pm} & \approx & t_{d\uparrow0}\pm gt_{d\uparrow1}.\label{eq:tup_pm}
\end{eqnarray}

\end{subequations}Collecting the zeroth-order in $g$ contributions
in Eqs. (\ref{eq:constraint1-1})--(\ref{eq:constraint6-2}), we derive
an implicit equation for $\Omega_{0}$ which coincides with the standard
textbook result for a three-layer system in the absence of an external
magnetic field \cite{maier},

\begin{equation}
\chi_{0}^{2}=\Bigg(\frac{\alpha_{d0}\epsilon_{m0}+\alpha_{m0}\epsilon_{d}}{\alpha_{d0}\epsilon_{m0}-\alpha_{m0}\epsilon_{d}}\Bigg)\Bigg(\frac{\alpha_{d0}\epsilon_{org}+\alpha_{org0}\epsilon_{d}}{\alpha_{d0}\epsilon_{org}-\alpha_{org0}\epsilon_{d}}\Bigg),\label{eq:chi_0}
\end{equation}
where $\alpha_{d0}$, $\alpha_{m0}$, $\alpha_{org0}$, and $\epsilon_{m0}$
are all functions of $\Omega_{0}$. This equation is the analogue
of Eqs. (\ref{eq:k_Omega_0}) and (\ref{eq:Omega_0}) for three layers.
It is not possible to explicitly solve for $\Omega_{0}$, but one
can readily compute it numerically from such implicit equation (see
Fig. \ref{fig:SP-dispersion-energy-three-layers}). The formula for
$\Omega_{2}$ is too long to display it here and is, anyway, not relevant
on its own. However, we make use of it to obtain our final results.

Next, we compute the coefficients in Eqs. (\ref{eq:tdown_pm}) and
(\ref{eq:tup_pm}) in analogy to the two-layer case in Eqs. (\ref{eq:t10})
and (\ref{eq:t11}). At zeroth-order in $g$,

\begin{subequations}

\begin{eqnarray}
t_{d\downarrow0} & = & \frac{\alpha_{m0}\epsilon_{d}-\alpha_{d0}\epsilon_{m0}}{4\alpha_{m0}\epsilon_{d}},\label{eq:td0_pm}\\
t_{d\uparrow0} & = & \frac{\alpha_{m0}\epsilon_{d}+\alpha_{d0}\epsilon_{m0}}{4\alpha_{m0}\epsilon_{d}},\label{eq:tu0_pm}
\end{eqnarray}

\end{subequations}while their first-order in $g$ corrections are,

\begin{subequations}

\begin{eqnarray}
t_{d\downarrow1} & = & \alpha_{d1}\Big(a\alpha_{d0}^{3}(\chi_{0}^{2}-1)\epsilon_{m0}\nonumber \\
 &  & -\alpha_{d0}^{2}\{\chi_{0}^{2}[a\epsilon_{m0}(\alpha_{m0}+\alpha_{org0})+a\alpha_{m0}\epsilon_{d}+\epsilon_{m0}]+a\alpha_{m0}\epsilon_{d}-a\alpha_{m0}\epsilon_{m0}+a\alpha_{org0}\epsilon_{m0}-\epsilon_{m0}\}\nonumber \\
 &  & +\alpha_{d0}\{a\alpha_{m0}^{2}(\chi_{0}^{2}+1)\epsilon_{d}+\alpha_{m0}(a\alpha_{org0}+1)[\chi_{0}^{2}(\epsilon_{d}+\epsilon_{m0})-\epsilon_{d}+\epsilon_{m0}]+2\alpha_{org0}\epsilon_{m0}\}\nonumber \\
 &  & -\alpha_{m0}^{2}(\chi_{0}^{2}-1)\epsilon_{d}(a\alpha_{org0}+1)\Big)/\nonumber \\
 &  & \{4\alpha_{m0}\epsilon_{d}[\chi_{0}^{2}(\alpha_{m0}-\alpha_{d0})(\alpha_{d0}-\alpha_{org0})+(\alpha_{d0}+\alpha_{m0})(\alpha_{d0}+\alpha_{org0})]\}\label{eq:td1}
\end{eqnarray}
and

\begin{eqnarray}
t_{d\uparrow1} & = & \Big[\alpha_{d1}\chi_{0}^{2}\Big(\alpha_{d0}\epsilon_{m0}\{a\alpha_{d0}^{2}-\alpha_{org0}[a(\alpha_{d0}+\alpha_{m0})+2]+a\alpha_{d0}\alpha_{m0}+\alpha_{d0}-\alpha_{m0}\}\nonumber \\
 &  & -\alpha_{m0}\epsilon_{d}(\alpha_{d0}+\alpha_{m0})[a(\alpha_{d0}-\alpha_{org0})-1]\Big)\nonumber \\
 &  & -\alpha_{d1}(\alpha_{d0}+\alpha_{m0})[a(\alpha_{d0}+\alpha_{org0})+1](\alpha_{d0}\epsilon_{m0}+\alpha_{m0}\epsilon_{d})\Big]/\nonumber \\
 &  & \{4\alpha_{m0}\epsilon_{d}[\chi_{0}^{2}(\alpha_{m0}-\alpha_{d0})(\alpha_{d0}-\alpha_{org0})+(\alpha_{d0}+\alpha_{m0})(\alpha_{d0}+\alpha_{org0})]\}.\label{eq:tu1}
\end{eqnarray}

\end{subequations}Hence, at zeroth order, $t_{MO\downarrow}^{+}+t_{MO\downarrow}^{-}=2t_{d\downarrow0}$
and $t_{MO\uparrow}^{+}+t_{MO\uparrow}^{-}=2t_{d\uparrow0}$, and
these total coefficients for exponentially decreasing and increasing
fields become identical to the textbook results for SP modes in the
three-layer setup in the absence of an external magnetic field.

\subsubsection{Collecting the expressions for the fields}

For reference, the zeroth-order fields are given by,

\begin{subequations}

\begin{eqnarray}
\vec{E}_{org0} & = & (E_{k,org0},0,E_{z,org0})\eta_{0}e^{-\alpha_{org0}z},\label{eq:Eorg0_3}\\
\vec{B}_{org0} & = & (0,B_{\theta,org0},0)\eta_{0}e^{-\alpha_{org0}z},\label{eq:Borg0_3}\\
\vec{E}_{d0} & = & 2t_{d\downarrow0}(1,0,E_{z,d\downarrow0})\eta_{0}e^{-\alpha_{d0}z}+2t_{d\uparrow0}(1,0,E_{z,d\uparrow0})\eta_{0}e^{\alpha_{d0}z},\label{eq:Ed0_3}\\
\vec{B}_{d0} & = & 2t_{d\downarrow0}(0,B_{\theta,d\downarrow0},0)\eta_{0}e^{-\alpha_{d0}z}+2t_{d\uparrow0}(0,B_{\theta,d\uparrow0},0)\eta_{0}e^{\alpha_{d0}z},\label{eq:Bd0_3}\\
\vec{E}_{m0} & = & (1,0,E_{z,m0})\eta_{0}e^{-\alpha_{m0}z},\label{eq:Em0_3}\\
\vec{B}_{m0} & = & (0,B_{\theta,m0},0)\eta_{0}e^{-\alpha_{m0}z},\label{eq:Bm0_3}
\end{eqnarray}

\end{subequations}with each of the components being,

\begin{subequations}

\begin{eqnarray}
E_{k,org0} & = & \frac{\alpha_{m0}(\chi_{0}^{2}+1)\epsilon_{d}-\alpha_{d0}(\chi_{0}^{2}-1)\epsilon_{m0}}{2\alpha_{m0}\chi_{0}\epsilon_{d}},\label{eq:Erorg0_3}\\
E_{z,org0} & = & -\frac{ik[\alpha_{d0}(\chi_{0}^{2}+1)\epsilon_{m0}+\alpha_{m0}(\epsilon_{d}-\chi_{0}^{2}\epsilon_{d})]}{2\alpha_{d0}\alpha_{m0}\chi_{0}\epsilon_{org}},\label{eq:Ezorg0_3}\\
B_{\theta,org0} & = & \frac{i\{2\alpha_{d0}\alpha_{m0}\alpha_{org0}\chi_{0}\epsilon_{org}+k^{2}[\alpha_{d0}(\chi_{0}^{2}+1)\epsilon_{m0}+\alpha_{m0}(\epsilon_{d}-\chi_{0}^{2}\epsilon_{d})]\}}{2\alpha_{d0}\alpha_{m0}\chi_{0}\Omega_{0}c\epsilon_{org}},\label{eq:Bthorg0_3}\\
E_{z,d\downarrow0} & = & -E_{z,d\uparrow0}=\frac{ik}{\alpha_{d0}},\label{eq:Ezd0_3}\\
B_{\theta,d\downarrow0} & = & -B_{\theta,d\uparrow0}=-\frac{i(k^{2}-\alpha_{d0}^{2})}{\alpha_{d0}\Omega_{0}c},\label{eq:Bthd0_3}\\
E_{z,m0} & = & -\frac{ik}{\alpha_{m0}},\label{eq:Ezm0_3}\\
B_{\theta,m0} & = & \frac{i(k^{2}-\alpha_{m0}^{2})}{\alpha_{m0}\Omega_{0}c}.\label{eq:Bthm0_3}
\end{eqnarray}

\end{subequations}The expressions for the $O(g)$ fields at each
layer are also cumbersome; we only show the electric field in the
organic layer, as it is the one associated with the coupling with
excitons (see Eq. (\ref{eq:E_BIG_expansion})),
\begin{equation}
\vec{E}_{org}\approx\vec{E}_{org0}+g\vec{E}_{org1}.\label{eq:Eorg_approximation}
\end{equation}
 The first order in $g$ contribution is

\begin{eqnarray}
\vec{E}_{org1} & = & \boldsymbol{E}_{org1}\eta_{0}e^{-\alpha_{org0}z},\label{eq:E_org_1}
\end{eqnarray}
where $\boldsymbol{E}_{org1}$ is purely tangential and purely-imaginary
valued,

\begin{eqnarray}
\boldsymbol{E}_{org1}\cdot\hat{\boldsymbol{\theta}} & = & -i\Omega_{0}^{2}\Big(\alpha_{m0}\epsilon_{d}\{a[\chi_{0}^{2}(\alpha_{d0}-\alpha_{org0})+\alpha_{d0}+\alpha_{org0}]-\chi_{0}^{2}+1\}\nonumber \\
 &  & +\epsilon_{m0}\{\chi_{0}^{2}[a\alpha_{d0}(\alpha_{org0}-\alpha_{d0})+\alpha_{org0}]+a\alpha_{d0}(\alpha_{d0}+\alpha_{org0})-\alpha_{org0}\}\Big)/\nonumber \\
 &  & 2\alpha_{m0}\epsilon_{d}[\chi_{0}^{2}(\alpha_{m0}-\alpha_{d0})(\alpha_{d0}-\alpha_{org0})+(\alpha_{d0}+\alpha_{m0})(\alpha_{d0}+\alpha_{org0})].\label{eq:Eorg_1_val}
\end{eqnarray}
Compared with Eq. (\ref{eq:Ed1}), the $z$-dependence of $\vec{E}_{org1}$
is purely exponential, as the lowest order correction of $\alpha_{org}$
to $\alpha_{org0}$ is $O(g^{2})$.

\subsection{Quantization and normalization of modes}

Equipped with these results, we use Eqs. (\ref{eq:constraint1-1})--(\ref{eq:constraint6-2})
to compile expressions for the fields in each layer. First, we aim
to compute the vertical normalization length $L_{\boldsymbol{k}}$
for each mode. In analogy to Eq. (\ref{eq:C_w}), we obtain,

\begin{eqnarray}
L_{\boldsymbol{k}} & = & \sum_{i}\Bigg[\epsilon_{0}\epsilon_{org}|E_{i,org}|^{2}+\frac{1}{\mu_{0}\mu}|B_{i,org}|^{2}\Bigg]\frac{1}{\epsilon_{0}}\int_{a}^{\infty}dze^{-(\alpha_{org}^{*}+\alpha_{org})z}\nonumber \\
 &  & +\sum_{ij}\sum_{\gamma,\delta\in\{+,-\}}\sum_{u,v\in\{\uparrow,\downarrow\}}(t_{MOu}^{\gamma})^{*}t_{MOv}^{\delta}\Bigg[\epsilon_{0}(\epsilon_{MO,ij}E_{j,MO}^{\gamma})^{*}E_{i,MO}^{\delta}+\frac{1}{\mu_{0}\mu}(B_{i,MO}^{\gamma})^{*}B_{i,MO}^{\delta}\Bigg]\nonumber \\
 &  & \,\,\,\,\,\,\,\,\,\,\,\,\,\,\,\,\,\,\,\,\,\,\,\,\,\,\,\,\,\times\frac{1}{\epsilon_{0}}\int_{0}^{a}dze^{[(\mbox{sgn}(u)(\alpha_{MO}^{\gamma})^{*}+\mbox{sgn}(v)\alpha_{MO}^{\delta}]z}\nonumber \\
 &  & +\sum_{i}\Bigg[\epsilon_{0}\frac{d(\omega\epsilon_{m}(\omega))}{d\omega}|E_{i,m}|^{2}+\frac{1}{\mu_{0}\mu}|B_{i,m}|^{2}\Bigg]\frac{1}{\epsilon_{0}}\int_{-\infty}^{0}dze^{(\alpha_{m}^{*}+\alpha_{m})z},\label{eq:C_k_3_layers}
\end{eqnarray}
where $\mbox{sgn}(\uparrow)=-\mbox{sgn}(\downarrow)=1$ denote the
exponentially increasing or decreasing fields, respectively. At $O(g)$,
$L_{\boldsymbol{k}}\approx L_{\boldsymbol{k}0}$ as in Eq. (\ref{eq:Lk_calculation}),
and the anisotropy of the MO layer is unimportant. Carrying out the
integrations explicitly (and taking care of the possibility that $\alpha_{d0}$
is complex-valued),

\begin{eqnarray}
L_{\boldsymbol{k}0} & = & \sum_{i}\Bigg[\epsilon_{0}\epsilon_{org}|E_{i,org0}|^{2}+\frac{1}{\mu_{0}\mu}|B_{i,org0}|^{2}\Bigg]\frac{e^{-2\Re\alpha_{org0}a}}{2\epsilon_{0}\Re\alpha_{org0}}\nonumber \\
 &  & +4\sum_{i}\Big|t_{d\uparrow0}\Big|^{2}\Bigg[\epsilon_{0}\epsilon_{d}|E_{i,d\uparrow0}|^{2}+\frac{1}{\mu_{0}\mu}|B_{i,d\uparrow0}|^{2}\Bigg]\Bigg(\frac{e^{2\Re\alpha_{d0}a}-1}{2\epsilon_{0}\Re\alpha_{d0}}\Bigg)\nonumber \\
 &  & +4\sum_{i}\Big|t_{d\downarrow0}\Big|^{2}\Bigg[\epsilon_{0}\epsilon_{d}|E_{i,d\downarrow0}|^{2}+\frac{1}{\mu_{0}\mu}|B_{i,d\downarrow0}|^{2}\Bigg]\Bigg(\frac{1-e^{-2\Re\alpha_{d0}a}}{2\epsilon_{0}\Re\alpha_{d0}}\Bigg)\nonumber \\
 &  & +\Bigg\{4\sum_{ij}(t_{d\uparrow0})^{*}(t_{d\downarrow0})\Bigg[\epsilon_{0}\sum_{i}\epsilon_{d}(E_{i,d\uparrow0})^{*}E_{i,d\downarrow0}+\frac{1}{\mu_{0}\mu}(B_{i,d\uparrow0})^{*}B_{i,d\downarrow0}\Bigg]\Bigg(\frac{a}{\epsilon_{0}}\Bigg)+\mbox{c.c.}\Bigg\}\nonumber \\
 &  & +\sum_{i}\Bigg[\epsilon_{0}\frac{d(\omega\epsilon_{m}(\omega))}{d\omega}|E_{i,m0}|^{2}+\frac{1}{\mu_{0}\mu}|B_{i,m0}|^{2}\Bigg]\frac{1}{2\epsilon_{0}\Re\alpha_{m0}},\label{eq:C_k_3_layers-1}
\end{eqnarray}
One may numerically compute $L_{\boldsymbol{k}}$ by plugging Eqs.
(\ref{eq:td0_pm}), (\ref{eq:tu0_pm}), and (\ref{eq:Eorg0_3})--(\ref{eq:Bthm0_3})
into Eq. (\ref{eq:C_k_3_layers-1}). We do not display the resulting
analytical expression, which anyhow, is lengthy and not particularly
illuminating.

\section{Exciton-exciton and exciton-SP couplings\label{sec:Coupling-of-excitons}}

In Secs. \ref{sec:Interface-between-a} and \ref{sec:3layers}, we
solved for the electromagnetic profile of the magneto-SP modes in
a two- and three-layer setup. We are now ready to describe the organic
superlattice. We regard the latter to be either ``embedded'' in
the MO layer (in the two layer setup) or in its separate third layer
(in the three-layer one). As explained in the following paragraphs,
the superlattice consists of a monoclinic array of organic aggregate
nanopillars. For simplicity, we take each of the nanopillars to be
a rectangular parallelepiped of volume $W_{x}W_{y}W_{z}$ (here, $W_{i}$
is the width of the nanopillar along the $i$-th axis). If the nanopillar
density is $\rho_{np}$, it contains $N_{np}=\rho_{np}W_{x}W_{y}W_{z}$
chromophores. Furthermore, the three-dimensional positions of the
individual chromophores constituting each nanopillar are denoted,

\begin{equation}
\boldsymbol{r}_{\boldsymbol{m}s}=\underbrace{(m_{x}\delta_{x}\hat{\boldsymbol{x}}+m_{y}\delta_{y}\hat{\boldsymbol{y}})}_{\equiv\boldsymbol{r}_{\boldsymbol{m}}}+\underbrace{s\delta_{z}}_{\equiv z_{s}}\hat{\boldsymbol{z}},\label{eq:rms}
\end{equation}
where $\delta_{i}$ is the spacing between chromophores along the
$i$-th direction,

\subsection{Dipolar couplings between nanopillars\label{sub:Dipolar-couplings-between}}

We shall first study the energetic contribution due to the excitons
alone. We model this as,

\begin{equation}
H_{exc}=\sum_{\boldsymbol{n},s}\omega_{\boldsymbol{n}}\sigma_{\boldsymbol{n}}^{\dagger}\sigma_{\boldsymbol{n}}+\sum_{\boldsymbol{n}\neq\boldsymbol{n'}}(J_{\boldsymbol{n}\boldsymbol{n'}}\sigma_{\boldsymbol{n}}^{\dagger}\sigma_{\boldsymbol{n'}}+\mbox{h.c.}),\label{eq:H_exc}
\end{equation}
where $\omega_{\boldsymbol{n}}$ is the bare energy of the $\boldsymbol{n}$-th
collective nanopillar dipole, which we take in the ideal case to be
$\omega_{\boldsymbol{n}}=\bar{\omega}$. The exciton hopping amplitude
between the $\boldsymbol{n}$-th and $\boldsymbol{n'}$-th nanopillars
is approximated as a near-field dipolar coupling,

\begin{equation}
J_{\boldsymbol{n}\boldsymbol{n'}}=\frac{\eta}{\epsilon|\boldsymbol{r}_{\boldsymbol{n}}-\boldsymbol{r}_{\boldsymbol{n'}}|^{3}}\Big[\boldsymbol{\mu}_{\boldsymbol{n}}\cdot\boldsymbol{\mu}_{\boldsymbol{n'}}-3(\boldsymbol{\mu}_{\boldsymbol{n}}\cdot\boldsymbol{e}{}_{\boldsymbol{nn'}})(\boldsymbol{\mu}_{\boldsymbol{n'}}\cdot\boldsymbol{e}{}_{\boldsymbol{nn'}})\Big],\label{eq:J_nn'}
\end{equation}
where $\eta=0.625\,\mbox{meV}(\mbox{nm}^{3}/\mbox{D}^{2})$, $\boldsymbol{e}_{\boldsymbol{nn'}}=\frac{\boldsymbol{r}_{\boldsymbol{n}}-\boldsymbol{r}_{\boldsymbol{n'}}}{|\boldsymbol{r}_{\boldsymbol{n}}-\boldsymbol{r}_{\boldsymbol{n'}}|}$,
where $\boldsymbol{r}_{\boldsymbol{n}}$ is the average in-plane location
of the $\boldsymbol{n}$-th nanopillar, and we take $\epsilon\approx1$,
the dielectric permittivity in the medium surrounding the nanopillars.
Eq. (\ref{eq:J_nn'}) implicitly relies on a separation of energy
scales, namely, that the coupling between chromophores is much stronger
within a single nanopillar than between different ones. Hence, we
start with the collective superradiant nanopillar transitions which
scale as $\boldsymbol{\mu}_{\boldsymbol{n}}=\sum_{\boldsymbol{m}s}\boldsymbol{p}_{\boldsymbol{m}s}\approx\sqrt{N_{np}}\boldsymbol{p}_{\boldsymbol{n}}$,
where $\boldsymbol{p}_{\boldsymbol{m}s}$ is the transition dipole
moment of the $\boldsymbol{m}s$-th chromophore in the nanopillar,
the sum is over all $\boldsymbol{m}s$ values associated with the
$\boldsymbol{n}$-th nanopillar, and $\boldsymbol{p}_{\boldsymbol{m}s}=\boldsymbol{p}_{\boldsymbol{n}}$,
that is, we take the dipole to be equal for all chromophores within
the corresponding nanopillar. This approximation should provide a
semiquantitative description of the dispersion of the organic superlattice
alone. A more refined description would rely on the coupled-dipole
method \cite{devoe1,devoe2,valleau_electromagnetic}, but is beyond
the scope of this work, as this simplified model illustrates the essence
of the problem.

\begin{figure}
\begin{centering}
\includegraphics[scale=0.4]{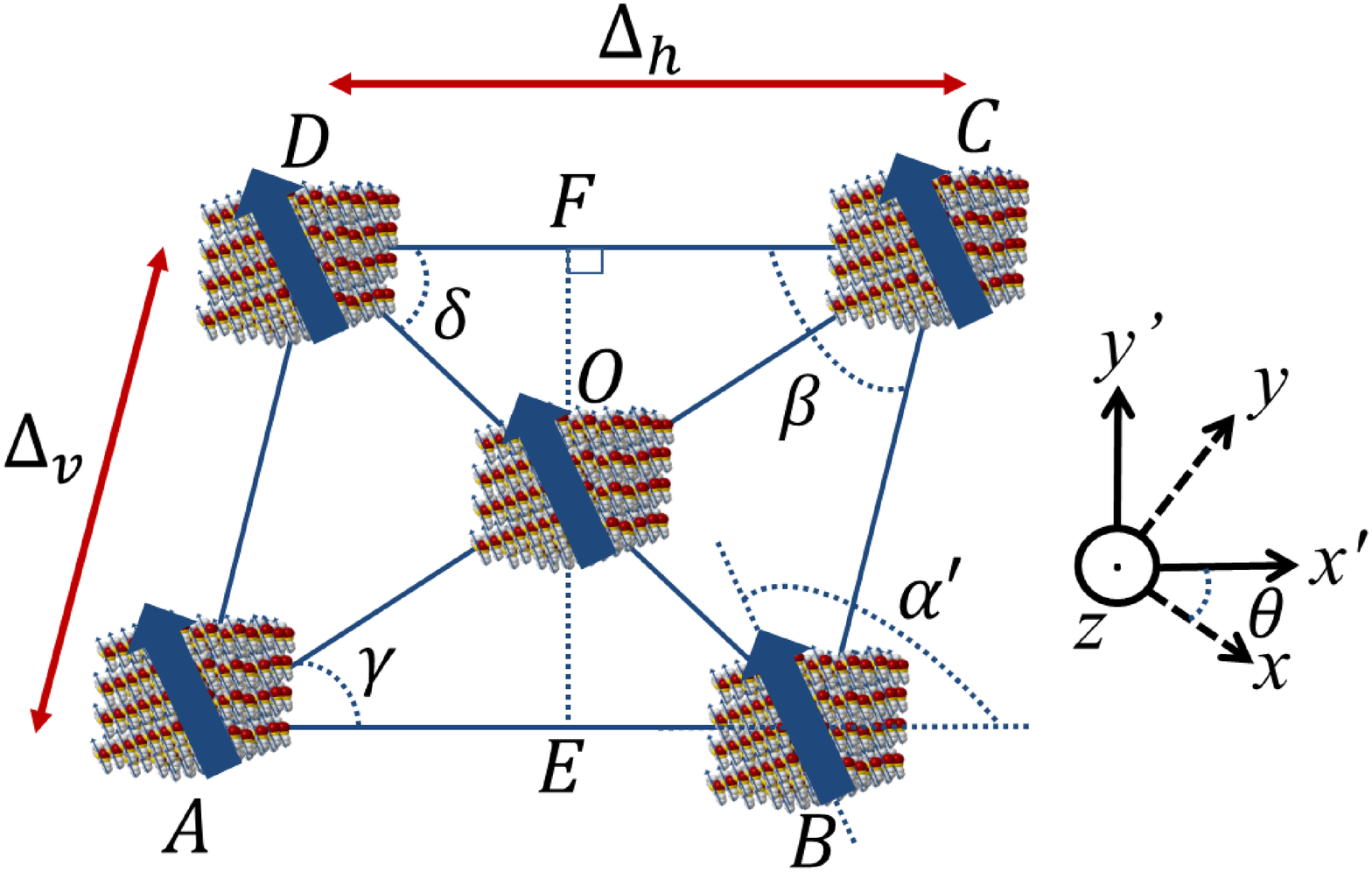}
\par\end{centering}

\protect\caption{\emph{Geometry of two-dimensional monoclinic superlattice of organic
nanopillars.} $x'y'$ denotes a temporary Cartesian coordinate system
which is rotated at $\theta$ from the original $xy$ system. The
collective transition dipole moments of the nanopillars make an angle
$\alpha'$ with respect to $\hat{\boldsymbol{x}}\boldsymbol{'}$ (or
$\alpha\equiv\alpha'+\theta$ with respect to $\hat{\boldsymbol{x}}$).
The horizontal and vertical distances $\Delta_{h}$ and $\Delta_{v}$,
together with the angles $\beta$, $\gamma$, and $\delta$ fully
define the superlattice. \label{fig:unit_cell_oblique}}
\end{figure}

We shall now consider a general two-dimensional monoclinic superlattice
with unit cell defined by vectors $\vec{OD}$ and $\vec{OC}$ depicted
in Fig. \ref{fig:unit_cell_oblique}. For convenience, we temporarily
adopt the $\theta$-rotated coordinate system $x'y'$ which, with
respect to the original $xy$ system, is defined by,

\begin{equation}
\Bigg[\begin{array}{c}
\hat{\boldsymbol{x}}\\
\hat{\boldsymbol{y}}
\end{array}\Bigg]=\Bigg[\begin{array}{cc}
\mbox{cos}\theta & -\mbox{sin}\theta\\
\mbox{sin}\theta & \mbox{cos}\theta
\end{array}\Bigg]\Bigg[\begin{array}{c}
\hat{\boldsymbol{x}'}\\
\hat{\boldsymbol{y'}}
\end{array}\Bigg].\label{eq:rotation_coordinates}
\end{equation}
We will later explain how to obtain $\theta$. We take two sides of
the parallelogram ($AB$ and $CD$) to be parallel to $\hat{\boldsymbol{x'}}$.
For simplicity, the nanopillars are taken to be rectangular parallelepipeds.
Their transition dipoles $\boldsymbol{\mu}_{\boldsymbol{n}}=\boldsymbol{\mu}$
are fixed in the $x'y'$ plane and make an angle $\alpha'$ with respect
to $\hat{\boldsymbol{x'}}$ (or $\alpha\equiv\alpha'+\theta$ with
respect to $\hat{\boldsymbol{x}}$). Notice that all sites are equivalent.
We only account for nearest-neighbor (NN) and next-nearest-neighbor
interactions (NNN). We classify the interactions as horizontal NNN
($AB$, $CD$), vertical NNN ($AD$, $BC$), diagonal type A NN ($OA$,
$OC$), and diagonal type B NN ($OB$,$OD$), respectively,

\begin{subequations}

\begin{eqnarray}
J_{h} & = & \eta\mu^{2}\frac{(1-3\mbox{cos}^{2}\alpha')}{\Delta_{h}^{3}},\label{eq:Jx}\\
J_{v} & = & \eta\mu^{2}\frac{[1-3\mbox{cos}^{2}(\alpha'-\beta)]}{\Delta_{v}^{3}},\label{eq:Jy}\\
J_{diagA} & = & \eta\mu^{2}\frac{[1-3\mbox{cos}^{2}(\alpha'-\gamma)]}{\Big(\frac{\Delta_{h}^{2}+2\Delta_{h}\Delta_{v}\mbox{cos\ensuremath{\beta}}+\Delta_{v}^{2}}{4}\Big)^{3/2}},\label{eq:Jdiag}\\
J_{diagB} & = & \eta\mu^{2}\frac{[1-3\mbox{cos}^{2}(\alpha'+\delta)]}{\Big(\frac{\Delta_{h}^{2}-2\Delta_{h}\Delta_{v}\mbox{cos\ensuremath{\beta}}+\Delta_{v}^{2}}{4}\Big)^{3/2}}.\label{eq:JdiagB}
\end{eqnarray}

\end{subequations}\noindent Here, $|\boldsymbol{\mu}|=\mu$ and the
side lengths are $\overline{AB}=\overline{CD}=\Delta_{h}$ and $\overline{BC}=\overline{AD}=\Delta_{v}$.
We have also conveniently introduced the angles $\beta\equiv\angle BCD=\angle DAB$,
as well as the following,

\begin{subequations}\label{eq:gamma,delta}

\begin{eqnarray}
\gamma & = & \mbox{atan}\frac{\Delta_{v}\mbox{sin}\beta}{\Delta_{h}+\Delta_{v}\mbox{cos}\beta},\label{eq:gamma}\\
\delta & = & \mbox{atan}\frac{\Delta_{v}\mbox{sin}\beta}{\Delta_{h}-\Delta_{v}\mbox{cos}\beta}.\label{eq:delta}
\end{eqnarray}

\end{subequations}

Assuming that all site energies are equal, $\omega_{\boldsymbol{n}}=\bar{\omega}$,
we may rewrite Eq. (\ref{eq:H_exc}) in $\boldsymbol{k}$ space, $H_{exc}=\sum_{\boldsymbol{k}}H_{exc,\boldsymbol{k}}$,
where $H_{exc,\boldsymbol{k}}=\omega_{exc,\boldsymbol{k}}\sigma_{\boldsymbol{k}}^{\dagger}\sigma_{\boldsymbol{k}}$,
where $\boldsymbol{k}=k_{x'}\hat{\boldsymbol{x}'}+k_{y'}\hat{\boldsymbol{y}'}$
and

\begin{eqnarray}
\omega_{exc,\boldsymbol{k}} & = & \bar{\omega}+2J_{h}\mbox{cos}(k_{x'}\Delta_{h})+2J_{v}\mbox{cos}\Bigg[k_{x'}\Delta_{v}\mbox{cos}\beta+k_{y'}\Delta_{v}\mbox{sin}\beta\Bigg]\nonumber \\
 &  & +2J_{diagA}\mbox{cos}\Bigg[k_{x'}\frac{\Delta_{h}+\Delta_{v}\mbox{cos}\beta}{2}+k_{y'}\frac{\Delta_{v}\mbox{sin\ensuremath{\beta}}}{2}\Bigg]\nonumber \\
 &  & +2J_{diagB}\mbox{cos}\Bigg[k_{x'}\frac{\Delta_{h}-\Delta_{v}\mbox{cos}\beta}{2}-k_{y'}\frac{\Delta_{v}\mbox{sin\ensuremath{\beta}}}{2}\Bigg]\label{eq:E_exc_k}
\end{eqnarray}
is the resulting dispersion relation for the excitons alone.

As explained in the main text, we would ideally like to design a plexciton
dispersion which features a global gap in the bulk. This requires
a superlattice with an ``H-aggregate'' dispersion along all wavevector
directions. Mathematically, this means that the dispersion $\omega_{exc,\boldsymbol{k}}$
should be a maximum at $\boldsymbol{k}=0$. It turns out that this
is not possible in a rectangular lattice ($\beta=\frac{\pi}{2}$),
as the resulting J-couplings ($J_{i}<0$) arising from the geometric
constraints end up dominating the H-couplings ($J_{i}>0$) at least
along one direction, yielding a minimum, or at best a saddle point
for $\omega_{exc,\boldsymbol{k}}$ at $\boldsymbol{k}=0$. Hence,
we proceed in a more systematic fashion. Taylor expanding Eq. (\ref{eq:E_exc_k})
up to quadratic order in $k_{i'}$,

\begin{eqnarray}
\omega_{exc,\boldsymbol{k}} & \approx & \bar{\omega}_{eff}'+[\begin{array}{cc}
k_{x'} & k_{y'}\end{array}]\Bigg[\begin{array}{cc}
M_{x'x'} & M_{x'y'}\\
M_{y'x'} & M_{y'y'}
\end{array}\Bigg]\Bigg[\begin{array}{c}
k_{x'}\\
k_{y'}
\end{array}\Bigg],\label{eq:Taylor}
\end{eqnarray}
where the constant offset is obtained by evaluating $\omega_{exc,\boldsymbol{k}}$
at $\boldsymbol{k}=0$,

\begin{equation}
\bar{\omega}_{eff}'=\bar{\omega}+2J_{h}+2J_{v}+2J_{diagA}+2J_{diagB}.\label{eq:zeroth_order_Eo'}
\end{equation}
Here, $M_{i'j'}=\frac{1}{2}\frac{\partial^{2}E_{exc,\boldsymbol{k}}}{\partial k_{i'}\partial k_{j'}}\Bigg|_{\boldsymbol{k}=0}$
denotes a Hessian matrix, which can be readily diagonalized,

\begin{equation}
\Bigg[\begin{array}{cc}
M_{x'x'} & M_{x'y'}\\
M_{y'x'} & M_{y'y'}
\end{array}\Bigg]=\Bigg[\begin{array}{cc}
S_{xx'} & S_{yx'}\\
S_{xy'} & S_{yy'}
\end{array}\Bigg]\Bigg[\begin{array}{cc}
m_{x} & 0\\
0 & m_{y}
\end{array}\Bigg]\Bigg[\begin{array}{cc}
S_{xx'} & S_{xy'}\\
S_{yx'} & S_{yy'}
\end{array}\Bigg].\label{eq:diagonalization}
\end{equation}
where $S_{ij'}$ is a unitary matrix, and ultimately yields the result,

\begin{subequations}

\begin{eqnarray}
\omega_{exc,\boldsymbol{k}} & \approx & \bar{\omega}_{eff}'+m_{x}k_{x}^{2}+m_{y}k_{y}^{2}\label{eq:E_exck_approx_first_row}\\
 & \approx & \bar{\omega}_{eff}+2J_{x}\mbox{cos}k_{x}\Delta_{x}+2J_{y}\mbox{cos}k_{y}\Delta_{y},\label{eq:E_exck_approx}
\end{eqnarray}

\end{subequations}\noindent where $k_{i}=S_{ix'}k_{x'}+S_{iy'}k_{y'}$,
and for our simulation, we choose (arbitrary) effective unit cell
dimensions $\Delta_{i}$ such that $J_{i}=-\frac{m_{i}}{\Delta_{i}^{2}}$
for $i=x,y$, and $\bar{\omega}_{eff}'=\bar{\omega}_{eff}+2J_{x}+2J_{y}$;
this identification renders Eqs. (\ref{eq:E_exck_approx_first_row})
and (\ref{eq:E_exck_approx}) equal up to quadratic order in $k_{i}$.
We have thus arrived at a very convenient expression; Eq. (\ref{eq:E_exck_approx})
shows that the oblique lattice renders the same long-wavelength physics
as a much simpler rectangular lattice with only NN interactions. This
approximation is insightful in that it exposes the physical origin
of the global gap; it also remains valid for our purposes as the topological
phenomena of our interest occurs at small $k_{i}$.

For a fixed value of $\Delta_{h}$, we Monte Carlo sample through
the parameters $r\equiv\frac{\Delta_{v}}{\Delta_{h}}\in[0,2]$, $\alpha\in[0,\pi]$,
and $\beta\in[0,\pi]$ and record those which yield $m_{x},m_{y}<0$.
We observe that only \textasciitilde{}8\% of the parameter space satisfies
the H-aggregate condition we are looking for. One such set of parameters
is $\alpha'=1.19(\sim68.5\,\mbox{degrees})$, $\beta=0.23(\sim13.1\,\mbox{degrees})$,
$r=1.2$, yielding $(m_{x},m_{y})=-(1.49,0.44)J_{0}\Delta_{h}^{2}$,
where $J_{0}\equiv\frac{\eta\mu^{2}}{\Delta_{h}^{3}}$ sets the energy
scale of the dipolar interactions. The associated eigenvector matrix
is,

\begin{eqnarray}
\Bigg[\begin{array}{cc}
S_{xx'} & S_{yx'}\\
S_{xy'} & S_{yy'}
\end{array}\Bigg] & = & \Bigg[\begin{array}{cc}
-0.91 & -0.41\\
0.41 & -0.91
\end{array}\Bigg]\nonumber \\
 & = & \Bigg[\begin{array}{cc}
\mbox{cos}\theta & -\mbox{sin}\theta\\
\mbox{sin}\theta & \mbox{cos}\theta
\end{array}\Bigg]\label{eq:rot_matrix-1}
\end{eqnarray}
where we obtain $\theta=2.72(\sim155.7^{o})$, which defines the angle
of rotation of our temporary coordinates $x'y'$ with respect to the
original ones $xy$; then, $\alpha=\alpha'+\theta=3.9(\sim223^{o})$.
Now, each nanopillar has a collective transition dipole moment value
of $\mu_{0}=\sqrt{N_{np}}|\boldsymbol{p}_{\boldsymbol{n}}|=\sqrt{N_{np}}\times10\,\mbox{D}=\sqrt{W_{x}W_{y}W_{z}\rho_{np}}\times10\,\mbox{D}$.
Choosing the nanopillars to be separated from one another by $\Delta_{h}$,
we get a value for the energy scale of $J_{0}=\frac{\eta W_{x}W_{y}W_{z}\rho_{np}}{\Delta_{h}^{3}}\times100\,\mbox{D}^{2}$.
Taking $\rho_{np}=37\,\mbox{molecules/nm}^{3},$ we obtain,

\begin{eqnarray*}
J_{x}=-\frac{m_{x}}{\Delta_{x}^{2}} & = & 3.45\times10^{3}\,\mbox{meV}\times\zeta_{x}\\
J_{y}=-\frac{m_{y}}{\Delta_{y}^{2}} & = & 1.02\times10^{3}\,\mbox{meV}\times\zeta_{y}
\end{eqnarray*}
where $\zeta_{i}=\frac{W_{x}W_{y}W_{z}}{\Delta_{h}\Delta_{i}^{2}}$
are dimensionless ratios which govern the effective dispersion of
the superlattice. By choosing the physically reasonable parameters
$\Delta_{h}=100\,\mbox{nm}$, $W_{x}=7.5\,\mbox{nm}$, $W_{y}=50\,\mbox{nm}$,
$W_{z}=70\,\mbox{nm}$, and $\bar{\omega}_{eff}'=2.15\,\mbox{eV}$
as well as the effective simulation parameters $\Delta_{x}=\Delta_{y}=50\,\mbox{nm}$,
we obtain $\Delta_{v}=r\Delta_{h}=88\,\mbox{nm}$, $\zeta_{x}=\zeta_{y}=0.105$,
$J_{x}=362\,\mbox{meV}$, $J_{y}=107\,\mbox{meV}$, and $\bar{\omega}_{eff}=3.09\,\mbox{eV}$.
We emphasize that $J_{x}$ and $J_{y}$ can have arbitrary values
which depend on our choice of parameters $\Delta_{x}$ and $\Delta_{y}$.
The latter set the spatial resolution of our real space simulations,
and hence, the size of the systems we can computationally study. These
simulations are carried out in order to calculate edge states as well
as understand the effects of disorder.

We note that when choosing parameters, we need to make sure that (a)
the nanopillars do not yuxtapose each other and (b) the number of
chromophores in the organic layer is large enough to achieve considerable
coupling with the SPs (see Eqs. (\ref{eq:J(k)}) and (\ref{eq:Jbar_k})).
Condition (a) is easily checked computationally and graphically. With
respect to condition (b), we note that the surface area of the $ABCD$
parallelogram in Fig. \ref{fig:unit_cell_oblique} is $\Delta_{h}\Delta_{y}\mbox{sin}\beta$.
It contains two nanopillars of surface area $W_{x}W_{y}$. The surface
coverage fraction of the organic layer is hence,
\begin{equation}
f=\frac{2W_{x}W_{y}}{\Delta_{h}\Delta_{v}\mbox{sin}\beta}.\label{eq:surface_fraction}
\end{equation}
For our chosen parameters, $f=0.38$. In general, we need both $f$
and $W_{z}$ to not be very small ($f>\sim0.2$, $W_{z}>\sim40\,\mbox{nm}$).

In the next subsections, we shall work with the effective rectangular
superlattice of $N_{x}\times N_{y}$ nanopillars (where $N_{i}$ is
the number of nanopillars along the $i$-th direction) instead of
the original monoclinic one. This is a good approximation not only
for the interactions between the various nanopillars, but also for
the exciton-SP couplings, as long we use the average density of the
original monoclinic lattice (see Eqs. (\ref{eq:J(k)}) and (\ref{eq:Jbar_k})).

\subsection{Exciton-SP couplings\label{sub:Exciton-magneto-SP-couplings}}

We are now ready to discuss the effective interaction between SPs
and a single nanopillar. Consider the dipole operator $\hat{\boldsymbol{p}}_{\boldsymbol{m}s}=\boldsymbol{p}_{\boldsymbol{m}s}(b_{\boldsymbol{m}s}^{\dagger}+b_{\boldsymbol{m}s})$,
where $b_{\boldsymbol{m}s}^{\dagger}$($b_{\boldsymbol{m}s}$) creates
(annhilates) an exciton at the $\boldsymbol{m}s$-th chromophore of
some nanopillar. The time-independent electric field operator is $\hat{\vec{\mathcal{E}}}'(\boldsymbol{r})\equiv\sum_{\boldsymbol{k}}\sqrt{\frac{\omega(\boldsymbol{k})}{2\epsilon_{0}SL_{\boldsymbol{k}}}}a_{\boldsymbol{k}}\vec{E}'(\boldsymbol{k})+\mbox{h.c.}$,
which results from transforming Eq. (\ref{eq:operator_E}) from the
Heisenberg to the Schrodinger picture by removing the dynamical phases
$e^{-i\omega(\boldsymbol{k})t}$ (see Eqs. (\ref{eq:E_BIG}) and (\ref{eq:E_m})),
i.e., $\vec{E}'(\boldsymbol{k})\equiv\vec{E}(\boldsymbol{k})e^{i\omega(\boldsymbol{k})t}$.
Using Eq. (\ref{eq:Eorg_approximation}) and (\ref{eq:E_org_1}),
the dipolar coupling between the $\boldsymbol{n}$th nanopillar and
the SP modes is given by

\begin{eqnarray}
H_{exc-SP}^{(\boldsymbol{n})} & = & -\sum_{\boldsymbol{m},s}\hat{\boldsymbol{p}}_{\boldsymbol{m}s}\cdot\hat{\vec{\mathcal{E}}}'(\boldsymbol{r})\nonumber \\
 & = & \sum_{\boldsymbol{k},\boldsymbol{m},s}\mathcal{J}_{\boldsymbol{k},\boldsymbol{m}s}a_{\boldsymbol{k}}b_{\boldsymbol{m}s}^{\dagger}e^{i\boldsymbol{k}\cdot\boldsymbol{r_{m}}}+\mbox{h.c.},\label{eq:H_exc_SP_n}
\end{eqnarray}
where the sum over $\boldsymbol{m}s$ is restricted to the chromophores
in the $\boldsymbol{n}$-th nanopillar. We have also used the rotating-wave
approximation to discard far-off-resonant terms of the form $a_{\boldsymbol{k}}\sigma_{\boldsymbol{n}s}$
and $a_{\boldsymbol{k}}^{\dagger}\sigma_{\boldsymbol{n}s}^{\dagger}$.
Using Eq. (\ref{eq:operator_E}), the corresponding coupling is given
by,

\begin{equation}
\mathcal{J}_{\boldsymbol{k},\boldsymbol{m}s}=-\sqrt{\frac{\omega(\boldsymbol{k})}{2\epsilon_{0}SL_{\boldsymbol{k}}}}e^{-\alpha(\boldsymbol{k})z_{s}}\boldsymbol{p}_{\boldsymbol{m}s}\cdot\boldsymbol{E}(\boldsymbol{k},z_{s}),\label{eq:g_k_zs}
\end{equation}
where, depending on whether we use the two-layer or three-layer setup
results, we make the following substitutions,

~

\begin{center}
\begin{tabular}{|c|c|c|}
\hline
 & \multicolumn{2}{c|}{For two-layer setup}\tabularnewline
\hline
\hline
$L_{\boldsymbol{k}}$ & $L_{\boldsymbol{k}0}$ & Eq. (\ref{eq:C_w})\tabularnewline
\hline
$\alpha$ & $\alpha_{d0}$ & Eq. (\ref{eq:alphad0})\tabularnewline
\hline
$\boldsymbol{E}(\boldsymbol{k},z_{s})$ & $\boldsymbol{E}_{MO}(\boldsymbol{k},z_{s})=\boldsymbol{E}_{d0}(\boldsymbol{k})+g\boldsymbol{E}_{d1}(\boldsymbol{k},z_{s})$ & Eqs. (\ref{eq:Ed0}), (\ref{eq:Ed1})\tabularnewline
\hline
\end{tabular}
\par\end{center}

~

\begin{center}
\begin{tabular}{|c|c|c|}
\hline
 & \multicolumn{2}{c|}{For three-layer setup}\tabularnewline
\hline
\hline
$L_{\boldsymbol{k}}$ & $L_{\boldsymbol{k}0}$ & Eq. (\ref{eq:C_k_3_layers-1})\tabularnewline
\hline
$\alpha$ & $\alpha_{org0}$ & Eq. (\ref{eq:alpha_org_0})\tabularnewline
\hline
$\boldsymbol{E}(\boldsymbol{k},z_{s})$ & $\boldsymbol{E}_{org}(\boldsymbol{k},z_{s})=\boldsymbol{E}_{org0}(\boldsymbol{k})+g\boldsymbol{E}_{org1}(\boldsymbol{k},z_{s})$ & Eqs. (\ref{eq:Eorg0_3}), (\ref{eq:Erorg0_3}), (\ref{eq:Ezorg0_3}),
(\ref{eq:Eorg_1_val}).\tabularnewline
\hline
\end{tabular}
\par\end{center}

~

\noindent Eq. (\ref{eq:g_k_zs}) exposes the 3D nature of our problem
with its $z_{s}$ dependence: there is an exponential contribution
$e^{-\alpha(\boldsymbol{k})z_{s}}$ from the SP evanescent field,
and even a linear correction in $z_{s}$ due to $\boldsymbol{E}_{d1}$
for the two-layer setup. In any case, it will prove convenient to
derive an effective 2D description for our model. We have two ways
to do so.

\subsubsection{Mean-field approximation (MFA) \label{sub:Mean-field-approximation-(MFA)}}

Using Eqs. (\ref{eq:H_exc_SP_n}) and (\ref{eq:g_k_zs}),

\begin{subequations}

\begin{eqnarray}
H_{exc-SP}^{(\boldsymbol{n})} & = & -\sum_{\boldsymbol{k}}\sqrt{\frac{\omega(\boldsymbol{k})}{2\epsilon_{0}SL_{\boldsymbol{k}}}}e^{-\alpha(\boldsymbol{k})\bar{z}(\boldsymbol{k})}a_{\boldsymbol{k}}\boldsymbol{p}_{\boldsymbol{m}s}\cdot\Bigg[\sum_{\boldsymbol{m}s}\boldsymbol{E}(\boldsymbol{k},z_{s})e^{i\boldsymbol{k}\cdot(\boldsymbol{r}_{\boldsymbol{m}}-\boldsymbol{r}_{\boldsymbol{n}})-\alpha_{0}(\boldsymbol{k})(z_{s}-\bar{z}(\boldsymbol{k}))}b_{\boldsymbol{m}s}^{\dagger}\Bigg]e^{i\boldsymbol{k}\cdot\boldsymbol{r}_{\boldsymbol{n}}}+\mbox{c.c.}\label{eq:H_exc-SP-first-line}\\
 & \approx & -\sum_{\boldsymbol{k}}\sqrt{\frac{\omega(\boldsymbol{k})}{2\epsilon_{0}SL_{\boldsymbol{k}}}}e^{-\alpha(\boldsymbol{k})\bar{z}(\boldsymbol{k})}\boldsymbol{E}(\boldsymbol{k},\bar{z}(\boldsymbol{k}))\cdot\boldsymbol{\mu_{n}}a_{\boldsymbol{k}}\sigma_{\boldsymbol{n}}^{\dagger}e^{i\boldsymbol{k}\cdot\boldsymbol{r}_{\boldsymbol{n}}}+\mbox{c.c.},\label{eq:H_mol_SP_mean_field}
\end{eqnarray}

\end{subequations}\noindent where we have formally taken $\bar{z}(\boldsymbol{k})$
to be an average ($\boldsymbol{k}$-dependent) vertical position for
the chromophores in the nanopillar (we will discuss how to compute
this parameter later, see Eq. (\ref{eq:constraint in z_bar(k)})),
made the MFA that $e^{i\boldsymbol{k}\cdot(\boldsymbol{r}_{\boldsymbol{m}}-\bar{\boldsymbol{r}})-\alpha_{0}(\boldsymbol{k})(z_{s}-\bar{z})}\approx1$,
assumed that the dipoles $\boldsymbol{p}_{\boldsymbol{m}s}=\boldsymbol{p}_{\boldsymbol{n}}$
for all the chromophores in the $\boldsymbol{n}$-th nanopillar, and
defined the collective exciton operator,
\begin{equation}
\sigma_{\boldsymbol{n}}^{\dagger}=\frac{1}{\sqrt{N_{np}}}\sum_{\boldsymbol{m}s}b_{\boldsymbol{m}s}^{\dagger},\label{eq:modes}
\end{equation}
such that its corresponding transition dipole is superradiantly enhanced
at $\boldsymbol{\mu}_{\boldsymbol{n}}=\sqrt{N_{np}}\boldsymbol{p}_{\boldsymbol{n}}$.

Having addressed the effective interaction between a single nanopillar
and the SP modes, we can move on to the description of the superlattice,
$H_{exc-SP}=\sum_{\boldsymbol{n}}H_{exc-SP}^{(\boldsymbol{n})}$.
If $\boldsymbol{\mu}_{\boldsymbol{n}}=\boldsymbol{\mu}$ and we assume
periodic boundary conditions (PBCs), we can construct Fourier $\boldsymbol{k}$
modes for the excitons too, $\sigma_{\boldsymbol{k}}^{\dagger}=\frac{1}{\sqrt{N_{x}N_{y}}}\sum_{\boldsymbol{n}}\sigma_{\boldsymbol{n}}^{\dagger}e^{i\boldsymbol{k}\cdot\boldsymbol{r}_{\boldsymbol{n}}}$.
Then, we arrive at the Hamiltonian, $H_{exc-SP}=\sum_{\boldsymbol{k}}H_{exc-SP,\boldsymbol{k}}=\sum_{\boldsymbol{k}}\mathcal{J}(\boldsymbol{k})a_{\boldsymbol{k}}\sigma_{\boldsymbol{k}}^{\dagger}+\mbox{h.c.}$,
where

\begin{equation}
\mathcal{J}(\boldsymbol{k})=\sqrt{\Bigg(\frac{N_{x}N_{y}}{S}\Bigg)\Bigg(\frac{\omega(\boldsymbol{k})}{2\epsilon_{0}L_{\boldsymbol{k}0}}\Bigg)}e^{-\alpha_{0}(\boldsymbol{k})\bar{z}(\boldsymbol{k})}\boldsymbol{\mu}\cdot\boldsymbol{E}(\boldsymbol{k}),\label{eq:J(k)}
\end{equation}
and $\boldsymbol{k}$ runs for all the allowed discretized wavevectors
$k_{i}=-\frac{\pi}{\Delta_{i}}+\frac{2\pi}{N_{i}\Delta_{i}}q_{i}$
for $q_{i}=0,1,\cdots,N_{i}-1$.

Within the MFA, we have achieved to represent each nanopillar as a
single collective transition dipole $\boldsymbol{\mu}_{\boldsymbol{n}}$
associated with the operator $\sigma_{\boldsymbol{n}}^{\dagger}$.
There is, however, an ambiguity in this approximation, namely, the
criterion to optimize the parameter $\bar{z}(\boldsymbol{k})$. We
will discuss this in Subsec. (\ref{sub:Comparison}).

\subsubsection{Beyond the MFA}

Going back to Eq. (\ref{eq:H_exc-SP-first-line}) and assuming $\boldsymbol{p}_{\boldsymbol{m}s}=\boldsymbol{p}_{\boldsymbol{n}}$,
we can consider alternative nanopillar modes. We follow González-Tudela,\emph{
et al} \cite{gonzalez_tudela}. Let us define the new modes (distinguished
from the others by the overbar notation),

\begin{equation}
\bar{\sigma}_{\boldsymbol{n}}^{\dagger}(\boldsymbol{k})=\frac{1}{\sqrt{\mathcal{N}}}\sum_{\boldsymbol{m}s}\boldsymbol{p}_{\boldsymbol{n}}\cdot\boldsymbol{E}(\boldsymbol{k},z_{s})e^{i\boldsymbol{k}\cdot(\boldsymbol{r}_{\boldsymbol{m}}-\boldsymbol{r}_{\boldsymbol{n}})-\alpha_{0}(\boldsymbol{k})z_{s}}b_{\boldsymbol{m}s}^{\dagger},\label{eq:new_modes}
\end{equation}
where the corresponding normalization is given by,

\begin{eqnarray}
\mathcal{N} & = & \sum_{\boldsymbol{m}s}\Big|\boldsymbol{p}_{\boldsymbol{n}}\cdot\boldsymbol{E}(\boldsymbol{k},z_{s})e^{i\boldsymbol{k}\cdot(\boldsymbol{r}_{\boldsymbol{m}}-\boldsymbol{r}_{\boldsymbol{n}})-\alpha_{0}(\boldsymbol{k})z_{s}}\Big|^{2}\nonumber \\
 & \approx & \frac{N_{np}}{W_{z}}\int_{z_{0}}^{z_{f}}dz|\boldsymbol{p}_{\boldsymbol{n}}\cdot\boldsymbol{E}(\boldsymbol{k},z)|^{2}e^{-2\Re\alpha_{0}(\boldsymbol{k})z}.\label{eq:normalization_new_modes}
\end{eqnarray}
where we identified $z_{f}-z_{0}=W_{z}$ as the vertical thickness
of each nanopillar. Introducing the corresponding $\boldsymbol{k}$
mode $\bar{\sigma}_{\boldsymbol{k}}^{\dagger}=\frac{1}{\sqrt{N_{x}N_{y}}}\sum_{\boldsymbol{n}}\bar{\sigma}_{\boldsymbol{n}}^{\dagger}(\boldsymbol{k})e^{i\boldsymbol{k}\cdot\boldsymbol{r}}$,
Eq. (\ref{eq:H_exc-SP-first-line}) becomes $H_{exc-SP}=\sum_{\boldsymbol{k}}H_{exc-SP,\boldsymbol{k}}\approx\sum_{\boldsymbol{k}}\bar{\mathcal{J}}(\boldsymbol{k})a_{\boldsymbol{k}}\bar{\sigma}_{\boldsymbol{k}}^{\dagger}+\mbox{h.c.}$,
where,

\begin{equation}
\bar{\mathcal{J}}(\boldsymbol{k})\approx\sqrt{\rho\Bigg(\frac{\omega(\boldsymbol{k})}{2\epsilon_{0}L_{\boldsymbol{k}0}}\Bigg)}\sqrt{\int_{z_{0}}^{z_{f}}dze^{-2\Re\alpha_{0}(\boldsymbol{k})z}|\boldsymbol{p}_{\boldsymbol{n}}\cdot\boldsymbol{E}(\boldsymbol{k},z)|^{2}},\label{eq:Jbar_k}
\end{equation}
where we have identified
\begin{equation}
\rho=\frac{N_{x}N_{y}N_{np}}{SW_{z}}=\underbrace{\frac{(N_{x}W_{x})(N_{y}W_{y})}{S}}_{<1}\rho_{np}\label{eq:rho_average}
\end{equation}
as the \emph{average} density of chromophores in the organic superlattice
which, due to the ``void space'' between nanopillars, is lower than
$\rho_{np}$. Except for a different convention in the phases of our
exciton modes $\bar{\sigma}_{\boldsymbol{k}}^{\dagger}$, this solution
has the same structure as the one presented in \cite{gonzalez_tudela},
even though the latter deals with an organic layer of uniform density.

\subsubsection{Comparison\label{sub:Comparison}}

When we wrote $H_{exc-SP}\approx\sum_{\boldsymbol{k}}\mathcal{J}(\boldsymbol{k})a_{\boldsymbol{k}}\sigma_{\boldsymbol{k}}^{\dagger}+\mbox{h.c.}$,
we made a MFA to Eq. (\ref{eq:H_exc-SP-first-line}) by invoking definitions
for the mode $\sigma_{\boldsymbol{n}}^{\dagger}$ and the coupling
$\mathcal{J}(\boldsymbol{k})$ (Eqs. (\ref{eq:modes}) and (\ref{eq:J(k)})).
The essence of this approximation is the exponential factor $e^{-\alpha_{0}(\boldsymbol{k})\bar{z}(\boldsymbol{k})}$
in $\mathcal{J}(\boldsymbol{k})$, which implies that when each nanopillar
interacts with the $\boldsymbol{k}$-th electromagnetic mode, it behaves
as a collective dipole placed at the effective height $z=\bar{z}(\boldsymbol{k})$.
On the other hand, when going beyond the MFA, we introduced $\bar{\sigma}_{\boldsymbol{n}}^{\dagger}(\boldsymbol{k})$
and $\bar{\mathcal{J}}(\boldsymbol{k})$ (Eqs. (\ref{eq:new_modes})
and (\ref{eq:Jbar_k})) and showed that $H_{exc-SP}\approx\sum_{\boldsymbol{k}}\bar{\mathcal{J}}(\boldsymbol{k})a_{\boldsymbol{k}}\bar{\sigma}_{\boldsymbol{k}}^{\dagger}+\mbox{h.c.}$
is an exact representation of Eq. (\ref{eq:H_exc-SP-first-line})
(notwithstanding the excellent approximations of converting the sums
over chromophores to integrals and the PBCs). Hence, $\bar{\sigma}_{\boldsymbol{k}}^{\dagger}$
(and not $\sigma_{\boldsymbol{k}}^{\dagger}$) is the natural exciton
mode which couples to the SP mode $a_{\boldsymbol{k}}^{\dagger}$.

The solution beyond MFA might be a more convenient description if
one is interested in a careful description of the energetics of the
problem. However, for purposes of the topological characterization
of the plexciton bandstructure, it is more pertinent to adopt the
MFA description. Notice from Eq. (\ref{eq:new_modes}) that $\bar{\sigma}_{\boldsymbol{n}}^{\dagger}(\boldsymbol{k})$
depends explicitly on $\boldsymbol{k}$, so that the Fourier modes
$\bar{\sigma}_{\boldsymbol{k}}^{\dagger}=\frac{1}{\sqrt{N_{x}N_{y}}}\sum_{\boldsymbol{n}}\bar{\sigma}_{\boldsymbol{n}}^{\dagger}(\boldsymbol{k})e^{i\boldsymbol{k}\cdot\boldsymbol{r}}$
have an additional dependence on $\boldsymbol{k}$ beyond the phase
factor $e^{i\boldsymbol{k}\cdot\boldsymbol{r}}$. This introduces
a technicality for the numerical computation of the Chern number for
each plexciton band, which we wish to avoid at present. This complication
does not occur in the MFA, where $\sigma_{\boldsymbol{n}}^{\dagger}$
uniformly sums over the excitons operators $b_{\boldsymbol{m}s}^{\dagger}$
for a given nanopillar regardless of $\boldsymbol{k}$. Thus, we make
a compromise: we formally use the structure of the MFA, but heuristically
make the energetic approximation $\Big|\mathcal{J}(\boldsymbol{k})\Big|=\bar{\mathcal{J}}(\boldsymbol{k})$.
Comparing Eqs. (\ref{eq:J(k)}) and (\ref{eq:Jbar_k}), this identity
requires that $\bar{z}(\boldsymbol{k})$ satisfies,

\begin{equation}
\Bigg|\boldsymbol{p}_{\boldsymbol{n}}\cdot\boldsymbol{E}(\boldsymbol{k},\bar{z}(\boldsymbol{k}))e^{-\alpha_{org0}(\boldsymbol{k})\bar{z}(\boldsymbol{k})}\Bigg|=\sqrt{\frac{1}{W_{z}}\int_{z_{0}}^{z_{f}}dz|\boldsymbol{p}_{\boldsymbol{n}}\cdot\boldsymbol{E}(\boldsymbol{k},z)|^{2}e^{-2\Re\alpha_{org0}(\boldsymbol{k})z}}.\label{eq:constraint in z_bar(k)}
\end{equation}
This constraint has the appealing physical content of computing the
mean-field effective position $\bar{z}(\boldsymbol{k})$ of the collective
nanopillar dipole by averaging interaction of the SP with respect
to the interval $z\in[z_{i},z_{f}]$. Operationally, however, it is
not necessarily to solve for $\bar{z}(\boldsymbol{k})$, as having
the value of $\bar{\mathcal{J}}(\boldsymbol{k})$ in Eq. (\ref{eq:Jbar_k})
suffices for our calculations.

To remind ourselves, when dealing with the dispersion of the organic
layer alone, we have coarse grained it into an effective rectangular
superlattice. However, when computing the exciton-SP coupling via
Eq. (\ref{eq:Jbar_k}), $\rho$ must be taken to be the density of
the original monoclinic superlattice. Eq. (\ref{eq:rho_average})
is undetermined as $N_{x}$ and $N_{y}$ are artificial simulation
parameters, instead, we shall write
\begin{eqnarray}
\rho & = & f\rho_{np},\label{eq:average_density_bis}
\end{eqnarray}
the average density is equal to the surface coverage fraction times
the original nanopillar density.

\subsubsection{Representative coupling values\label{sub:Representative coupling values}}

\begin{figure}
\begin{centering}
\includegraphics[scale=0.4]{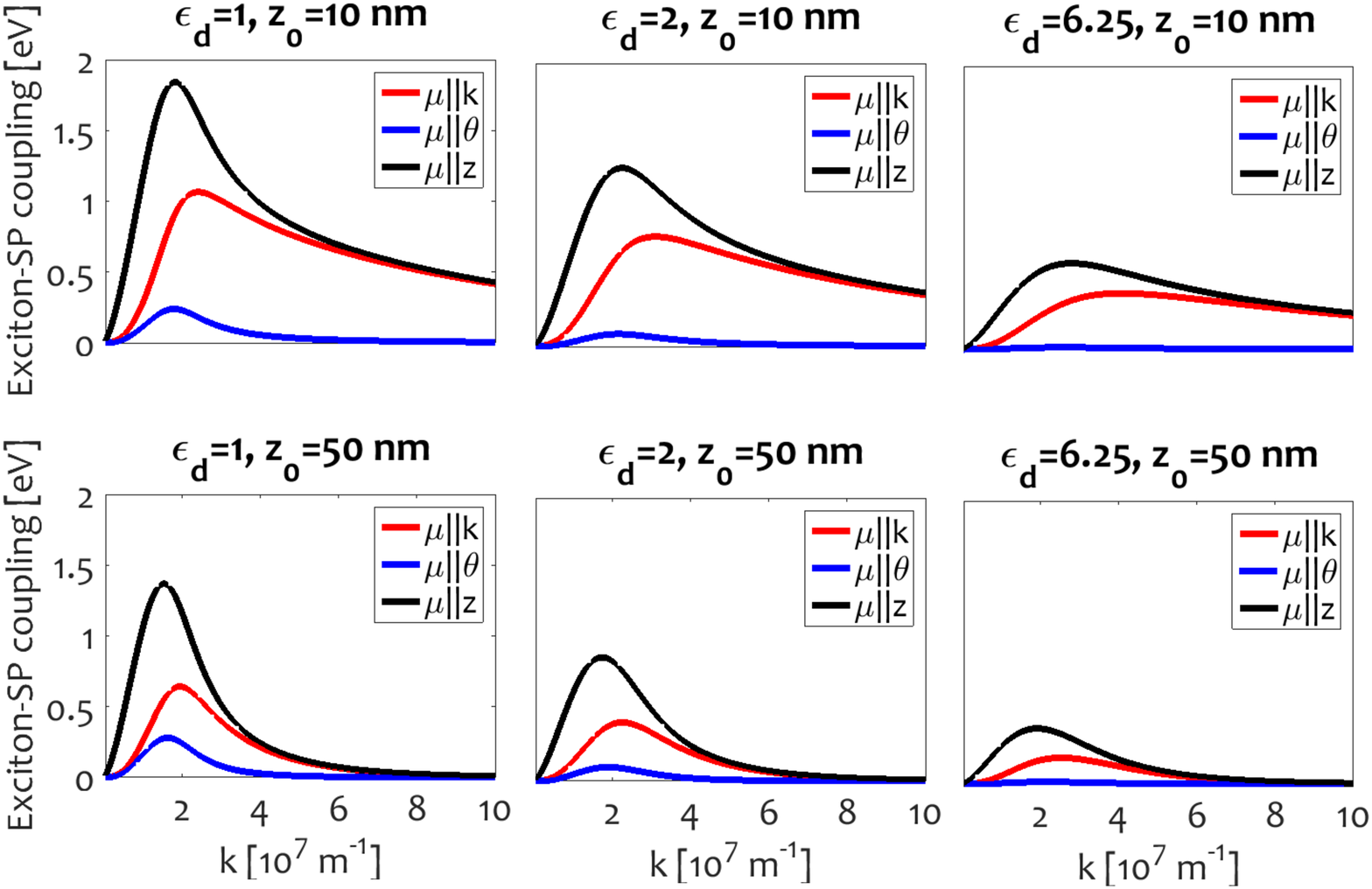}
\par\end{centering}

\protect\caption{\emph{Representative exciton-SP coupling values for two-layer (metal-MO
dielectric) setup.} The calculations have been carried out using Eq.
(\ref{eq:Jbar_k}). We display results for different orientations
of the transition dipoles of the nanopillars, when $\boldsymbol{\mu}\parallel\hat{\boldsymbol{k}}$,
$\boldsymbol{\mu}\parallel\hat{\boldsymbol{\theta}}$, and $\boldsymbol{\mu}\parallel\hat{\boldsymbol{z}}$.
Notice that all the curves have maxima at short wavevectors. The calculations
assume that $\epsilon_{\infty}\sim4$, $\omega_{P}\sim9$~eV, $g=0.3$,
and the height of the nanopillars being $W_{z}=70\,\mbox{nm}$. \label{fig:plots_two_layers}}
\end{figure}

\begin{figure}
\begin{centering}
\includegraphics[scale=0.35]{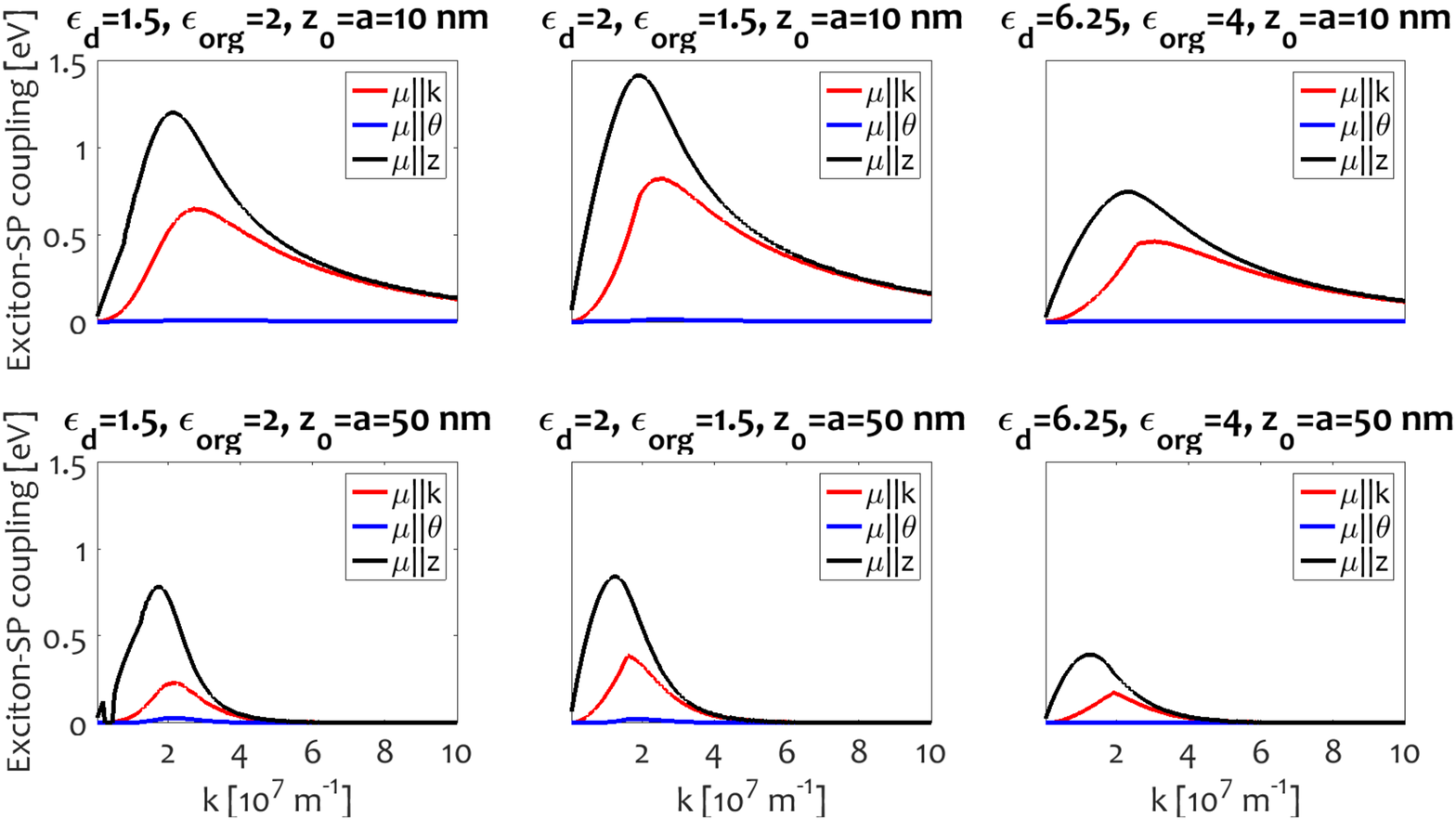}
\par\end{centering}

\protect\caption{\emph{Representative exciton-SP coupling values for three-layer (metal-MO
dielectric-organic) setup.} Just as with Fig. \ref{fig:plots_two_layers},
the calculations have been carried out using Eq. (\ref{eq:Jbar_k}).
We display results for different orientations of the transition dipoles
of the nanopillars, when $\boldsymbol{\mu}\parallel\hat{\boldsymbol{k}}$,
$\boldsymbol{\mu}\parallel\hat{\boldsymbol{\theta}}$, and $\boldsymbol{\mu}\parallel\hat{\boldsymbol{z}}$.
Notice that all the curves have maxima at short wavevectors. The calculations
assume that $\epsilon_{\infty}\sim4$, $\omega_{P}\sim9$~eV, $g=0.3$,
and the height of the nanopillars being $W_{z}=70\,\mbox{nm}$.\label{fig:plots_three_layers}}
\end{figure}

Figs. \ref{fig:plots_two_layers} and \ref{fig:plots_three_layers}
plot representative exciton-SP coupling values $\bar{\mathcal{J}}(\boldsymbol{k})$
using Eq. (\ref{eq:Jbar_k}) for the two-layer and three-layer setups,
respectively. We display calculations for different orientations of
the transition dipoles of the nanopillars, when $\boldsymbol{\mu}\parallel\hat{\boldsymbol{k}}$,
$\boldsymbol{\mu}\parallel\hat{\boldsymbol{\theta}}_{\boldsymbol{k}}$,
and $\boldsymbol{\mu}\parallel\hat{\boldsymbol{z}}$. Throughout the
plots, we have chosen silver Drude parameters $\epsilon_{\infty}\sim4$,
$\omega_{P}\sim9$~eV and $g=0.3$. Each of the panels displays results
corresponding to a particular dielectric permittivity $\epsilon_{d}$
and the base height of the nanopillars $z_{0}$, and fixing the same
organic layer height $W_{z}=70\,\mbox{nm}$. Taking the parameters
for the organic superlattice described in Subsec. \label{sub:Dipolar-couplings-between-1}
and assuming $\rho_{np}=37\,\mbox{chromophores/nm}^{3}$, the average
density using Eq. (\ref{eq:average_density_bis}) is $\rho=14\,\mbox{chromophores/nm}^{3}$.
Fig. \ref{fig:plots_two_layers} corresponds to the physical scenario
where the organic nanopillars are ``embedded'' in the MO dielectric
layer starting at the base height $z_{0}$, while Fig. \ref{fig:plots_three_layers}
assumes that the MO dielectric layer has a width $a$ and the base
height of the nanopillars is also at $a$.

As a reminder, in the absence of the MO effect, the electric field
of the $\boldsymbol{k}$-th SP mode has no tangential component (see
Eqs. (\ref{eq:Ed0}), (\ref{eq:Eorg0_3})). Thus, the blue curves
in the plots ($\boldsymbol{\mu}\parallel\hat{\boldsymbol{\theta}}$)
must vanish identically for $g=0$. For $g\neq0$, these couplings
scale linearly with $g$ (see Eqs. \ref{eq:Ed1}, \ref{eq:Eorg_1_val}),
so it is easy to predict these perturbative exciton-SP couplings for
other values of $g$. On the other hand, the red ($\boldsymbol{\mu}\parallel\hat{\boldsymbol{k}}$)
and black ($\boldsymbol{\mu}\parallel\hat{\boldsymbol{z}}$) curves
are independent of the value of $g$, as they are $O(g^{0})$. Notice
that all the curves peak at some short wavevector $k_{max}$. The
effective site energy $\bar{\omega}_{eff}$ (see Eq. (\ref{eq:zeroth_order_Eo'}))
was optimized so that the exciton dispersion $H_{exc,\boldsymbol{k}}$
and the SP dispersion $H_{SP,\boldsymbol{k}}$ become degenerate (Dirac
points) at wavevectors $\boldsymbol{k}^{*}$ satisfying $|\boldsymbol{k}^{*}|=k_{max}$.
This is carried out so that upon inclusion of the MO effect, the topological
anticrossing generated at the Dirac point $\boldsymbol{k}^{*}$ becomes
as large as possible. The latter implies that one should maximize
$\bar{\mathcal{J}}(\boldsymbol{k}^{*})$ for $\boldsymbol{\mu}\parallel\hat{\boldsymbol{\theta}}_{\boldsymbol{k}}$.

The simulations displayed in the main text correspond to the first
(upper left corner) panel of Fig. \ref{fig:plots_two_layers}, where,
yielding $\bar{\mathcal{J}}(\boldsymbol{k}^{*})=0.24\,\mbox{eV}$
for $\boldsymbol{\mu}\parallel\hat{\boldsymbol{\theta}}$, or equivalently,
a topological anticrossing gap of $2\bar{\mathcal{J}}(\boldsymbol{k}^{*})=0.48\,\mbox{eV}$.
This value becomes reasonably large compared to the linewidths of
the exciton and SP modes even if we reduce $g$ by a factor of three
or four. Note that quite often, the largest couplings occur when $\boldsymbol{\mu}\parallel\hat{\boldsymbol{z}}$,
reaching values which are comparable to the exciton site energies.
This regime, known as ultra-strong coupling \cite{PhysRevLett.105.196402},
is interesting in its own right and gives rise to novel effects, which
are beyond the scope of our work. Unfortunately, for our purposes,
we cannot exploit these large couplings, as they do not vanish for
any $\boldsymbol{k}$ and hence, does not yield Dirac points (see
main text).

A few interesting trends can be obtained from scanning through $\epsilon_{d}$
and $z_{0}$; some of these results are displayed in Figs. \ref{fig:plots_two_layers}
and \ref{fig:plots_three_layers}. First, couplings decrease as $z_{0}$
increases. This is not surprising, as owing to the evanescent nature
of the SP fields, the coupling should be strongest for the chromophores
that are closest to the interface at $z=0$. Second, couplings decrease
as the dielectric $\epsilon_{d}$ increases. Finally, notice that
the three-layer setup yields very weak values of $\bar{\mathcal{J}}(\boldsymbol{k}^{*})$
for $\boldsymbol{\mu}\parallel\hat{\boldsymbol{\theta}}_{\boldsymbol{k}}$.
We believe that these two problems are related to index mismatch between
the various interfaces. As mentioned in the main text, one possibility
to ameliorate this problem is to embed the MO material inside of a
low dielectric polymer.

An alternative solution to increase $\bar{\mathcal{J}}(\boldsymbol{k}^{*})$
for $\boldsymbol{\mu}\parallel\hat{\boldsymbol{\theta}}_{\boldsymbol{k}}$
is to use materials with large $g$ values at the UV/visible, which
is what we need (recall the crossing between the SP and exciton dispersions
at $\boldsymbol{k}^{*}$ happens at 3.1 eV in our calculation). Some
examples of the latter are Co alloy films \cite{temnov2010active},
orthoferrites \cite{orthoferrites}, or spinels \cite{spinels}. A
caveat about the latter is that they are also highly absorptive at
those same wavelengths (large imaginary part of $\epsilon_{d}$, which
we have neglected in this work). Ce substituted YIG has less of a
problem in that regard \cite{ross_acsphotonics}. We are currently
addressing all these possibilities, including different stacking geometries,
in order to induce strong MO effects.

\end{document}